# ANALYSIS OF DOWNWARD TERRESTRIAL GAMMA-RAY FLASHES USING A LARGE-AREA COSMIC RAY SCINTILLATION DETECTOR

by

Jackson Roe Remington

A dissertation submitted to the faculty of
The University of Utah
in partial fulfillment of the requirements for the degree of

Doctor of Philosophy

in

Physics

Department of Physics and Astronomy

The University of Utah

December 2021

# The University of Utah Graduate School

## STATEMENT OF DISSERTATION APPROVAL

This dissertation of

_________________________ Jackson Roe Remington _________________________

has been approved by the following supervisory committee members:

| | | |
|---|---|---|
| _________ John Belz _________ , Chair | _________ 2021 _________ |
| | Date Approved |
| _________ Douglas Bergman _________ , Member | _________ 2021 _________ |
| | Date Approved |
| _________ John Matthews _________ , Member | _________ 2021 _________ |
| | Date Approved |
| _________ Pearl Sandick _________ , Member | _________ 2021 _________ |
| | Date Approved |
| _________ David Schurig _________ , Member | _________ 2021 _________ |
| | Date Approved |

and by _________________ Christoph Boehme _________________ , Chair of

the Department of

_________________ Physics and Astronomy _________________

and by David B. Kieda, Dean of The Graduate School.

# ABSTRACT

The Telescope Array (TA) of central Utah was designed to detect ultra-high-energy cosmic rays (UHECRs) and is the largest of its kind in the Northern hemisphere. Its capabilities, however, are not limited to extra-terrestrial sources. Scintillation detectors (SDs) in the array are designed to measure energy deposit from charged cosmic ray secondaries, but have recently caught bursts of electromagnetic radiation that do not match the typical signature of cosmic ray air showers. After investigation, the bursts were tied to individual lightning strikes. In an effort to better understand the phenomenon, Langmuir Laboratory for Atmospheric Research at New Mexico Tech provided specialized lightning detectors across the array, eventually confirming the events as downward terrestrial gamma-ray flashes (TGFs) produced in the first few microseconds of lightning flashes.

After 20 identified TGFs and 2 publications, the lightning instrumentation at Telescope Array was upgraded in the summer of 2018 with the addition of a broadband interferometer (INTF) for high-resolution lightning mapping. This unique variety of lightning detectors has become one of the leading studies of downward TGFs. The analysis of these data are non-trivial due to the fact that the cosmic ray detectors record only downward particle showers, while the lightning detectors measure only electrical processes during lightning development. In addition, the different detectors have no direct data communication can be separated by up to 32 km.

In order to examine these relationships on a microsecond scale, I successfully developed a method using available time and location data from each detector to correlate TGF particle data with lightning breakdown activity. This dissertation is focused on the application of this method on four events from late 2018 and the results thereof. The individual TGFs lasted $<10$ $\mu$s and likely consisted of $\geq 10^{12}$ gamma-rays, with evidence of some energies $\geq 2.6$ MeV. I determined TGF source times, plan locations, and altitudes with average uncertainties of 0.6 $\mu$s, 140 m, and 25 m, respectively. This high resolution showed that TGFs in all four of these events occured during strong initial breakdown pulses and was driven by streamer-based fast negative breakdown in the first couple milliseconds of negative intracloud and cloud-to-ground lightning flashes.

# CONTENTS

# LIST OF FIGURES

# LIST OF TABLES

# CHAPTER 1

# INTRODUCTION

Perhaps the most obvious display of natural power, lightning has captivated and confounded humanity throughout its history. Whereas other, more energetic phenomena are seen using modern technology, none can so impress upon the naked senses like the flash and thunder of lightning strikes. That said, we have yet to fully understand the spectacle. It has been attributed to magic and mythology, but its true mechanism has remained a mystery until studies of the last few decades.

In May 1994, Fishman et al. (1994) reported a strange sight by the Burst and Transient Source Experiment (BATSE) on board the Compton Gamma Ray Observatory satellite. Rather than the cosmic gamma-ray bursts it was looking for, the experiment caught several brief flashes of energetic photons originating from Earth's atmosphere, earning the name Terrestrial Gamma Flashes (TGFs). Many more TGFs were measured over the next decade, primarily by orbiting gamma observatories. Results have shown that the radiation is sourced from lower altitudes inside thunderstorms, specifically during the initial stages of upward intracloud (IC) lightning.

This discovery revealed a mechanism by which to study the initial stages of lightning and shed light on the mysterious, energetic processes involved. Due to the symmetry of charge distributions in typical storm clouds, it was also theorized that TGFs could be produced downward. The recent observations at the Telescope Array (TA) in Central Utah show that this is indeed the case, and analysis on these events can provide a unique perspective. This dissertation reveals the research done, observations taken, and conclusions reached during the study of these anomalies.

## 1.1   Outline of the Dissertation

This dissertation will begin by introducing the stages, types, and processes of lightning which are important to this study (chapter 2), including a detailed accounting of previous TGF observations and interpretations (chapter 3). Following this will be documentation of the instruments used in this study, including both cosmic ray detectors (chapter 4) and those installed for the sole purpose of studying lightning (chapter 5).

After providing background, the observations at Telescope Array will be presented and discussed in detail in three parts. The first will review the data and conclusions of the previous TGF studies at TA which laid the foundation of the current work (chapter 6). The second will present the observations of this study on four TGF events from 2018 and give a detailed description on the analysis methods which were the focus of my work (chapter 7). Finally, the interpretation of these results will be discussed alongside their implications on TGF production and lightning discharge as a whole (chapter 8).

# CHAPTER 2

# LIGHTNING ANATOMY

Despite millennia of awe and study, the intricacies of lightning remain obscure. The chaotic thunderstorm environment makes direct measurement difficult; strong turbulence, radiation, and electric fields interfere with traditional observations, but breakthroughs of the last few decades have lead to insight into its inner workings. This study focuses on the examination of the powerful radiation emitted during the early stages of lightning breakdown, specifically terrestrial gamma-ray flashes (TGFs), and what they can tell us about lightning in general. However, we must first review the basics of lightning. This chapter introduces thunderstorm structure, electrification, types of electrical breakdown, and the resulting phenomena which are relevant in the study of TGFs.

## 2.1   Dielectric Breakdown

Ionization is the process of stripping (or adding) electrons to a neutral atom, leaving it with a net charge. By this process, a medium can more easily conduct electricity as charge carriers are free to move from one molecule to the next. There are a number of ways ionization can be initiated, including particle collisions and powerful electric fields. Dielectric breakdown involves subjecting an insulator to an electric field strong enough to separate electrons from their nuclei, called the dielectric strength (or dielectric breakdown) of the material. In nature, these electric fields are produced when two regions collect enough opposing charges to overcome a medium's dielectric strength, ionizing a channel between the regions and allowing the charges to neutralize.

Take the trivial example of getting shocked by a doorknob; when your hand develops a slight negative charge and nears a piece of conducting metal, a positive charge is induced and creates a potential difference between the two. As you approach the metal, the electric field increases until it breaches the dielectric strength of the air of ∼3,200 kV/m at sea level, also called conventional breakdown (Rigden (1996)). Electrons are then free to travel along the ionized air into the metal creating a brief, bright shock and neutralizing the charge difference.

At its core, lightning is an up-scaled version of the same process. Typically, it occurs in thunderclouds, but not always. For example, it can occur inside ash clouds created during volcanic activity, where dust grains rub against one another in the extreme turbulence (Arason et al. (2011)). This friction can strip electrons from one particle to the next, separating charge on a large enough scale to exceed the dielectric strength of the air and ash. In storm clouds, charge separation occurs as a result of a special interaction between ice crystals and graupel (Saunders & Brooks (1992)).

Volatile conditions inside thunderclouds makes observation difficult, but the basic mechanisms of storm electrification take place at temperatures between –15° and –25°C (Krehbiel (1986)). Light ice crystals are carried upward by air currents, while heavier, super-cooled droplets and graupel fall downward. As the two phases collide in their movements, the graupel strips outer electrons from the neatly-formed crystals, causing an excess of negative charges below and positive above (Figure 2.1). Because of the temperature dependence of these processes, the negative charge region is relatively constant between clouds and between storms (Krehbiel (1986)). For this study, in Central Utah, negative charge regions tended to develop between 2.5–5.5 km above ground level (4–7-km MSL). These strong charge imbalances also induce positive charge on the Earth's surface due to its relatively high conductivity. Theoretically, when the electric field between two charge regions surpasses the local dielectric strength (conventional breakdown), a channel of air between two regions would fully ionize, short out the potential difference, and cause lightning. However, measurements indicate that these fields do not exist in thunderclouds on a large scale.

Early attempts at measuring electric fields in storms revealed surprising results; the original methods used weather balloons carrying instruments into storm clouds, reporting field strengths far below the theoretical dielectric strength (Marshall & Rust (1991)). Rocket and aircraft observations report similar results up to only a few hundred kV/m, an order of magnitude less than conventional breakdown (Winn & Moore (1971); Winn et al. (1974)). There are a number of possible explanations, including the instruments' interference with ambient fields, small, unmeasured localization of strong fields, etc. The fact remains that this issue has not been empirically resolved, though simulations provide possible answers that will be discussed in subsequent sections (section 3.2).

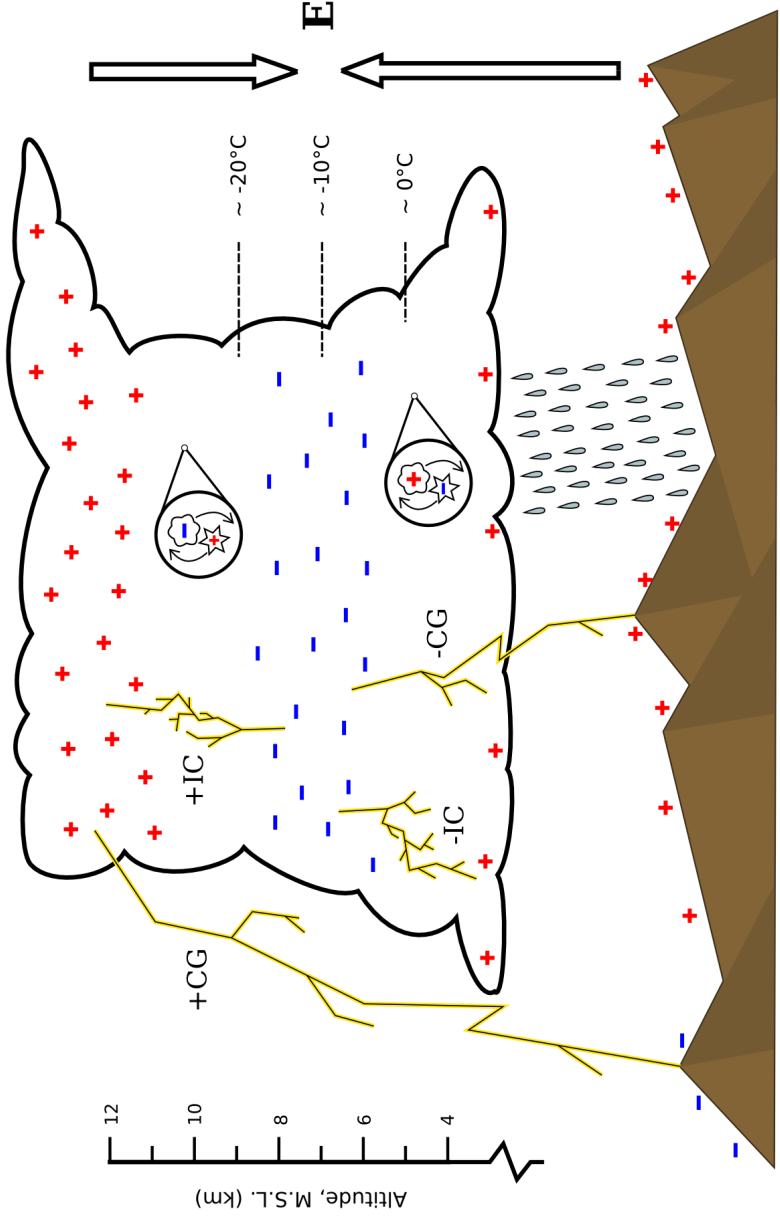

**Figure 2.1:** Main negative charge region is shown alongside upper, lower, and induced ground-level positive charge regions. Polarity of the resulting large-scale electric fields are given on the right. Inset circles illustrate charge separation via falling graupel and rising ice crystals alongside characteristic temperatures which result in the mid-level negative charge region. A typical altitude scale of these regions and temperatures is given along the left side according to Krehbiel (1986) (the space between ground and the cloud's lower edge is exaggerated). Positive and negative polarities of intracloud and cloud-to-ground lightning flashes are also shown, being the most commonly associated types with TGFs as discussed in section 2.2.

## 2.2   Types and Stages of Lightning

There are a number of different ways lightning can develop, each with its own properties. The two main classes of lightning are cloud-to-ground (CG) and cloud-to-cloud (CC), which can be distinguished by the naked eye. As the names suggest, CC lightning connects opposing charge regions in neighboring clouds or within the same cloud (intracloud or IC), while CG lightning bridges the gap between storm clouds and the ground. Within these two types, there are subcategories related to the charge and direction of the lightning's propagation; flashes can be either upward/downward and negative/positive depending on the direction of leader propagation and resulting field change, respectively. For example, a CG flash originating in a cloud's negative charge region discharges the positive electric field below the cloud, making it a negative (–CG) flash. Conversely, upward intracloud lightning from a negative charge region to a higher positive region would be discharging the negative field in between, resulting in a positive (+IC) flash. In this study, we examine TGF data from ground level in association with downward –CG and –IC flashes as well as comparisons to upward +IC flashes . Previous TGF observations associated with these types of flashes by other experiments are also discussed in section 3.1. Orientations of these different lightning polarities are illustrated in Figure 2.1 alongside the rare +CG variation, which has not been associated with TGFs.

The storm electrification process of section 2.1 results in the typical charge structure visualized in Figure 2.1. This drawing depicts the concentration of positive charge in upper cloud regions with negative charges that collect below, as well as a lesser positive charge that can develop at the very bottom of some clouds. The dominant charge in a particular region at low altitude also induces an opposing charge in Earth's conductive surface and results in a somewhat symmetrical tripole structure with electric fields directed toward the mid-level negative region (Krehbiel (1986)).

Each of these lightning configurations consist of two primary stages: stepped leaders followed by the main stroke(s) (Figure 2.2). The aptly-named leaders are the development of an ionized channel by which charge can flow more freely between two opposing charge regions (section 2.3). Once the channel is complete, charge carriers are exchanged extremely quickly, producing currents of tens of kA, occasionally reaching as high as many hundreds of kA. Since the channel is not a perfect conductor, incredible amounts of energy are produced in a flash of light, heat, and sound that we commonly identify as the lightning stroke and accompanying thunder. Additionally, after the main return stroke, charges rearrange themselves in the active regions and may trigger a series of decreasingly-powerful strokes until either the charge discrepancies are resolved or the ionized channel is neutralized (Cooray (1993)).

## 2.3   Stepped Leader Phase

The existence of leaders preceding natural lightning was presented via streak photography as early as Schonland & Collens (1934). These observations showed that the air between cloud and ground was ionized in a set of discrete 'steps' of length ∼50 m and lasting an average ∼10 ms in total (average speed of $4 \times 10^5$ m/s over 4 km). A quick, bright return stroke followed the leader phase, traveling up along the conducting channel, and was often followed by a series of subsequent strokes. Figure 2.2 illustrates this overall stepping process in its basic form.

As mentioned in section 2.1, measurements have not been able to confirm that electric fields in storms are strong enough to initiate conventional breakdown (3,200 kV/m at sea level), but even the early study of Malan et al. (1935) postulated that the leader process may be driven by negative streamers having a lower breakdown threshold of 1,250 kV/m (Pasko (2006)). These thresholds scale linearly with pressure, dropping to ≃1,600 and 625 kV/m respectively at 0.5 atm (∼5 km MSL). As discussed in section 2.1, however, large-scale fields within thunderclouds have still not been measured this high. Instead, small-scale field enhancement allows streamers to develop in the vicinity of sharp, conducting objects. In clouds, this is thought to occur near the corners of charged ice crystals and liquid hydrometeors where the field strength is between that of negative streamer formation and conventional breakdown. Under these conditions, streamers are able to develop while heating and ionizing the air into an isolated 'space stem' which continues to grow until reconnecting with the hydrometeor (da Silva & Pasko (2013a)). The built-up potential instantly transfers into the newly-formed, conducting leader channel. Again, the concentration of charges at the new leader tip creates a strong, local field and the process repeats for each subsequent step, known as the streamer-to-leader transition (illustrated in Figure 2.3). This process is also the basis for seeding TGFs under the cold runaway model, as discussed in subsection 3.2.3 and supported by the results of this study.

This model of leader step formation is reinforced by laboratory spark experiments (Gorin et al. (1976); Gallimberti et al. (2002)) and by increasingly-popular high-speed video of natural lightning (Biagi et al. (2010); Hill et al. (2011); Stolzenburg et al. (2013)), made possible by the high luminosity of the streamer-to-leader transition. These observations show leader development stall as the space leader clearly develops separately and ahead of the leader before reconnecting and producing a bright flash traveling back upward along the established leader channel. Installation of a similar camera is underway at Telescope Array to more clearly observe the relationship between leader development and TGF production (subsection 9.3.4).

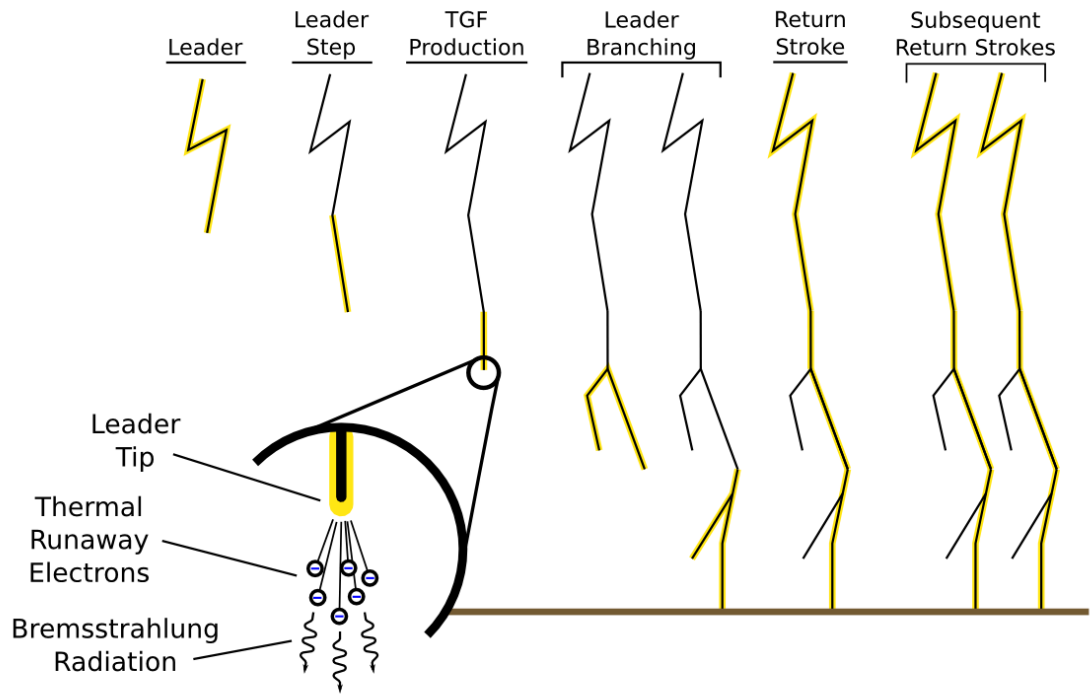

**Figure 2.2:** Lightning leader development illustrated for a TGF-producing, cloud-to-ground flash (leader step sizes are exaggerated for simplicity). More detailed information on the individual leader stepping process in given in section 2.3 and visualized in Figure 2.3. Basic TGF production is also shown according to the cold runaway model (subsection 3.2.3) during early leader steps. The process of thermal runaway electrons seeding RREA is omitted in this figure, but is discussed in subsection 3.2.1.

These cameras (as well as early streak photography) typically observe leader propagation at low elevations as they approach ground level, reporting leader step lengths of $\sim$3–50 m, $\sim$5–50 $\mu$s between steps, and overall average speeds of 1–30$\times$10$^5$ m/s (Malan et al. (1935); Berger (1967); Hill et al. (2011)). Higher-altitude IC leaders, however, differ by 1–2 orders of magnitude. A study of a 10 km IC flash by Edens et al. (2014) presents leader lengths >200 m, 4.3 ms between steps, and overall speed of 4.7$\times$10$^4$ m/s. These longer, slower steps are presumably due to the lower pressure, resulting in larger streamer zones and longer mean free path of electrons. The same effect is seen in the VHF data comparisons of this study between CG and high-altitude IC flashes at Telescope Array (subsection 8.4.1).

## 2.4 Discharge Processes

### 2.4.1 Initial Breakdown Pulses

TGFs have been understood to occur within the first 1–3 ms of development in upward IC lightning flashes (Stanley et al. (2006); Cummer et al. (2015)), and have been specifically tied to large-current sferic pulses called initial breakdown pulses (IBPs) (Brook & Kitagawa (1960); Lu et al. (2011); Cummer et al. (2011)). IBPs develop in the first few milliseconds of breakdown activity and consist of strong, bipolar E-field changes and subsequent slow, reverse-polarity 'overshoots' (lower panel of Figure 2.4 for example). The polarity of the initial pulse corresponds to that of the overall field change — +IC flashes begin with positive changes while –IC and CG flashes begin with negative changes, seen in the comparisons of section 8.4. In addition, the IBPs take place during steps in the early leader development, mimicking the difference in leader step duration and spacing for these flashes, likely connecting the two processes (Villanueva et al. (1994)). The fact that IBPs are seen only during initial breakdown stages indicates that the leader stepping process evolves over the course of flash development.

Following the step formation process of section 2.3, the leader-IBP connection implies that the streamer-to-leader transition causes the IBPs' initial polarity change. The space leader's reattachment to the existing channel, and resulting potential transfer into the new step, then produces the IBP's reverse-polarity overshoot and accompanying upward-propagating luminosity as in the optical measurements of Biagi et al. (2010); Hill et al.

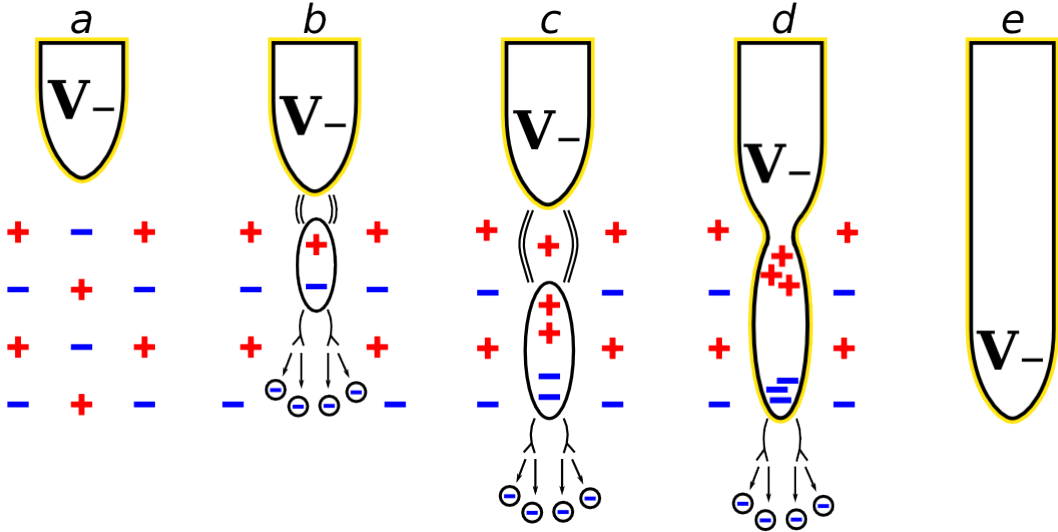

**Figure 2.3:** Streamer-to-leader transition illustrated for new step formation during negative leader development. *(a):* The conducting leader carries negative potential from the previous step. *(b):* The virgin air ahead of the existing leader tip begins to heat and ionize, creating one or more space stems/space leaders which are indirectly connected to the tip via streamers (positive behind, negative ahead). *(c):* As the air continues heating and the space leader grows, the propagating negative streamers accelerate thermal electrons forward, seeding RREA and a TGF if conditions allow. *(d):* The space leader recombines with the leader tip. Negative potential transfers to the new segment and current flows backward along the existing channel, producing the reverse polarity sferic characteristic of IBPs. *(e):* The new segment is now a fully-fledged conducting leader and the process repeats.

(2011); Stolzenburg et al. (2013).

In the first few milliseconds of initial breakdown, IBPs begin appearing with weak amplitude. Subsequent pulses gradually become stronger, accompanied by strengthening VHF pulses, until reaching a peak and then weakening again. This is especially noticeable in low-altitude IC and CG flashes wherein many IBPs occur in close succession, but is also visible in the upward IC flashes recorded at TA containing only a few IBPs (section 8.4). Note in the accompanying figures that the strongest IBP of each flash corresponds to the leaders' transition from mostly linear development to increased branching. For cloud-to-ground flashes, these strong IBPs are typically of comparable strength to the pulse of the final return stroke (upper panel of Figure 2.4). For intracloud flashes without ground strokes, IBPs are by far the strongest impulsive events in the sferic.

A recent classification of IBPs was identified by Lyu et al. (2015), consisting of extremely high current (>200 kA) pulses called energetic in-cloud pulses (EIPs). Most, if not all, EIPs are thought to be producers of a class of TGFs (Lyu et al. (2016)), but the rarity of these events means that there are very few simultaneous observations. Conversely, not all TGFs are associated with EIPs, and no EIPs have yet been observed at Telescope Array. Although originally understood as a subset of IBPs, the recent study of Tilles (2020) suggests that the VHF activity of EIPs differs from that of IBPs, perhaps resulting in two distinct classes of associated TGFs.

Although not classified as IBPs, a similar sferic signal is produced by narrow bipolar events (NBEs), identified by LeVine (1980) as the strongest source of VHF radiation from lightning under the name of bipolar pulses. They have shorter durations of 10–20 $\mu$s and lack the sub-pulses characteristic of IBPs. NBEs have also been shown to be separated from previous breakdown activity, instead initiating fast breakdown in virgin air as discussed in subsection 2.4.2 (Rison et al. (2016)). This breakdown occurs in the strong fields near hydrometeors, is caused by either positive streamer systems, and produces a sferic pulse with polarity depending on the direction of fast breakdown (Willett et al. (1989); Tilles et al. (2019)). For example, the NBE which initiated the parent flash of TGF A consisted of upward-propagating FPB and produced a negative sferic (chapter 7).

### 2.4.2   Fast Breakdown

Fast positive breakdown (FPB) was recently identified as the cause of narrow bipolar events (NBEs), a type of discharge that have been seen to initiate lightning breakdown at high altitudes (LeVine (1980); Rison et al. (1999)). These events are characterized by their VHF radiation, bipolar sferic signal, and disassociation from preceding breakdown.

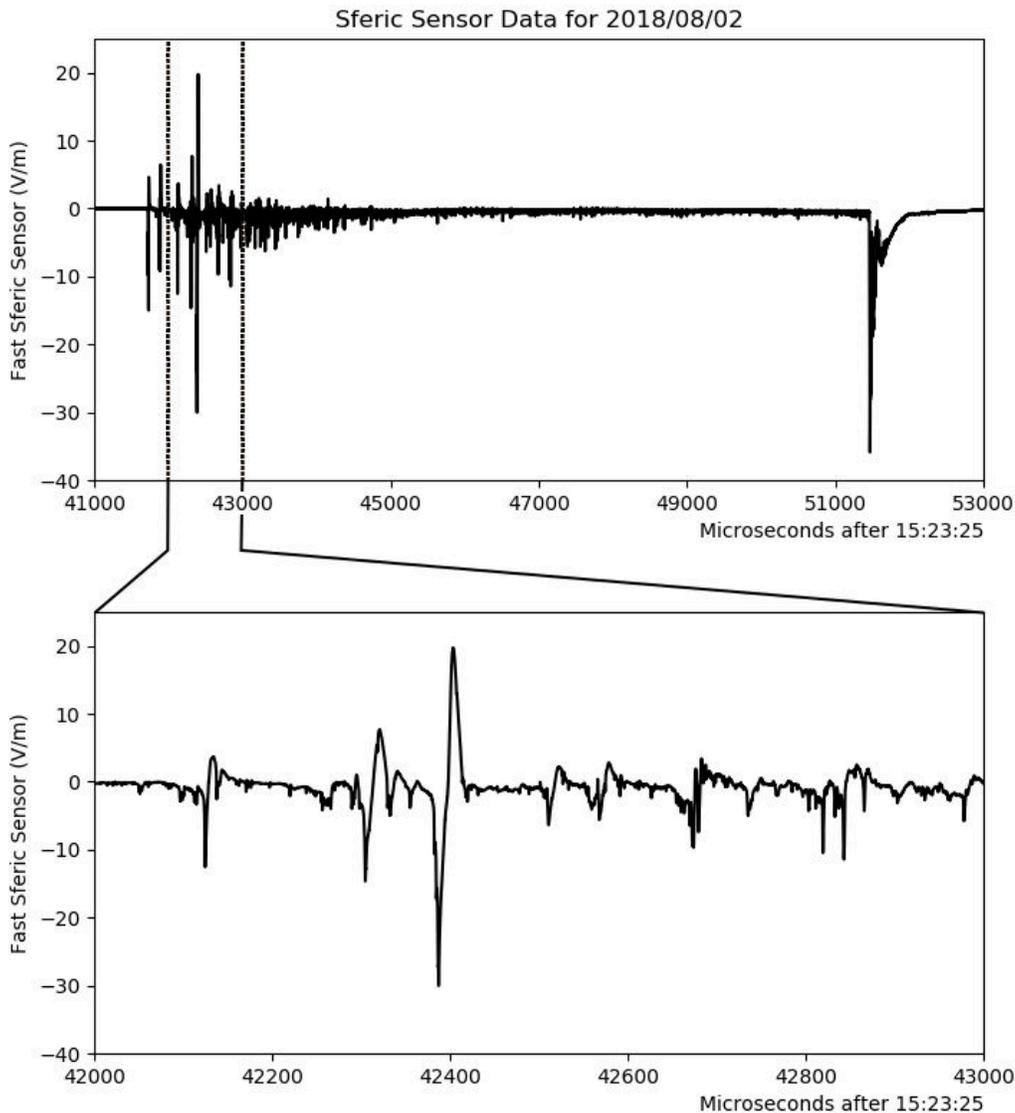

**Figure 2.4:** Fast electric field changes recorded by the fast antenna (section 5.2) in association with TGF B. Upper panel shows the signal of the entire flash, and the lower panel is a zoomed-in view of the IBPs in its initial stages. For this –CG flash, the sferic pulses begin with a sharp negative field change followed by a slower positive overshoot. In this case, the strongest IBP at 42400 $\mu$s was comparable in strength to the return stroke itself at 51500 $\mu$s.

The study of Rison et al. (2016) showed that these events propagate $\lesssim$500 m at speeds on the order of $\sim$10$^7$ m/s, ruling out the involvement of hot conducting leaders, which propagate 1–2 orders of magnitude slower (section 2.3). Instead, NBEs are caused by a breakdown of virgin air by a system of positive streamers initiated in the high-field regions surrounding sharp corners of hydrometeors. Once streamers develop, their enhanced electric fields become self-generating and facilitate classic leader development (Griffiths & Phelps (1976); Attanasio et al. (2019)). The parent flash of TGF A of this study was initiated by an NBE paired with FPB, discussed in chapter 7.

The opposite-polarity process involving negative streamers, fast negative breakdown (FNB), was observed more recently in Tilles et al. (2019). It is shown to behave similarly to the positive polarity version, greatly enhancing the ambient electric field ahead of its advancing streamer front. In terms of speed, the initiating FPB of TGF A's parent flash, for example, propagated upward at an average speed of $1.7\times10^7$ m/s compared to the later FNB's average speed of $1.5\times10^7$ m/s. FPB is likely a more common breakdown initiator due to positive streamers' lower stability field, requiring $\sim$5$\times10^5$ V/m to propagate at sea level compared to $\sim$12.5$\times10^5$ V/m for negative streamers (da Silva & Pasko (2013b)). These fields scale proportionally to atmospheric pressure, falling to $2.5\times10^5$ V/m and $6.3\times10^5$ V/m at 5 km altitude ($\sim$0.5 atm). The difference in these thresholds is possibly due to different electron drift speeds during streamer breakdown (Williams (2006)). Regardless, in cases where positive streamer development is suppressed, negative streamer breakdown may occur alone. In the case of the streamer-to-leader transition (Figure 2.3), both positive and negative streamers play a role. The flashes of this study consist of negative leader steps, meaning that the negative streamers advance ahead of the developing space leader while positive streamers connect back into the existing channel (Biagi et al. (2010)). In the events of this study (chapter 7), downward VHF development accompanied by a negative sferic indicates downward negative breakdown. At speeds of $1.6$–$2.7\times10^7$ m/s, these are instances of streamer-based FNB during negative leader step formation.

### 2.4.3  Other Radiative Discharge Phenomena

Many additional forms of energetic discharge in thunderstorms have also been known as producers of radiation, dating back to early predictions of Wilson (1924). The paper discussed discharges between upper thunderclouds and the atmosphere which would later be known as sprites. These high-altitude ($\sim$70 km) events extend tens of kilometers, last up to 100 ms, and usually result from +CG lightning (Rycroft (2006)). A similar type of event called elves can occur even higher ($\sim$90 km) above tropical storms. Both of these consist of visible red glows, and are therefore termed transient luminous events (TLEs). They also radiate in ELF and VLF radio bands, resulting from thermal discharge or relativistic runaway avalanching similar to TGFs (subsection 3.2.1). Consequently, TLEs were mistaken as the source of TGFs in early studies (Fishman et al. (1994); Inan et al. (1996)).

Gamma-ray glows, also called thunderstorm ground enhancements (TGEs), are another source of thundercloud radiation which can last several minutes (Wada et al. (2020); Chilingarian et al. (2011)). With individual gamma energies of up to ~10 MeV (Torii et al. (2011)), these events may also involve RREA shower development in large-scale thundercloud electric fields. However, they are often recorded without the occurrence of lightning, suggesting they have different seeding mechanisms than TGFs. In fact, these glows sometimes die out abruptly during lightning strikes, presumably due to rapid discharge of the field and resulting interruption to the avalanching process (Wada et al. (2019)).

# CHAPTER 3

# TERRESTRIAL GAMMA FLASHES

To quote the artist Bob Ross, "we don't make mistakes — we just have happy accidents." So too the serendipitous discovery of terrestrial gamma-ray flashes (TGFs) followed from a series of accidental discoveries. This began with the discovery of extra-galactic gamma-ray bursts, which were detected by the Vela satellites (Klebesadel et al. (1973)), originally designed to search for gamma radiation signatures of nuclear weapons in 1963. After the measurement of sixteen of these unexpected, high-energy events, any terrestrial origin was ruled out and the phenomenon birthed its own field of study. Several satellites were launched for their study, the first of which was the BATSE detector aboard the Compton Gamma Ray Observatory. Many more detectors were deployed to improve the efficiency and energy range of detections in order to better understand the gamma-ray bursts. These satellites were designed to measure extraterrestrial photons using a combination of calorimeters and scintillator detectors.

In decades of observations, high-energy gamma-ray bursts were recorded en masse, successfully leading to a basic understanding of their spectra and sources (Fishman et al. (1993)). Some events, however, had strange properties and stood out from the larger set, seeming to originate from the Earth itself, but not fitting the originally-expected signature of nuclear weapons testing. As already introduced, these would become known as terrestrial gamma flashes (TGFs). BATSE was the first to report TGFs in 1994 (Fishman et al. (1994)), and subsequent experiments improved greatly on its limited data. The satellite experiments of section 3.1 have now collected thousands of TGF events and have been able to compose thorough spectrum and timing analyses.

After their unexpected discovery, TGFs were thought to originate in high-altitude Earth atmosphere above thunderstorms (Fishman et al. (1994)), leading many to assume the flashes are given off by elves or sprites (Inan et al. (1996)), which are electromagnetic discharges high above storm clouds (section 2.4). Further observations revealed the sources to be much lower, even inside the thunderstorms themselves. In particular, they were shown to originate in the initial stages of upward negative IC flashes (Cummer et al. (2011, 2015);

Lu et al. (2010); Lyu et al. (2016); Mailyan et al. (2016); Shao et al. (2010); Stanley et al. (2006)), perhaps exposing the unknown origins of the lightning's breakdown process.

## 3.1  Satellite and Ground Observation

The early satellite observations of TGFs were exciting, but hardly conclusive. Studies lacked a direct connection to the leaders' breakdown due to timing uncertainties inherent in studying events on much shorter timescales than their intended subjects. Ideally, incident particles are completely absorbed by the crystal-based calorimeters in a series of interactions so that total energy is recorded, requiring substantial volume ((Fabjan 1991, Chapter 4) for calorimeter design in particle detectors). Unfortunately, this comes at the price of timing resolution. These restrictions are inherent to calorimeters, a result of the photon scattering time within crystalline structures as well as deadtime issues, which is discussed for individual instruments below. These are often supplemented by scintillator detectors optimized for timing measurements while lacking in spectral resolution. Each new experiment carried advanced instruments which produced different (and improved) results, forcing the definition of TGFs to evolve over time as detectors became more sensitive to shorter, brighter events.

The most recent measurements indicate upward TGFs typically have the following characteristics:

1. Source durations of 20–200 $\mu$s (Østgaard et al. (2019); Marisaldi et al. (2015)).

2. Photon spectra up to 40 MeV and roughly consistent with power law or RREA having characteristic energy of $\simeq$7.3 MeV (Dwyer (2012); Marisaldi et al. (2010, 2014); Smith et al. (2005)).

3. Occurring in the first 1–3 ms of leader development in +IC flashes (Cummer et al. (2015); Mailyan et al. (2016)).

4. A wide range of total photon fluences between $\simeq 10^{15}$–$10^{19}$ photons (model dependent, see Celestin et al. (2015); Xu et al. (2012)).

A popular topic was speculation on whether the bursts could be seen from below storms as well, which would allow for much closer, more direct observations (<10 km compared to $\simeq$500 km) as well as more consistent measurements of lightning development (electric field sensors, VHF source tracking, etc.). Theoretically they should also be produced downward due to the symmetric electric field configuration in storm clouds (chapter 2); charge separation between upper positive, lower negative, and positive-induced ground means that both upward and downward fields exist simultaneously, though upward and downward TGFs have not been observed in tandem.

The downside of downward TGF production is the much lower propagation distance, effectively reducing the field of view for ground-based detectors. For example, with an opening angle of $\simeq$30°–40°, TGFs can be detected by satellite within a radius of hundreds of kilometers. From below, a stationary detector array must be set up and wait for a lightning storm within a couple kilometers in which TGFs are still relatively rare. One solution is to install artificial triggering systems (Dwyer (2004); Hare et al. (2016)), thereby ensuring nearby lightning. In a few cases, downward TGFs have also been recorded alongside natural –CG discharges (Dwyer et al. (2012); Tran et al. (2015); Wada et al. (2019)). Unlike satellite observations, downward TGFs have thus far shown widely varying characteristics compared to one another, exacerbated by their relatively low statistics. The following sections outline the instruments used by each experiment and their results.

### 3.1.1 BATSE Events

The Burst and Transient Source Experiment (BATSE) was an instrument aboard the Compton Gamma Ray Observatory, launched in 1991. The experiment was equipped with eight detector modules at each corner, providing full-sky coverage (Fishman et al. (1989)). Each module consisted of a large-area detector and a spectroscopy detector, both of which were uncollimated NaI(Tl) crystal calorimeters. The former was optimized for directional analysis of photons between 30–1900 keV while the latter provided better energy resolution for x-rays as low as 15 keV up to gamma-rays of 2.7 MeV and higher, depending on detector gain. BATSE was designed for the search and study of celestial gamma-ray sources, but in 1994 reported the observation of at least a dozen gamma flashes with sources in Earth's atmosphere, later dubbed terrestrial gamma-ray flashes (Fishman et al. (1994)).

By the time of its deorbit in 2000, 76 confirmed TGFs had been observed and were found to be consistent with high-energy brehmssatrahlung emission above thunderstorms. The original report suggested source altitudes of 40–80 km, but subsequent studies have placed them in the 15–20 km range (Carlson et al. (2007)), closer to the altitude of upward intracloud lightning and consistent with other studies (Dwyer & Smith (2005)). The TGFs were seen to occur both singly and in groups, with each pulse lasting $\sim$1 ms. Grefenstette et al. (2008) showed the significant effect of dead-time losses on the durations, however, and the TGF sources were later estimated to typically last 200–270 $\mu$s (Gjesteland et al. (2010)).

Because the TGFs were recorded by the large-area detector, the energy resolution was poor compared to later studies. Photons were grouped into four channels with energy ranges of 25–50 keV, 50–100 keV, 100–300 keV, and >300 keV. While these are mostly high-energy x-rays, Nemiroff et al. (1997) showed that the TGFs were typically dominated by the two latter channels (100–1900 keV). The same study also found that the TGFs could be roughly fit to a hard power law spectrum, again dominated by the highest-energy bin. Another feature of the flashes was their tendency to 'soften' over the pulse duration, though Grefenstette et al. (2008) showed that this was likely exaggerated due to detector dead-time.

### 3.1.2   RHESSI Events

The Reuven Ramaty High Energy Solar Spectroscopic Imager (RHESSI), designed as a solar flare observatory, was launched in 2002 to study solar photons in the range of $\simeq$25 keV–17 MeV. As with BATSE, the experiment also detected TGFs — an early study of 86 events was released after its first 6 months of operation (Smith et al. (2005)). The satellite employed 9 germanium crystal detectors to indirectly measure photon energies via ejected electrons (Lin et al. (2002)). The detector's method of recording energy and arrival time within 1 $\mu$s of each photon, rather than BATSE's electronic triggering system, allowed for better efficiency at identifying and analyzing TGFs. As a result, most signals were single pulses with durations as low as 200 $\mu$s and up to 3.5 ms (Smith et al. (2005)).

The detectors' high spectral resolution led to the first detailed spectrum of TGFs, which was consistent with bremsstrahlung emission from 35 MeV electrons, with evidence of some gamma photons above 20 MeV. In addition, modeling provided the first TGF fluence estimates of $\simeq$3$\times$10$^{15}$ electrons (Smith et al. (2005)). A following study the same year showed the RHESSI TGF spectrum fit very well with the relativistic runaway electron avalanche model (RREA) with average photon energy of 7.3 MeV and produced at altitudes of 15 km (Dwyer & Smith (2005); Grefenstette et al. (2008); Dwyer (2012)). These results led to RREA becoming the primary suspect for TGF production in following years (This and other models are discussed in more detail in section 3.2). Further analysis of the next 6 years of data yielded a catalog of 820 TGFs (Grefenstette et al. (2009)), allowing refinement of the spectrum and finding no significant variations due to day/night occurrence, latitude, or TGF intensity. Additionally, simulations showed that the minimal dead-time issues (compared to BATSE) did not significantly affect the spectrum.

### 3.1.3   AGILE Events

The Italian satellite Astro-Rivelatore Gamma a Immagini Leggero (AGILE) was a gamma-ray observatory which began operation in 2007. Its three primary instruments cover a large, non-contiguous energy range; a silicon-based detector detects x-rays of 18–60 keV, a tungsten-silicon tracker optimized for temporal and angular resolution of high-energy gamma-rays between 30 MeV–50 GeV, and the CsI(Tl) minicalorimeter (MCAL) for broadband gamma-ray detection between 300 keV–100 MeV (Pittori & Tavani (2004)). The MCAL provided all the TGF data reported by AGILE, of which 34 were analyzed early in the run between June 2008 and March 2009 (Marisaldi et al. (2010)). These events had relatively long durations averaging 1.5 ms. As with other experiments, these observations are biased toward longer events due to dead-time effects. A further study of the effect led to an order of magnitude improvement in TGF sensitivity, particularly to those of short duration (Marisaldi et al. (2015)). The enhanced configuration resulted in a much lower median $t_{50}$ value (duration of 50% of photon counts) of 86 $\mu$s, compared to 290 $\mu$s for the standard configuration. These improvements have again shown that TGFs are shorter than previously thought, but also that instrument limitations may be preventing the measurement of the lower limit of TGF durations.

Perhaps due to AGILE's energy sensitivity, the reported TGFs contain photons higher than seen in previous studies. The first study gave data on TGFs with average photon energies 2–10 MeV, maximums of 5–43 MeV, and overall consistent with RHESSI TGF spectra (Marisaldi et al. (2010)). Later studies, however, showed MCAL measurements of photons reaching as high as 100 MeV, clearly deviating from the established RREA model (Tavani et al. (2011)). Instead, these observations of photons above 30 MeV better fit a broken power law. In following years, AGILE's classic TGFs were constrained to have maximum photon energy ≤30 MeV (Marisaldi et al. (2014, 2015)), but alternate leader/streamer models have provided a possible explanation for the high-energy variants (Celestin et al. (2012), subsection 3.2.3)

### 3.1.4 GBM Events

The Gamma-ray Burst Monitor on board the Fermi satellite was launched the following year in 2008 and remains in operation. The instrument is intended for the study of gamma-rays bursts of both solar and extra-galactic origin using an array of twelve NaI(Tl) detectors covering lower photon energies of 8–1000 keV as well as two bismuth germanate crystal detectors extending sensitivity up to ≃40 MeV (Meegan et al. (2009)). As with BATSE, the NaI(Tl) scintillation detectors are oriented such that relative counting rates and timing can be used to determine arrival direction.

The first set of 12 TGFs was reported in 2010 (Briggs et al. (2010)) with typical maximum photon energies near 30 MeV, one reaching as high as 38 MeV. However, two long-duration outlier events were also much softer in energy, and were later found to be detections of electron-positron beams following magnetic field lines, though likely still sourced by TGFs (Briggs et al. (2011)). Two of the 'normal' TGFs consisted of pairs of separated pulses while two other events were best fit by closely overlapping pulses. The average pulse duration was 740 $\mu$s, including the two longest pulses produced by electron-positron beams at 2.08 and 3.08 ms. Excluding these, the average duration was 430 $\mu$s. The GBM's nominal dead time per event is relatively low at only 2.6 $\mu$s per event, but still restricted the instrument's triggering of short duration TGFs and under-reported their energy deposit (Fishman et al. (2011)). Further development of analysis methods correct for most of this effect, increasing detection rate by a factor of 10 and enabling the observation of much shorter-duration TGFs (Briggs et al. (2013)). This study presented a median TGF duration of 425 $\mu$s, with $t_{90}$ values (duration of 90% of photon counts) ranging from 40 $\mu$s to 4 ms. Even with the improved methods, dead-time losses were reported to still likely prevent detection of the shortest TGFs.

### 3.1.5  ASIM Events

Unlike the previous satellites, the Atmosphere-Space Interactions Monitor (ASIM) is an experiment on board the International Space Station and is intended to study high-energy atmospheric photons. Launched in 2018, the experiment is composed of two instruments; the Modular X and Gamma Ray Instrument (MXGS) for photon spectral analysis and the Modular Multispectral Imaging Array (MMIA) for optical observations. TGF observations are made by the MXGS, which consists of two detectors. The Low Energy Detector is made of CdZnTe crystals sensitive to x-rays of 15–400 keV, recording both energy and trajectory. The High Energy Detector is made up of 12 BGO scintillators and is sensitive to gamma-rays with energies of 200 keV–>30 MeV (Østgaard et al. (2019); Neubert et al. (2019)).

The detector's dead-time has significantly improved over other experiments, being only $\simeq$550 ns. As such, ASIM has recorded even shorter TGFs than previously. Although they are rather incomplete metrics, $t_{50}$ values are commonly used for duration comparisons of TGFs, representing the time between 25% and 75% of photon counts. In ASIM's first ten months of operation, 217 TGFs were reported, half of which had $t_{50}$ values between 20–60 $\mu$s, with a median of 45.5 $\mu$s (Østgaard et al. (2019)). The distribution of $t_{90}$ values had a maximum between 60–120 $\mu$s. Additionally, the same study reports that the already low $t_{50}$ times could still be overestimated due to lingering dead-time effects and detector saturation during the brightest and shortest TGFs. The detection capability of ASIM also showed interesting comparisons with a TGF that was simultaneously detected by Fermi GBM. While Fermi reported a single, short-duration TGF, ASIM observed several pulses lasting $\simeq$2 ms overall. The observation suggests there is some TGF substructure, perhaps associated with upward leader steps.

### 3.1.6 ICLRT Events

The Thunderstorm Energetic Radiation Array (TERA) of the International Center for Lightning Research and Testing (ICLRT) at Camp Blanding, Florida consists of 24 detector stations covering 1 km$^2$. The stations contain NaI and NaI(Tl) scintillation detectors for gamma and x-ray detection with energies from 30 keV–$\geq$10 MeV (Dwyer et al. (2004); Saleh et al. (2009)). The center studies both natural and artificially-triggered lightning, which have produced comparable results in terms of energetic radiation. Although this study focuses on the development of natural lightning and TGFs, some relevant examples of energetic radiation from artificially-triggered lightning are briefly reviewed here

The method of rocket-triggering at ICLRT was found particularly useful for the reliable study of lightning in close proximity (Rakov (1998)), in which a spool of thin copper wire trails behind small rockets, creating a partial path between a thundercloud and the ground below. Studies by Dwyer (2003); Dwyer et al. (2004) reported x-ray detections from most of these flashes, consisting of short, intense bursts of photons in the $\sim$80 $\mu$s immediately preceding triggered return strokes. The bursts primarily consisted of short pulses of photons in the range of 30–250 keV, lasting about a microsecond each, and separated by 2–10 $\mu$s. One abnormal event, however, occurred early in the flash's initial breakdown during upward positive leader development and produced gamma-rays of energy >1 MeV. The main flux of gamma-rays lasted $\sim$75 $\mu$s and was followed by individual pulses separated by 5–30 $\mu$s leading up the return stroke 200 $\mu$s later. Despite being artificially triggered, the spectral and temporal characteristics appeared similar to the observations of Moore et al. (2001), which still remain unconfirmed as TGFs. A decade later, another event was detected following a rocket-triggered flash (Hare et al. (2016)). This TGF occurred during a relatively high-current pulse 13 ms after the return stroke lasting 290 $\mu$s. Scintillation detectors were quickly saturated by the high flux and signal pileup, but photons were estimated to have multi-MeV energies.

Dwyer et al. (2005) also reported on x-ray emission from two natural –CG flashes at ICLRT. The detections were similar to those of artificially-triggered lightning, beginning $\sim$1 ms before the return stroke as lone pulses separated by $\sim$200 $\mu$s and increasing in frequency before stopping abruptly as the leader made contact with ground. Just before the stroke, the pulses' separations were 5–20 $\mu$s, each correlated with individual leader steps as recorded by the array's electric field sensors. The final, closely-spaced set of pulses lasted $\simeq$300 $\mu$s. Individual photons within the pulses were estimated to have maximum energies of a few hundred keV, with the majority <150 keV. These lower x-ray energies suggested a different source of acceleration than the large-scale RREA which was considered to play a strong role in the production of upward TGFs at the time.

A single observation later presented by Dwyer et al. (2012) was originally considered to be 'TGF-like' and distinct from the previous x-ray bursts. The event lasted 52.7 $\mu$s, occurring 191 $\mu$s after the return stroke. While this timing was shorter and more delayed than other ground observations, similar events were later reported by Tran et al. (2015); Wada et al. (2019). Spectral analysis also appeared consistent with upward TGFs, consisting of gamma-rays up to 20 MeV or higher and consistent with the RREA energy scale (section 3.2).

### 3.1.7   TETRA Events

The TGF and Energetic Thunderstorm Rooftop Array (TETRA) was initially installed in 2010 for studying lightning-associated radiation following the observations of ICLRT and others. The array consisted of four detectors covering $\simeq$900 km$^2$, each consisting of three NaI(Tl) scintillators with sensitivity to photons in the 50 keV–2 MeV range. TETRA reported 24 event candidates in its first 2.5 years of operation (Ringuette et al. (2013)), though they were not confirmed as TGFs. These events exhibited a wide range of attributes, with $t_{90}$ durations of 24 $\mu$s–>4 ms and only 10 occurring within 100 ms of lightning triggers, from 6 ms before the main stroke to 80 ms after. This is in part due to the triggering software's timing uncertainty on the order of 2 ms, which would later improved by a factor of 10. TETRA's limited sensitivity to energetic gamma-rays makes spectral comparison difficult, but other characteristics were consistent with satellite observations.

The experiment was succeeded by TETRA-II in 2015, consisting of three small-scale arrays in Louisiana, Panama, and Puerto Rico. Detectors were changed to BGO scintillators with higher sensitivities up to 6 MeV. Individual pulses are recorded with 50 ns resolution, though spectral analysis is only available for the detection of single pulses within a 13 /$mu$s time window. The upgraded arrays reported 22 gamma-ray flashes in the first two years of operation (Pleshinger et al. (2019)), though three of these were not associated with any reported lightning strike data within 8 km. The others had familiar durations of 0.1–2 ms and all began $\leq$1.3 ms before their respective lightning return strokes, often arriving in bursts. This suggests the TGFs were produced during steps in the final stages of the leader development, similar to the previous x-ray observations of Moore et al. (2001); Dwyer & Smith (2005), except with longer durations and occasionally continuing for up to 1 ms after the return stroke.

### 3.1.8   Other Ground Observations

Aside from the collections of observations above, TGFs have only been detected a few times at ground level. Moore et al. (2001) reported early observations of energetic radiation from three nearby natural –CG flashes in New Mexico. Gamma-rays were detected early in the flash, continuing for a few milliseconds leading up to the return stroke, after which they ceased abruptly. The NaI scintillation detectors saturated at energies of $\simeq$1.2 MeV, and were unable to distinguish individual photons during main pulses leading up to the return strokes. The radiation was suspected to have originated during the development of negative stepped leaders, but could not be confirmed as TGFs due to instrumental restrictions.

A single TGF was reported in 2015 at the Lightning Observatory in Florida, which combines an NaI x-ray detector with a series of lightning instruments (Tran et al. (2015)). Although the detector is designed for x-rays, it has an upper-energy limit of $\simeq$5.7 MeV. In 2014, a –CG flash produced pulses of gamma-rays 200 $\mu$s after its return stroke and lasting 16 $\mu$s. It categorized as a TGF according to the following criteria: "(a) no sign of pile-up, characteristic of x-rays associated with leaders near ground, is seen in the recorded pulses, (b) the duration of the recorded pulse sequence is less than 1 ms, and (c) energy values for the largest pulses corresponding to individual photons exceed 1 MeV." The six pulses appeared to be produced by individual photons ranging from $\simeq$100 keV to a few MeV, with two saturating the detectors at >5.7 MeV. These measurements closely mimic the event of Dwyer et al. (2012), consistent with RREA production of high-energy gamma-rays following a flash's return stroke and distinct from events associated with individual leader steps.

The final example is that of a peculiar lightning flash detected in Japan which produced both gamma-ray glows as well as a TGF (Wada et al. (2019)). The BGO scintillators are sensitive to photon energies 0.4–20 MeV. The gamma-ray glow persisted for over a minute before ceasing abruptly with the return stroke, at which time a TGF was detected. The scintillators were saturated, suggesting maximum photon energies in excess of 20 MeV. The TGF's precise duration was unclear, but lasted no more than 1–2 ms followed by $\simeq$200 ms of amplified photon count rates. In addition, the two phenomena's sources were colocated, suggesting that electrons produced by the gamma-ray glow may have initiated the TGF by providing seed particles for RREA.

## 3.2   Models and Mechanisms

Even for upward TGFs, ground detection of the parent flashes can reveal a lot about gamma-ray production. Similar to lightning discharge itself, however, the details remain obscure. Analyses of Stanley et al. (2006) highlighted a connection between RHESSI TGFs and ground-level sferic pulses early in lightning development, leading to several followup investigations. Shao et al. (2010); Lu et al. (2010, 2011); Mailyan et al. (2016) showed that TGFs were likely generated in the first few milliseconds of upward +IC leader development at altitudes of ≃10–14 km, but could not confirm the mechanism(s) of TGF production.

Most leading models of shower development involve some form of relativistic runaway electron avalanches (RREA), a significant source of radiation in thunderstorms which require existing energetic electrons to develop (see subsection 3.2.1). These 'seed' electrons can be produced by cosmic-ray air showers, but 'cold runaway' has also been theorized as a possible origin (subsection 3.2.3). Either way, RREA does not solely explain the extremely high fluences implied by satellite observations. Relativistic feedback (RFD) is one viable solution to the problem, greatly amplifying existing RREA in an ambient electric field like those of a thunderstorm (Dwyer (2012), subsection 3.2.2). Many satellite observations suggest RFD enhancement while those at ground level may not. The different models are introduced in detail below alongside references to the observations from section 3.1.

### 3.2.1   Relativistic Runaway Electron Avalanches (RREA)

Although RHESSI was not the first satellite to observe TGFs, its detectors were more suited than BATSE for measuring photons over a few MeV. The resulting spectrum was consistent with the composition of relativistic runaway electron avalanche (RREA) from ≃25 keV up to 20 MeV with flux falling off quickly at high energies (Dwyer & Smith (2005); Grefenstette et al. (2009)). In addition to RHESSI, TGF analyses by the AGILE satellite yielded similar results up to 40 MeV (Marisaldi et al. (2010)), though their sensitive detectors have recorded excess gamma-rays at higher energies up to 100 MeV, suggesting there may be some power law component to the spectrum (Tavani et al. (2011)). Later ground observations by Dwyer et al. (2012); Tran et al. (2015) also showed agreement with RREA spectra.

The process was theorized to produce an extreme amount of photons from thunderclouds nearly a century ago in Wilson (1924, 1925) as a result of the decrease in stopping power (effective friction force) of electrons as they increase in energy between $10^2$–$10^6$ eV (see Figure 3.1). The minimum of 200 keV/m corresponds to the minimum ionizing case of an electron having 1.2 MeV. The local maximum of ∼24 MeV/m occurs at ∼160 eV, representing the thermal runaway regime (Diniz et al. (2019)). Between these two critical points, an electric field stronger than the corresponding friction force will accelerate electrons more than they are slowed via atmospheric interactions. Since large-scale electric fields in storm clouds often exceed the minimum ionizing case (in kV/m, see Marshall et al. (1995)), the cascade of electrons in continuously intensified while emitting gamma- and x-rays. This mechanism has been modeled in detail analytically (Gurevich et al. (1992)) and numerically for simulation (Moss et al. (2006); Dwyer (2012)), but a brief review is given here. Additional information on electromagnetic showers and particle interactions in general is given in subsection 3.2.4.

Although the friction force is only ∼200 keV/m for a minimum-ionizing electron having 1.2 MeV, the simulated threshold for RREA propagation is slightly higher at ≳284 kV/m due to the electrons' tendency to scatter away from trajectories antiparallel with the applied field (Symbalisty et al. (1998); Dwyer (2003); Babich et al. (2004)). Given a field of this strength, electrons must exist in the atmosphere with energies above the friction curve (≳200 keV) to seed an avalanche. Alongside its formulation of the process, the study of Gurevich et al. (1992) postulated that cosmic ray secondaries may provide seed particles, being relatively common at these energies and altitudes. From Daniel & Stephens (1974), energetic electrons are present in excess of 1000 $(m^2 \text{ s. sr.})^{-1}$ between 10–15 km above sea level, although this varies by an order of magnitude depending on altitude and location. Another possible source lies in the cold runaway regime at lower energies. Slow electrons near the local friction maximum require incredibly strong electric fields to run away (Gurevich (1961)), only theorized to occur naturally in streamer systems at the tips of lightning leaders (see subsection 3.2.3). These strong, localized fields would accelerate electrons very quickly to several MeV, subsequently seeding RREA cascades in the presence of weaker, yet more stable fields (Moss et al. (2006); Celestin & Pasko (2011)).

However RREA is initiated, the resulting showers tend to self-regulate due to the stable equilibrium (minimum) in dynamic friction force, consistently producing electrons with average energy of ≃7.2 MeV under a wide range of electric fields (Dwyer (2004)). These showers exhibit the expected composition consistent with electromagnetic showers free of an applied field (subsection 3.2.4), differing only in their overall intensity. The defining characteristic of RREA showers is the avalanche (E-folding) length $\lambda$ which can be well approximated by:

$$\lambda = \frac{7.2 \text{ MeV}}{eE - 275 \text{ kV/m}} \tag{3.1}$$

where $E$ is the applied electric field (Dwyer (2003, 2012)). Consequently, the number of runaway electrons produced by RREA with $N_0$ seed electrons after developing over a distance $z$ is then given by:

$$N = N_0 e^{z/\lambda}, \tag{3.2}$$

such that a shower with $10^6$ seed particles is amplified to $1.5 \times 10^8$ over 5 avalanche lengths or $2.2 \times 10^{10}$ over 10 avalanche lengths. Relativistic feedback can additionally intensify these showers by up to an additional factor of $10^9$, consistent with values implied by satellite measurements (see subsection 3.2.2)

From Figure 3.1, note that the total friction force rises above the RREA threshold near $\sim$20 MeV, meaning that particles with energy above a few tens of MeV are suppressed as much as those below 100 keV, depending on strength of the applied field. Following the discoveries of Gurevich et al. (1992) and Fishman et al. (1994), the RHESSI spectra published in 2005 exhibited this effect and were best fit by RREA modeling, making it a staple in TGF theory (Smith et al. (2005); Dwyer & Smith (2005)).

Aside from emitting EM radiation, these avalanches also provide a means of discharging the built-up potential in storm clouds. The conventional breakdown of air occurs at $\sim$32 kV/cm at sea level (or about half that at 6 km altitude), but fields of this strength have only rarely been measured in thunderstorms, typically being on the order of only a few kV/cm (Marshall & Rust (1991); Marshall et al. (1995)). Lightning obviously occurs more often than these measurements would suggest, possibly due to the lower field requirement of streamer breakdown as described in subsection 2.4.2. Regardless, electric fields are likely kept below the conventional breakdown threshold by RREA, where resulting electron cascades help to equalize the charge differential between regions of the cloud (Marshall et al. (1995)). This discharge process has a characteristic duration of a few seconds, comparable to the rate of storm electrification (Dwyer (2003)).

### 3.2.2 Relativistic Feedback Model

The relativistic feedback (RFD) model was developed into its modern form by Dwyer (2003) to explain the massive fluence of photons implied by TGF observations of energetic radiation from lightning. In addition, the study attempts to place an upper limit on the possible strength of large-scale thunderstorm fields, above which they would rapidly discharge due to enhanced RREA (subsection 3.2.1).

As an avalanche grows exponentially in a large enough volume, scattered particles can seed secondary avalanches, greatly speeding up and amplifying the process. This can happen via back-scattered photons or by positrons accelerated in the direction of the electric field until scattering or annihilating, from which produced electrons begin the runaway process again (Figure 3.2). This feedback process greatly speeds up discharge of the electric potential and can repeat as long as the large-scale electric field persists. Therefore, RFD enhancements rely on the field stability, which can last many seconds with standard RREA. Once an avalanche has developed with active feedback, however, fields are estimated to discharge in <1 ms (Dwyer (2003)). RFD also greatly inflates the number of photons in a shower, helping to explain TGF observations, while maintaining the energy distribution expected from RREA. Given enough space to cascade, Dwyer et al. (2003) shows that this enhancement increases the particle flux of Equation 3.2 exponentially over time, potentially by factors of up to $\sim 10^9$ and reaching the incredibly high photon fluences of $10^{15}$–$10^{19}$ implied by simulation Celestin et al. (2015).

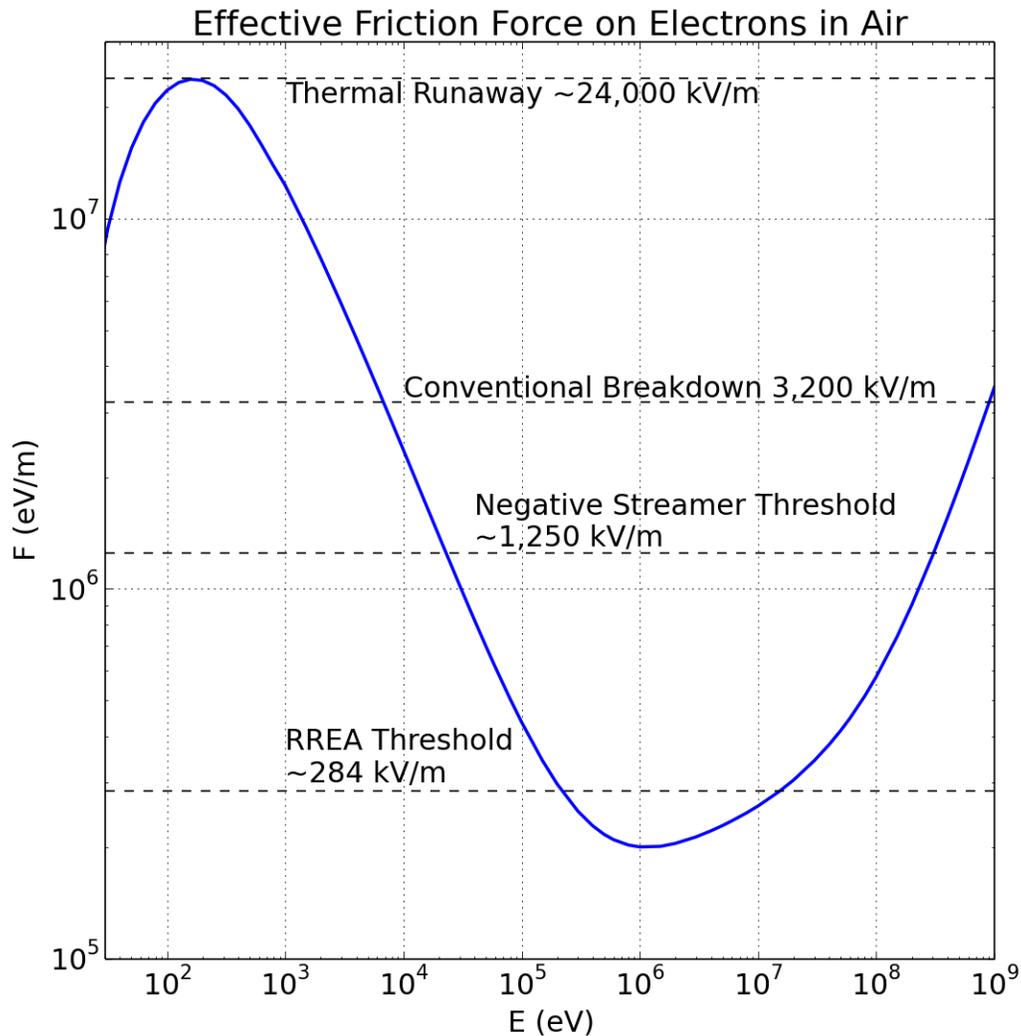

**Figure 3.1:** The energy loss of an electron of energy E per distance traveled in dry air at sea level, also called the stopping power or effective friction force. The two critical points at 160 eV and 1.2 MeV are referred to as the cold (or thermal) runaway and minimum ionizing particle cases, respectively. Several thresholds that are discussed in this section are labeled in units of the electric field strength required to exceed the friction curve. Data for energies $10^3$–$10^9$ eV were obtained via NIST database (Berger et al. (2017)). Data for 50–1000 eV were calculated following the examples of Moss et al. (2006); Diniz et al. (2019) and using the scattering cross section data from the Phelps database on lxcat (Phelps (retrieved July 20, 2021)).

### 3.2.3 Cold Runaway Model

Like RREA, the cold (or thermal) runaway model involves electric fields accelerating electrons above the friction curve of Figure 3.1. In the regime close to the local maximum at 160 eV, however, extreme fields in excess of $10^7$ V/m are required to accelerate electrons above their stopping power. Of course, large-scale electric fields have not been seen at this strength in thunderclouds. Instead, the model suggests that the locally-enhanced fields near streamer tips are sufficient to accelerate low-energy electrons to the relativistic regime.

As discussed in section 2.3, lightning leaders are hot, conducting channels that develop in discrete steps. Individual steps are formed via streamers, which heat up and ionize the virgin air ahead of the existing channel (da Silva & Pasko (2013a) and Figure 2.3). Simulations of the streamer zones near leader tips show that fields can be amplified as high as $10^7$ V/m, and can even briefly reach sufficient strength for thermal runaway during branching (Moss et al. (2006)), producing electrons up to 100 keV (Celestin & Pasko (2011)). At these energies, leader tip field enhancements on the order of $10^5$–$10^6$ V/m are more than enough to continue facilitating electron runaway to the MeV regime (Celestin et al. (2015)). These studies show that the fluence of runaway electrons would be as high as $10^{18}$–$10^{19}$ /s during streamer development. With durations of 1 $\mu$s–1 ms, this results in $10^{12}$–$10^{16}$ per leader step.

Further acceleration of these electrons can continue into high-energy x-rays and gamma-rays depending on the available potential difference surrounding the developing leaders. Electrons of $\gtrsim 200$ keV also become candidates for seeding traditional RREA in ambient electric fields of $\geq 2.8 \times 10^5$ V/m. Therefore, the cold runaway model is also consistent with traditional satellite TGF detections. In these cases, the runaway electrons emitted from streamer zones would subsequently seed RREA to match the observed hard spectra and possibly produce relativistic feedback to account for the extremely high fluences. In fact, cold runaway may also unify the x-ray observations of Dwyer et al. (2005) under a single model. The bursts were linked to low-altitude leader steps leading up to the parent flash's return stroke, but recorded spectra were too low energy for classic RREA. This may be explained as high fluxes of thermal runaway electrons accelerated by the remaining ambient field, but without developing into fully-fledged RREA.

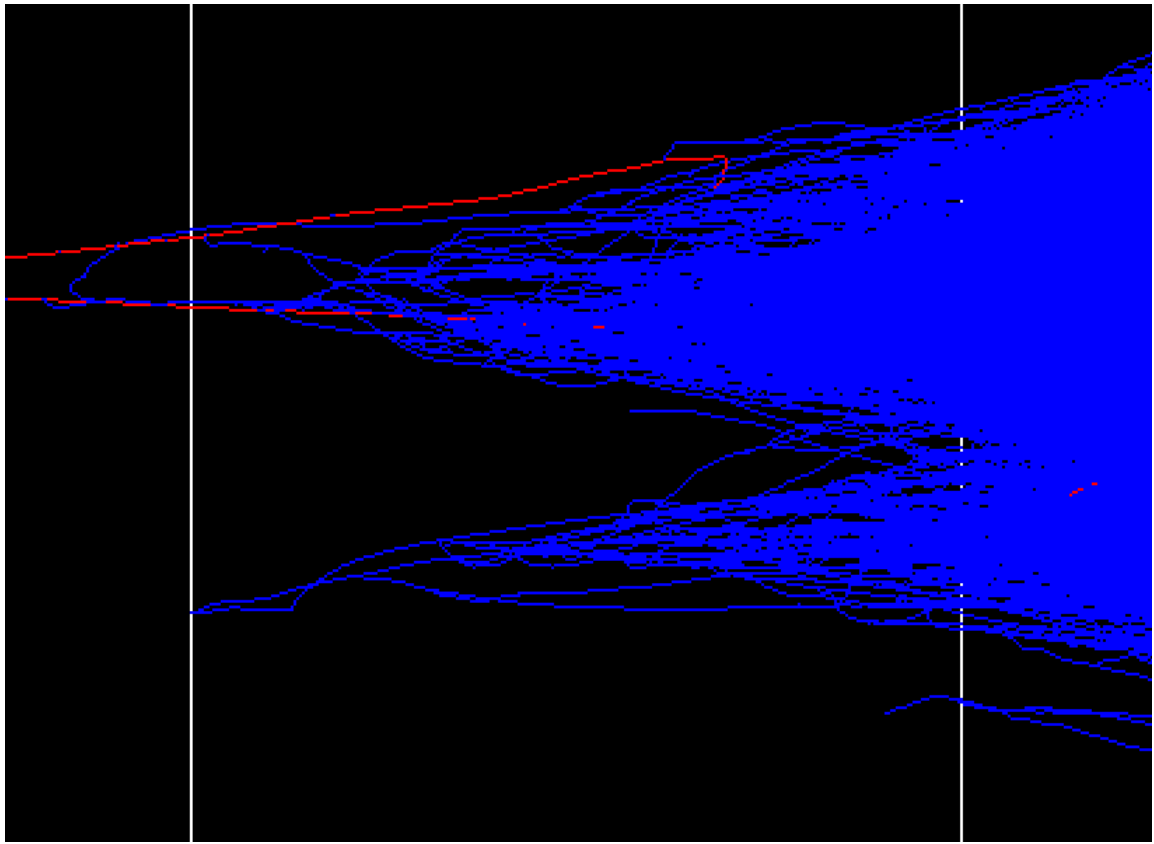

**Figure 3.2:** Particle tracking of RREA simulation using GEANT4. A single 1 MeV electron (blue tracks) was injected into an electric field of 1,500 kV/m 42 m long by 37 m wide. The RREA shower began in the lower left moving to the right against the field. A positron (red track) was emitting during the cascade and accelerated in the direction of the field. Pair annihilation cascaded into a secondary EM shower (upper left), greatly enhancing the RREA. The Simulation was performed by LeVon (2019).

### 3.2.4   Electromagnetic Showers

Now that we have established TGFs and their production, it is important to discuss the propagation and development of electromagnetic (EM) showers in general. To start, the term "gamma flash" is a bit of a misnomer; the phenomenon was originally observed via gamma detectors, but of course the shower composition itself is not purely photons. Figure 3.3 shows the particle composition of an EM shower without an applied electric field. Note that for both electron and photon-seeded showers, the composition becomes dominated by ∼90% photons, ∼7% electrons, and ∼3% positrons after only a few radiation lengths and remains constant. When subjected to an electric field, however, electrons in a shower remain at high energy longer, multiply faster, and lose less energy to bremsstrahlung photons. As a result, the ratio of photons to electrons is significantly lower. For strong electric fields between 400–2,000 kV/m, this ratio ranges between 1–0.1, resulting in a shower dominated by electrons rather than photons (Skeltved et al. (2014)). This relative composition remains constant after stabilizing, but returns to that of a standard EM shower after exiting the applied field.

Once a shower is initiated, its electrons, positrons, and photons collide with atmospheric molecules and introduce new particles into the shower via two dominant, high-energy processes: bremsstrahlung radiation and pair production ((Zyla et al. 2020, Section 34.4), Figure 3.4). These result in twice as many particles in the shower with 50% energy on average (minus 1.022 MeV rest mass energy for pair production). Therefore, after traveling $n$ avalanche lengths ($\lambda$), the shower fluence is multiplied by $2^{n\lambda}$ and average particle energy is reduced by $2^{-n\lambda}$. This growth continues until the average energy drops below the minimum ionizing level (1.2 MeV in air) and particles begin losing more and more with each interaction (see Figure 3.1). At these lower energies, the processes above become dominated by ionization, Compton scattering, and the photoelectric effect. For the case of a shower developing in an applied electric field such as RREA, electrons are continually accelerated after each interaction, keeping energy losses low and hastening shower multiplication according to Equation 3.2. An RREA shower seeded by cold runaway (subsection 3.2.3) is visualized in Figure 3.4 alongside the dominant particle interactions.

As discussed in section 4.2, the Telescope Array surface detectors are designed to detect charged particles only, meaning that gamma-rays are only recorded indirectly via secondary particles. Compton-scattering photons produce electrons which, if energetic enough, penetrate the SD enclosure and at least one layer of the scintillator to deposit energy. Even though these detectors cannot identify particles by design, we can use this inefficiency to our advantage. In a high-energy cosmic ray extensive air shower (EAS), the dominant component is muonic at ground-level ((Sokolsky 1989, Chapter 3)), for which the SDs have maximum efficiency (section 4.2). As such, the responses of the detectors' upper and lower layers are nearly identical as the vast majority of particles penetrate and deposit energy in both layers. Scattered electrons, on the other hand, are often detected by only a single layer and result in distinct signals (subsection 6.2.2), hinting at the electromagnetic composition passing through.

Another tool for identification is numerical simulation (Figure 3.5). These detectors are simulated in their entirety in GEANT4 in order to test their response to different particle types at various energies. An interesting part of these simulations is the ratio of energy deposit in the upper vs. lower scintillators, which can be used to identify an individual particle's type and/or energy. Unfortunately, the SDs have poor sensitivity to electrons $\lesssim$10 MeV where upper/lower ratios converge. Since RREA showers produce electrons with an average energy of 7.2 MeV (Dwyer (2004)), this becomes an unreliable source of identification. As expected, these observed ratios are indeed consistent with unity in the observations of this study, except for a few low-fluence exceptions (chapter 7).

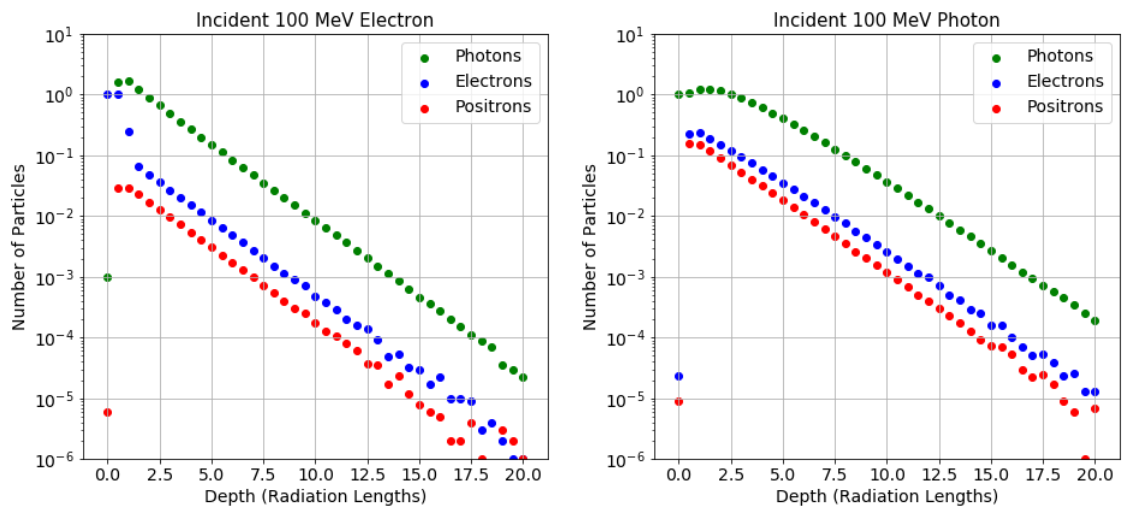

**Figure 3.3:** Particle populations in standard electromagnetic showers versus propagation depth. Showers were initiated by a single 100 MeV primary (electron in left panel, photon in right panel). Note that for both electron-seeded and photon-seeded showers, relative populations stabilize after only a couple radiation lengths ($\lambda$) at ∼90% photons, ∼7% electrons, and ∼3% positrons. Simulations were performed by /citesmout2020

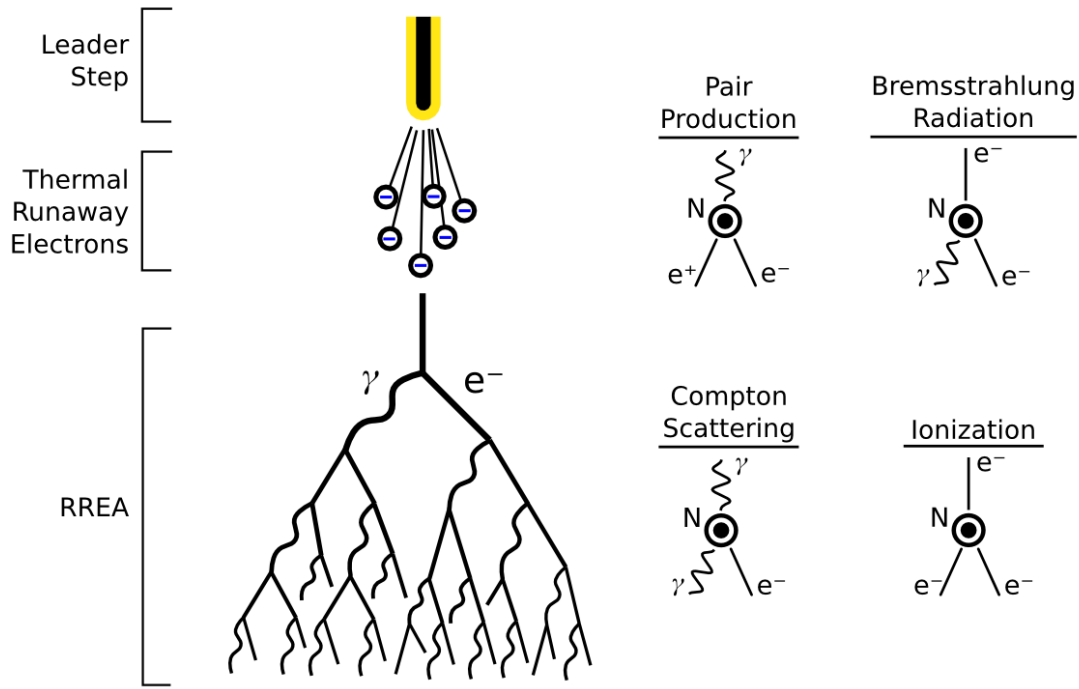

**Figure 3.4:** illustration of an electromagnetic shower fueled by RREA (subsection 3.2.1) and seeded via the cold runaway model of subsection 3.2.3. Under this model, thermal electrons are ejected during the leader-to-streamer transition in early leader steps (see Figure 2.3). In a sufficient electric field, these electrons seed a runaway cascade of primarily photons, electrons, and positrons which multiply by the four dominant processes shown. After losing energy via bremsstrahlung radiation or ionization, electrons are re-accelerated by the ambient field and continue to multiply. Under these conditions, additional rrea showers may be initiated by back-scattered photons and positrons as outlined in subsection 3.2.2.

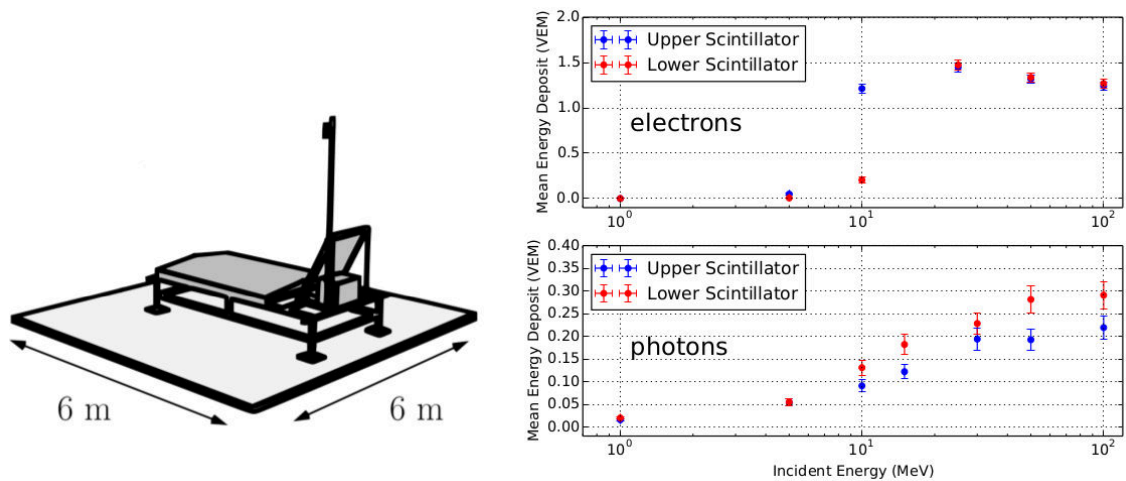

**Figure 3.5:** *Left:* Illustration of the detector model which was simulated in GEANT4 by Ivanov (2012). The full support structure, sensitive materials, and other components were was included in the model (section 4.2). *Right:* Resulting energy deposit in each layer of scintillator versus incident energy of individual photons and electrons.

# CHAPTER 4

# THE TELESCOPE ARRAY EXPERIMENT

The Telescope Array (TA) Experiment in Western Utah is made up of two components for the detection of cosmic rays. This chapter will focus on the physical instrumentation used for lightning detection, although a brief overview of cosmic rays is given in section 4.1. The first component, the surface detector array (TASD), is made up of 507 plastic scintillator detectors (SDs) spaced on a 1.2 km grid (Figure 4.1) and numbered as XXYY, where XX represents detectors' easterly positions and YY represents northerly positions. The original array covers 680 km$^2$, but is in the process of expanding to quadruple its size (see subsection 9.3.1). These detectors are designed to be sensitive to the charged particles produced in a cosmic ray shower. In this study, however, the SDs are indirectly detecting lightning-produced neutral gamma-rays that interact with the enclosure or the atmosphere overhead to produce sufficiently-energetic electrons. The details of these observations will be explored in chapter 7.

The second component, fluorescence detectors (FDs), are located at three stations around the array and consist of several telescopes overlooking the surface detectors. These detectors are sensitive to the UV light produced as cosmic ray air showers excite air molecules. Because of the telescopes' sensitivity, they are operated only during dark nights with no moon and no visible lightning which could overload the cameras' photomultiplier tubes (PMTs). For this reason, fluorescence data is not used in lightning analysis, but will be briefly reviewed in section 4.3.

## 4.1 Ultra-High Energy Cosmic Rays

Cosmic rays were first discovered in 1912 by V.F. Hess after measuring that electroscope discharge rates increased with altitude, suggesting that extraterrestrial radiation was being absorbed by the atmosphere (Hess (1912)). Since then, the rays (now known to be primarily charged nuclei) have been studied intensively. The energy spectrum follows an approximate broken power law with regions defined by $dN/dE \sim E^{-\alpha}$, where $\alpha$ dictates the slope (Sokolsky & Thomson (2020)). At relatively low energies of $10^7$ eV, the proton flux is $1/m^2$-sec-MeV and the spectrum is consistent with $\alpha = 2.6$ up to $\sim 10^{15}$ eV at the 'knee'. Between $10^{15}$ eV–$10^{18.5}$ eV, the slope steepens to $\sim 3$ and ends with the 'ankle'. $\alpha$ then levels off slightly, returning to $\sim 2.5$ (see TA results of Bergman (2020)).

$10^{18}$ eV (1 EeV) marks the lower bound of ultra-high energy cosmic rays (UHECRS), for which specialized, large-scale instrumentation is required to achieve any meaningful statistics — at these very high energies approaching $10^{20}$ eV, proton flux is as low as 1 per km$^2$ per century. The final feature in the spectrum is the GZK cutoff at $10^{19.7}$, named after its theorists Greisen, Zatsepin, and Kuzmin (Greisen (1966); Zatsepin & Kuzmin (1966)). The cutoff is predicted under the assumption of proton primaries, which interact with CMB photons to produce pions via $\Delta$ resonance. The extremely low flux at these energies requires any attempt to study UHECRs to be performed on a massive scale.

There are four basic methods for the detection of cosmic rays. Direct measurements involve satellite-borne instruments such as calorimeters which record the deposited energy from low-energy cosmic ray primaries. Above $10^{15}$ eV, however, the flux is too low to yield significant results. The three main methods of indirect measurement (ground arrays, fluorescence detectors, and Cherenkov detectors) make use of extensive air showers (EASs) to more easily detect individual events. When cosmic ray primaries enter Earth's atmosphere, they interact with atmospheric nuclei and produce billions of energetic secondary particles. These secondaries continue scattering and cascading, eventually generating full showers consisting of both hadronic and electromagnetic components. EM showers develop as discussed in subsection 3.2.4 without a driving electric field. The hadronic component is concentrated at the shower's core, seeding EM showers and producing muons via pion and kaon decay (Sokolsky & Thomson (2020)). These muons are vital in the study of cosmic rays — once produced early in the shower, muons have small cross-sections and are more likely to reach detectors at ground level than other particles. As such, a shower's muon distribution carries information about the earliest particle interactions, including the primary cosmic ray.

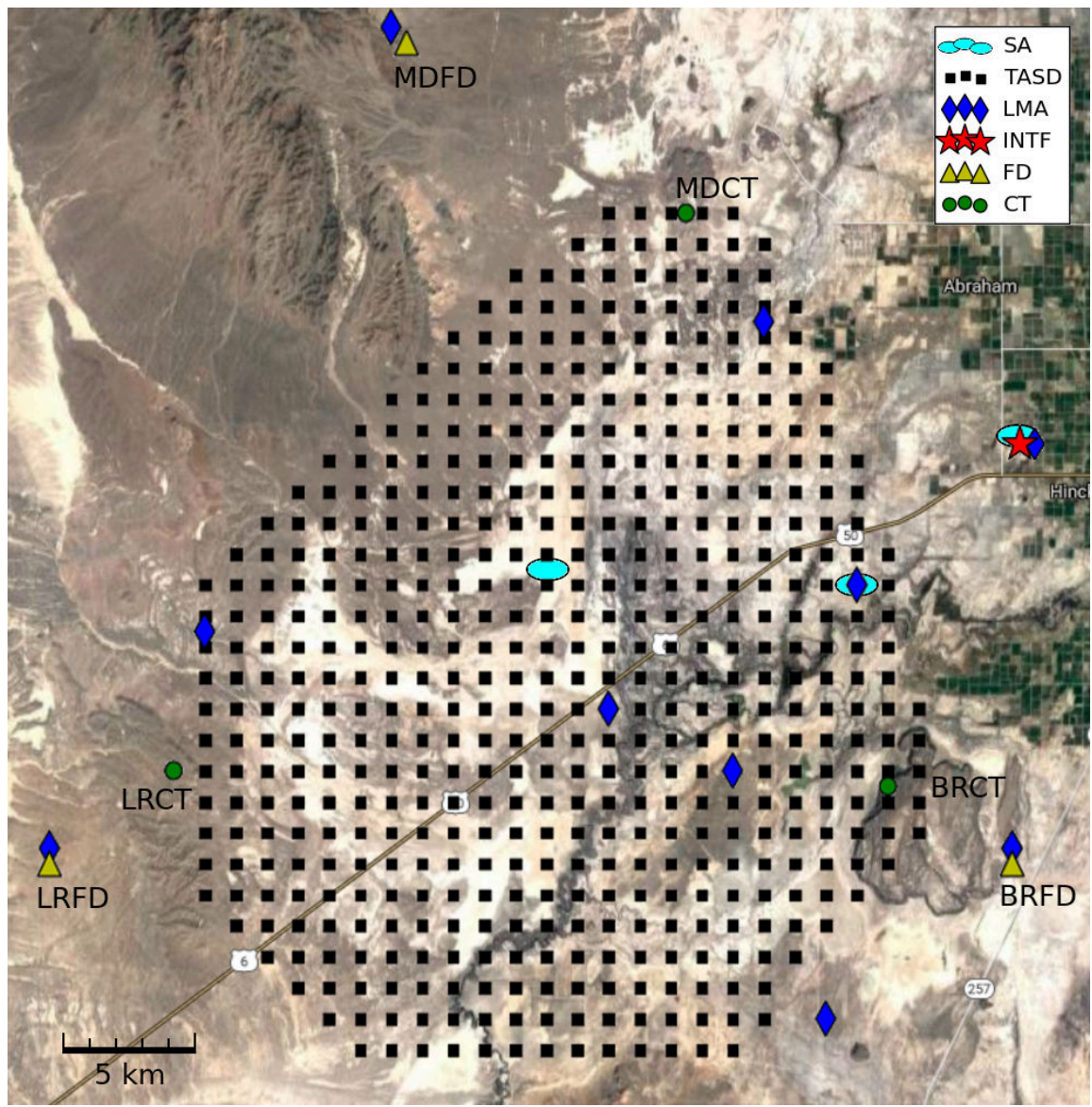

**Figure 4.1:** Map of the Telescope Array in Western Utah. The TASD locations (squares) show only the main array as it was in 2018 before TAx4 expansion and not including the low energy expansion (TALE). LMA detectors are shown as diamonds, the INTF station (with accompanying fast sferic sensor) is located at the star, and slow sferic sensors (SAs) are marked by ovals. TA fluorescence stations (FDs) and control towers (CTs) are also given as triangles and circles (respectively) and labeled.

EASs can spread to many tens of km$^2$ at ground level, and are best measured using an array of many widely-spaced detectors in order to maximize total area. These ground arrays are usually comprised of many scintillation detectors or water tanks to measure the energy deposit of shower secondaries that penetrate the atmosphere. Early arrays were pioneered by the Rossi group, showing UHECR detections of over $10^{18}$ eV in a $10^5$ m$^2$ array (Clark et al. (1961)). In the following decades, many robust arrays have been built all over the world with areas up to thousands of km$^2$. The original Telescope Array surface detector array (TASD, section 4.2) covered over 680 km$^2$, but is currently expanding to four times the size to study the highest-energy cosmic rays (subsection 9.3.1). The largest current ground array is the Pierre Auger Observatory, covering about 3,000 km$^2$ and utilizing Cherenkov water tanks instead of scintillation detectors (Aab et al. (2015)). Note that these detectors should not be confused with atmospheric Cherenkov detectors (see below). In addition to their large size, ground arrays are able to operate with ~100% uptime, unlike fluorescence detectors.

The air fluorescence technique uses the atmosphere itself as a calorimeter, detecting the energy lost by a cosmic ray to the air (section 4.3). As an EAS develops, secondary particles excite atmospheric nitrogen and oxygen, emitting UV light which can be measured by sensitive telescopes. This method allows observation of the entire shower development rather than just those particles which arrive at the ground. Fluorescence detectors (FDs) have very large detection volumes, but are restricted to run only on clear, moonless nights. They can be operated alone or in tandem with surface arrays, such as the Telescope Array.

As opposed to the above techniques, atmospheric Cherenkov detectors are typically used for lower-energy cosmic rays, like the VERITAS gamma-ray observatory (Weekes et al. (2002)). Similar to air fluorescence, this method uses the atmosphere as its medium, measuring the emitted light of cosmic rays interacting with air molecules. As the name suggests, detectors measure Cherenkov radiation emitted by the speeding shower secondaries rather than fluorescence. Because this radiation is only emitted in a narrow forward cone, Cherenkov detectors have a much smaller field of view, only detecting showers directed toward them and are not suited to the study of UHECRs.

## 4.2   Surface Detectors

Full documentation of Telescope Array surface detectors (Figure 4.2) can be found in Abu-Zayyad et al. (2013); Ivanov (2012), though an overview is presented here which is pertinent to the study of TGFs. The TASDs have three main elements: the sensitive materials and protective structure, the electronics and communication equipment, and the software for operation and triggering logic. Two layers of plastic scintillator are used to detect only charged particles that pass through, causing fluorescence in the sensitive material. Generated light then travels through embedded optical fibers before being captured by photomultiplier tubes and producing an analog signal, at which point the electronic components take over. The analog signal is then "counted" digitally as a tally of energy deposit in the detector. Unlike calorimeter-type detectors, the SDs do not necessarily capture 100% of particle energy nor track particle trajectories. Their advantage lies in their high time-resolution measurements of energy deposition, which is vital in the analysis of cosmic ray showers and, as it turns out, TGFs.

### 4.2.1 Scintillator and Enclosure

The TASD plastic scintillator (Abu-Zayyad et al. (2013)) is composed of two layers of base material polyvinyl toluene ($C_9H_{10}$, 1.032 g/cm$^2$), which fluoresces when exposed to charged, ionizing particles. From Figure 4.3, the layers are divided into two pieces, each 1 m x 1.5 m (3 m$^2$ total). Each piece of plastic scintillator is 1.2 cm thick and has 1.2 mm-diameter grooves, separated by 2 cm, which house the Y-11 Kurray wavelength-shifting optical fibers to reduce the light's attenuation on its way to the photomultiplier tubes (PMTs). Both pieces of each layer are fed into the two PMTs providing a gain of $\sim$2$\times$10$^6$ per photoelectron and are read out by onboard electronics for processing (see following sections). The two layers are separated by a 1 mm thick stainless steel plate and wrapped in tyvek sheeting and encased in a final light-tight stainless steel box with thickness 1.2 mm.

The scintillator box sits on a steel frame covered by a 1.2 mm steel roof (Figure 4.2). The frame also supports its local electronics, power supply, and communication antenna. The entire unit is powered by a single 12 V deep-cycle battery operating at 5 W and is fully charged by a 1 m$^2$, 125 W solar panel mounted on the frame, making for $\sim$100% uptime. Tall poles are also attached for antenna support. Each of these components have been combined in a GEANT4 model for accurate simulation of the detector's response to incident particles (discussed in relation to the events of chapter 6). The SDs' focus on charged particles, which dominate fully-developed EASs, means an inefficiency in detecting incident neutral photons. Because of this, all gamma-ray detections are actually indirect and rely on Compton scattering in the detector's vicinity, steel enclosure, or in the scintillators themselves. The subject of particle detection and calibration is discussed further in subsection 4.2.3.

### 4.2.2 Electronics and Communication

The scintillators' fiber optic output is fed constantly into a pair of PMTs which are read out by independent flash analog-to-digital converters (FADCs). The FADCs are sampled at 50 MHz, integrating energy deposit into 20 ns bins. Each bin records this deposit on a 12-bit scale from 0–4096, called the FADC count. The true energy equivalent of an FADC count is recalibrated in ten-minute intervals (subsection 4.2.3). As a rough reference, 18 FADC counts corresponds to ~1 MeV. These FADC responses are grouped into series of 128 bins (2.56 $\mu$s) called waveforms and are used to determine automatic triggering, divided into levels 0, 1, and 2.

A level-0 trigger is defined when both upper and lower waveforms report an integrated signal over 15 FADC counts (above the background pedestal). Waveforms at this level are stored locally and can be sent to the communication towers if requested. A level-1 trigger is defined as ten times that of level-0, namely that both layers report 150 FADC counts integrated above the pedestal. Level-1 trigger times are always reported to communication towers and are used as the criteria for level-2 triggers. The 507 TASDs operate independently, but are presided over by three communication towers positioned at, and named after, high points in the local geography (Figure 4.1). These are Smelter Knolls (SK), Blackrock Mesa (BR), and Longridge (LR); the latter two are colocated with fluorescence stations overlooking the array. The towers facilitate remote communication and handle group triggering protocols via WLAN modems and 2.4 GHz directional radio antennas (Figure 4.2).

A level-2 (or 'event') trigger is called when three adjacent detectors report level-1 triggers within the same 8 $\mu$s window. When called, level-0 waveforms ($\geq$15 FADC counts) from all detectors within 32 $\mu$s are requested and transferred to long-term storage for later analysis.

### 4.2.3 Detector Response and Calibration

TASDs are not designed to reconstruct individual particle energies, instead recording energy deposit as a count of the number of equivalent particles which pass through. We define this unit as the vertical equivalent muon (VEM), representing the expected energy deposit in one layer of plastic scintillator by a vertical muon at the minimum-ionizing energy, being the most abundant energetic particle capable of penetrating the TASD enclosure. This process is simulated in GEANT4 for muons up to 1 GeV (Figure 4.4). The minimum deposit occurs at 300 MeV, defined as the minimum ionizing particle (MIP). The distribution of possible energy deposit by an MIP then follows the approximate Landau distribution (Landau (1944)) shown in Figure 4.5. The most probable value (MPV) of this distribution for a vertical MIP incident on a TASD is ∼2.05 MeV, defining a VEM.

After signals are digitized by the FADC in the previous section, they are stored as waveforms, one for each layer. Since FADC counts can fluctuate and do not have an intrinsic energy equivalent, the TASDs record pulses from level-0 triggers every ten minutes for later calibration. Since muons with minimum-ionizing energy are the most common incident particles, level-0 pulses are tallied and compared to the MIP energy deposit distribution of Figure 4.5 to define a relationship between FADC counts per VEM. This is performed during later analysis by simulating the TASDs' responses to atmospheric particles and adjusting gain to produce the same distribution reported by each detector (Taketa et al. (2009)). Figure 4.6 shows the typical range of this relationship across the array with an average of ∼36 FADC counts per VEM (or ∼18 FADC counts per MeV). In addition to FADC calibration, the background pedestal level is recorded from each waveform and adjusted to keep the level-0 trigger rate at 750 MHz, equal to the expected rate of random incident muons in a single TASD. A more in-depth documentation of TASD calibrations can be found in Ivanov (2012).

## 4.3   Fluorescence Detectors

The Telescope Array fluorescence detectors (FD) work on a different premise. As cosmic rays collide with the upper atmosphere and develop showers, secondary particles excite atmospheric nitrogen and oxygen, which subsequently emit UV light. This emission persists throughout the cascade, producing a characteristic profile of the EAS. FDs are positioned at three stations surrounding Telescope Array (Figure 4.1) and consist of 4–14 telescopes using large mirrors to focus these UV emissions onto a 16x16 array of sensitive PMTs. Because of the great sensitivity of these instruments, operation is restricted to dark, moonless nights with minimal light pollution. The detector's low uptime is countered by its field of view covering an incredibly large volume above the Telescope Array (Tokuno et al. (2012); Abbasi et al. (2016)). In addition to single-telescope (monocular) observations, when an event is detected by at least two stations, the intersection of the shower planes uniquely determines the 3D position and orientation of a given EAS (Tameda et al. (2009)). This stereo method of reconstruction provides more detailed characteristics, rather than merely a silhouette.

The FD stations at LR (Southwest) and BR (Southeast) have 12 telescopes each overlooking the array. The largest FD station North of the array at Middle Drum (MD) has 14 primary telescopes (Figure 4.7) in addition to 10 telescopes adapted to observe lower-energy cosmic rays that may not penetrate deep enough in the atmosphere to be seen by the primary mirrors. Accompanied by a dense subset of TASDs, these are known as the TALE extension (Abbasi et al. (2021b)), designed to extend TA's energy sensitivity below $10^{16}$ eV. Both MD and BR stations have also been upgraded with new sets of four mirrors each in order to overlook the newly-expanded Northern and Southern lobes of TAx4 (see subsection 9.3.1).

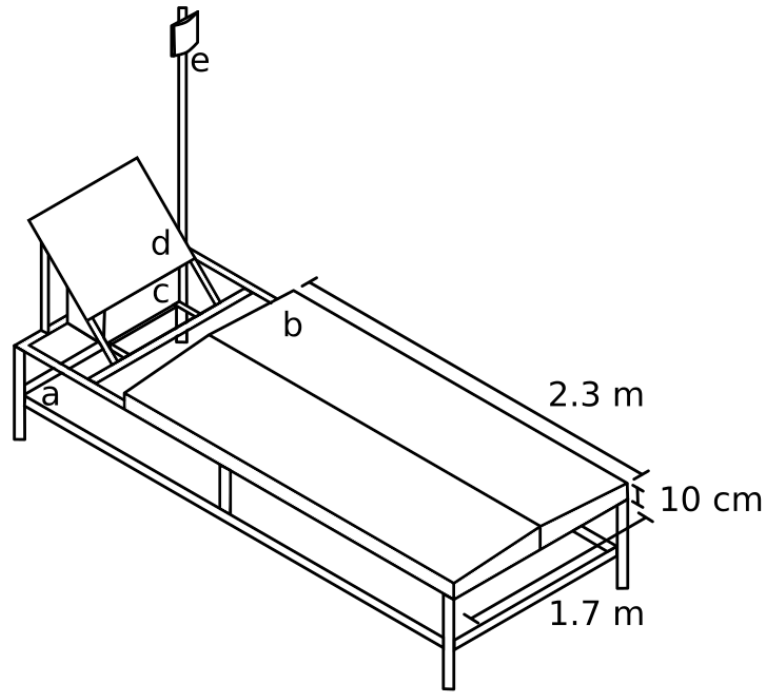

**Figure 4.2:** Diagram of a TASD with labeled components. *a)* Steel frame supporting all other components. *b)* Scintillator box (Figure 4.3) mounted on the frame and covered by a 1.2 mm-thick iron roof. *c)* Electronics box containing computer and battery. *d)* Solar panel providing power and charging the battery. *e)* Directional antenna for connection with communication towers.

As mentioned in this chapter's introduction, air fluorescence detectors are not currently used for this study of lightning, and are not addressed further. More information on these detectors and their findings can be found (Tokuno et al. (2012); Abbasi et al. (2016, 2021b)).

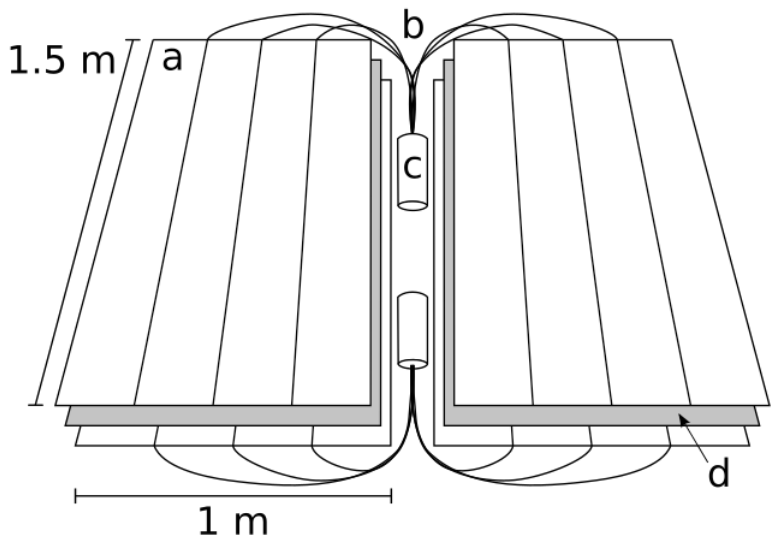

**Figure 4.3:** Schematic of the sensitive materials in each TASD. Two layers of scintillator (a) are separated by a steel plate (d), each with embedded wavelength-shifting fibers (WLSFs) (b) which carry light to two photomultiplier tubes (PMTs) (c) which read out the signal to the electronics.

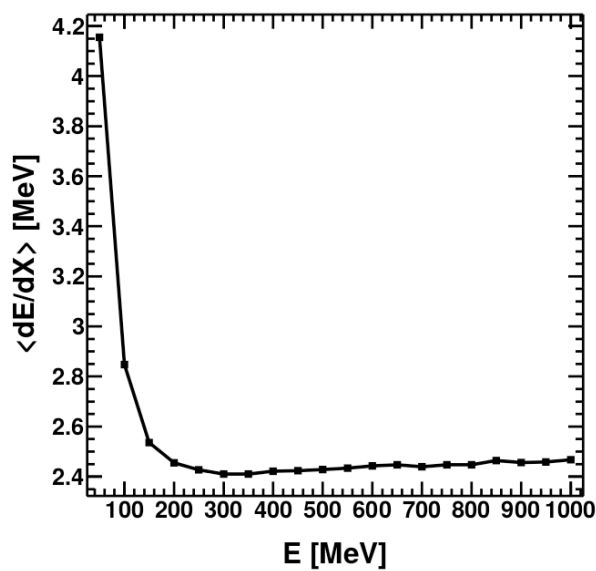

**Figure 4.4:** Results of GEANT4 modeling of incident muons on a TASD. Energy deposit is given versus incident particle energy. The minimum ionizing energy occurs at 300 MeV, defined as the minimum-ionizing particle (MIP). The expected energy deposit of MIPs is shown in Figure 4.5. Simulations and figures were produced by Ivanov (2012).

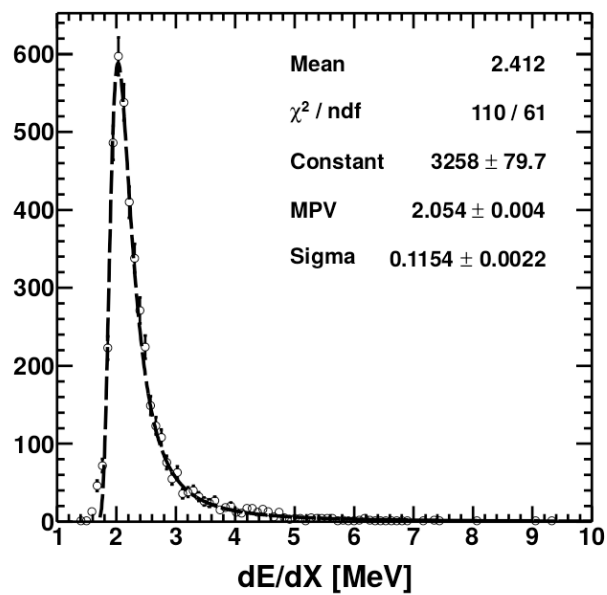

**Figure 4.5:** Results of GEANT4 modeling of incident vertical 300 MeV muons (MIP from Figure 4.4) on a TASD, following an approximate Landau distribution. The most probable value is at the histogram's peak of ~2.05 MeV, defined as a vertical equivalent muon (VEM). Simulations and figures were produced by Ivanov (2012).

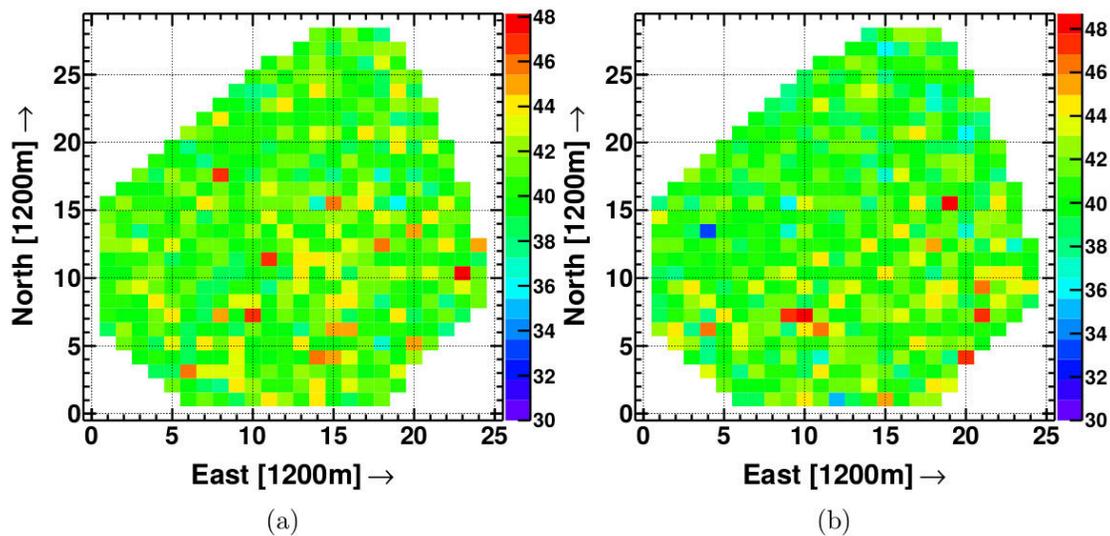

**Figure 4.6:** Calibration results showing mean FADC counts per VEM calculated for each surface detector on a typical day, (a) for upper scintillator layers and (b) for lower layers. On average, this relationship is approximately 36 FADC counts per VEM. Simulations and figures were produced by Ivanov (2012).

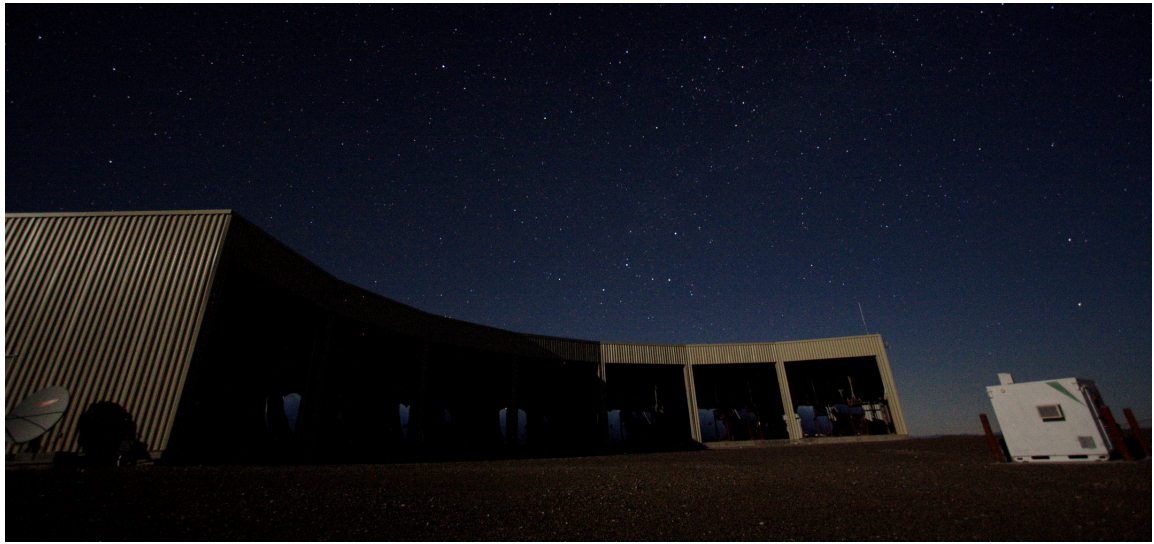

**Figure 4.7:** The 14 primary telescopes (2 per bay) of the Middle Drum fluorescence station at Telescope Array.

# CHAPTER 5

# LIGHTNING DETECTORS

Chapter 6 describes the initial detection of burst-type events at the Telescope Array (Abbasi et al. (2017)). Once the phenomenon's association with lightning strikes was realized, Telescope Array began a collaboration with Langmuir Labs at New Mexico Tech for their expertise in studying energetic radiation from lightning. TA's potential for TGF observations quickly became clear and was supplemented with the lightning mapping array (LMA, section 5.1) and slow sferic sensors (section 5.2) in the summer of 2014. The longer study of burst events using the new instrumentation was published in 2018 (Abbasi et al. (2018), section 6.2). The study concluded that the events were indeed downward TGFs, and additional, substantial upgrades were added to the growing array in summer of 2018, including a broadband interferometer (section 5.3) and fast sferic sensor (section 5.2). This chapter shall lay out the design and implementation of these varied instruments and their use in observing relevant lightning mechanisms.

## 5.1   Lightning Mapping Array

The Lightning Mapping Array (LMA) was the first lightning upgrade installed at Telescope Array after the initial discovery of burst events at the site (section 6.1). The LMA is a series of detectors (Figure 5.1) each consisting of an omnidirectional antenna, computer for processing/storage, and solar power supply system to ensure ∼100% uptime.

The antennas are tuned to VHF frequencies centered at 63 MHz, corresponding to the peak radiation produced by positive bipolar breakdown (Rison et al. (1999)). This type of breakdown has been shown to occur in the mid-level negative and upper positive charge regions in thunderclouds, producing VHF radiation ∼30 dB above that of other lightning processes. The resulting narrow, bipolar pulses are measured at each LMA station within range — these differing arrival times are then correlated between detectors and used in a time-of-arrival (TOA) analysis technique to determine 3-dimensional locations of the individual VHF pulse sources (e.g. Figure 5.2). In most cases, these strong pulses have well-defined peaks such that the time recorded at each station corresponds to the same source. It should be noted, however, that the LMA is susceptible to mislocating point sources during non-impulsive or VHF-noisy signals in which stations disagree on the timing of the signal's peak (Tilles et al. (2019)). This can occur during fast breakdown in NBEs (section 2.4) and during high-power, initial leader steps.

Using the locations and detection times $(x_i, y_i, z_i, t_i)$ of four different LMA stations $i$, the following equation can be solved for the pulse's source location and time $(x, y, z, t)$:

$$(c(t - t_i))^2 = (x - x_i)^2 + (y - y_i)^2 + (z - z_i)^2 \tag{5.1}$$

where $c$ is the speed of light. Although the solution to Equation 5.1 requires a minimum of only four participating stations, LMA observations require at least two degrees of freedom ($\geq$6 stations) to ensure accuracy. With ten LMA stations deployed at Telescope Array, overhead lightning flashes often produce hundreds or thousands of data points per flash.

The accuracy of LMA solutions depends on the geometry of each deployment, though uncertainties in previous studies (Rison et al. (1999); Thomas et al. (2004)) were reported as 40–50 ns in timing, 10–50 m in horizontal position, and 20–100 m in vertical position for overhead lightning signals, scaling according to $r^2$ at distances $r$ from the array's center. Solution accuracy and alternate methods of source determination are explored in greater detail in Thomas et al. (2004). Similar tests with known sources were not performed at TA, but the smaller 30–40 km diameter of LMA stations (Figure 4.1) would likely result in larger errors of individual source locations. Note, however, that the overall charge structure grouping of lightning data recorded at TA (e.g. Figure 5.3) is comparable to that of the studies above. For the analysis of events at Telescope Array published in 2018 (Abbasi et al. (2018)), the accuracy was sufficient to show that the timing of TGFs detected by the TASD were produced during the initial few milliseconds of lightning leader activity overhead and occurred within 1–3 km (approximate radius of particle detections at ground level). LMA observations during these initial stages produced point solutions at a rate of 1–10 per millisecond, meaning that mislocated solutions could be easily identified. These observations and their implications are discussed further in section 6.2.

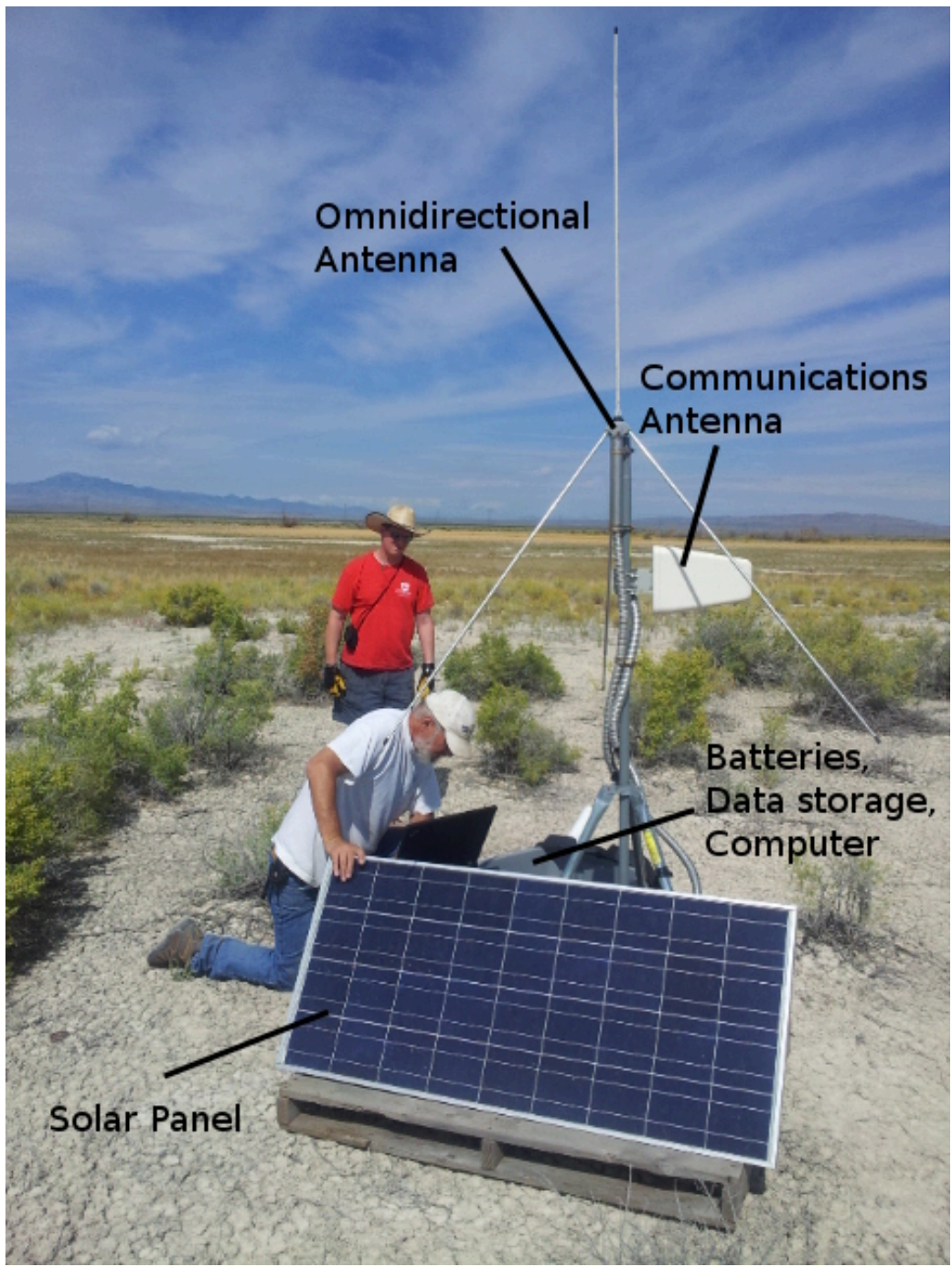

**Figure 5.1:** Bills Hanlon and Rison (shown in red and white respectively) working on one of nine stations of the lightning mapping array (LMA) colocated with the Telescope Array. The omnidirectional antenna records local VHF source amplitude and time for reconstruction. The communications antenna runs of a cell modem and is directed toward a nearby cell tower for remote communication and data transfer. The Electronics box contains the LMA computer, batteries, and hard drives for data storage. Batteries are connected to the solar panel for ∼100% detector uptime.

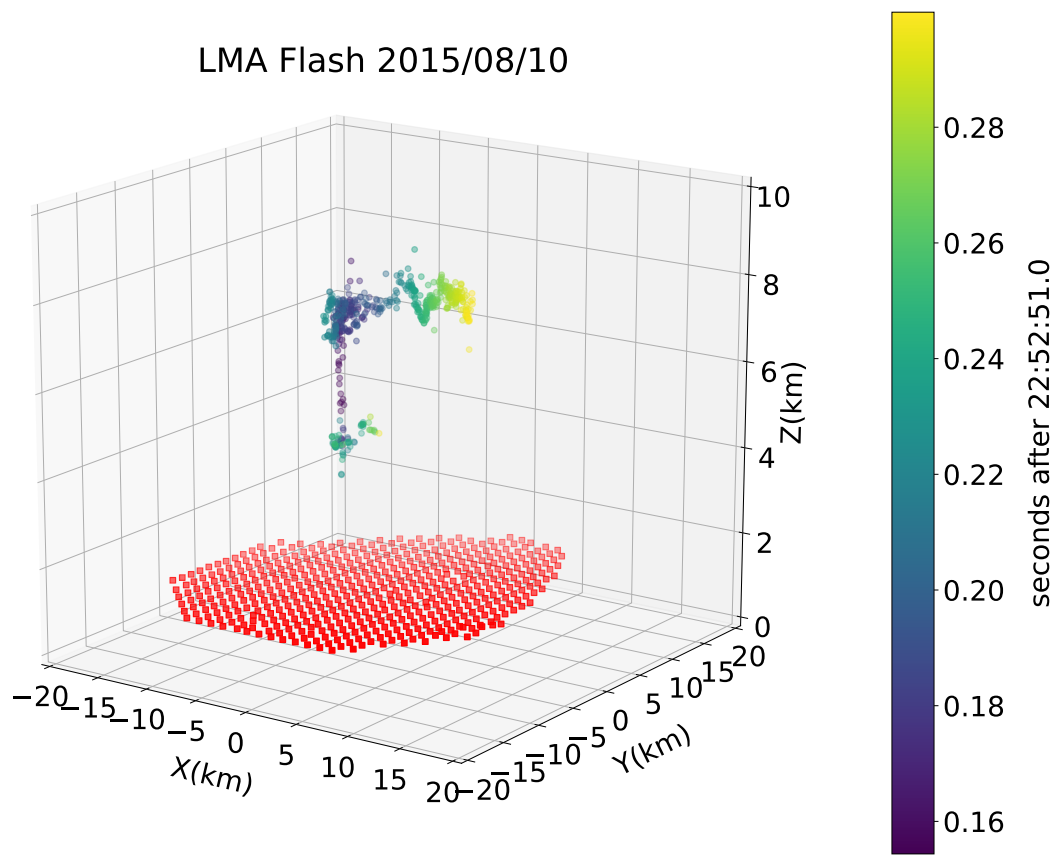

**Figure 5.2:** LMA data of an upward +IC flash from 2015/08/10. VHF pulses detected by ≥6 stations reconstructed as 3-dimensional point sources. Red squares at ground level represent TASDs with coordinates (0,0) representing the center of the array and altitude measured above sea level. The VHF sources are shown to develop upward very quickly from the mid-level negative region at 6 km to the upper positive region at 8.5 km in the first 10 ms, followed by increased activity in both regions and horizontal branching in the upper positive.

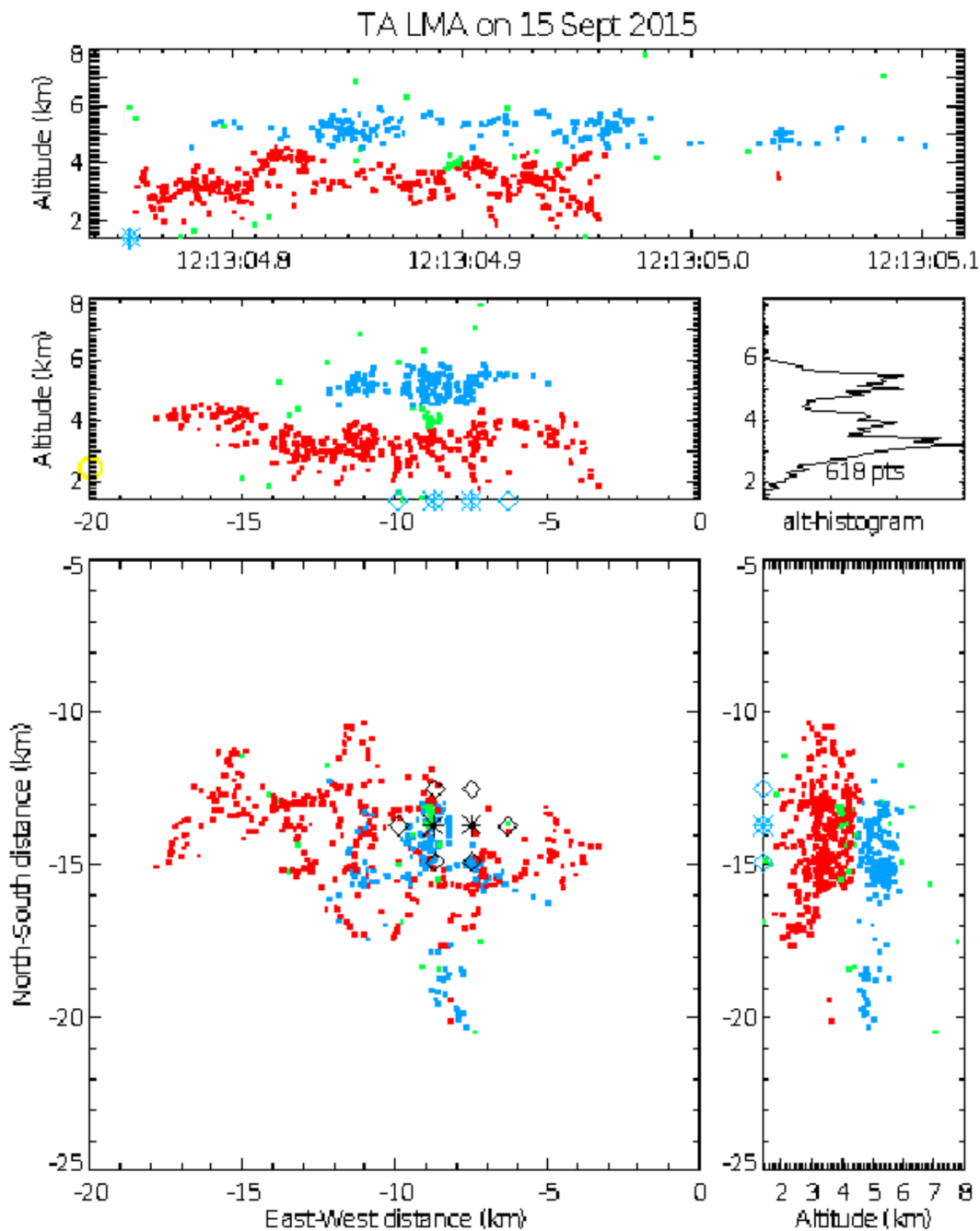

**Figure 5.3:** LMA observations of FL1 at 12:13:04 on 2015/09/15. Blue (red) points indicate LMA point source solutions occurring in the mid-level negative (lower positive) charge regions, while green points indicate sources between or outside of the approximate charge regions. Breakdown activity began at ~6 km and quickly progressed downward to the lower positive region, at which time the TASDs recorded multiple triggers (detectors with high energy deposit are marked by asterisks while others are marked with diamonds). Lightning activity continued in both charge regions for ~200 ms before the flash terminated in a –CG stroke.

## 5.2    Sferic Sensors

The term 'sferic' comes from 'radio atmospheric', the EM pulses produced by lightning flashes. They are produced by the changes in charge and current during flash development and consist of a wide band from ∼0.1 Hz to over 10 MHz (Uman (1987)), but are most intense in VLF (3–30 kHz) at large distances. Low attenuation at these frequencies means that lightning can be detected and measured from tens or hundreds of kilometers away. Perhaps the simplest method is that of a flat plate antenna, on which charge is induced by the electric field. Current to the plate can then be integrated to obtain a voltage proportional to both the induced charge and the field itself. An RC circuit is also included to slowly discharge the signal according to the time constant $\tau = RC$. The resistance $R$ and capacitance $C$ can then be modified to achieve the desired decay time — for example, an early version of the array in New Mexico used $\tau = 10$ s to examine the electric field development over the course of an entire flash (Krehbiel et al. (1979)).

Three slow antennas (SAs) with similar design were deployed at Telescope Array (Figure 5.4) alongside the LMA (section 5.1) in 2014. Compared to the original design, the newer sensors were designed to be easily relocated, as they have been over their >7-year deployment. All sferic sensor locations in Figure 4.1 are given as they were at the time of this study's 2018 events. As of 2021, these positions have not changed. The SAs are tuned to $\tau = 10$ s and sampled at 10 kHz in order to detect large-scale E-field changes during lightning flashes while remaining insensitive to short-duration pulses. These changes are recorded with a sensitivity range of $10^{-2}$–$10^4$ V/m with 24-bit sampling. During the upgrade of 2018, a fast antenna (FA) was deployed alongside the broadband interferometer (INTF). The FA is nearly identical in design to the SA, only with $\tau = 100$ $\mu$s and sampled at 180 MHz to more closely examine impulsive signals like IBPs during early leader development (subsection 2.4.1). Figure 5.5 shows example signals from both slow and fast sferic sensors during a single flash recorded on 2018/08/02.

## 5.3    Broadband Interferometer

The instrumentation upgrade in summer of 2018 included the broadband interferometer (INTF) and fast sferic sensor location at the station indicated by Figure 4.1 and based on the work of Stock et al. (2014). As discussed in section 5.2, sferic sensors are primarily sensitive to LF and VLF radiation (<30 kHz) produced by lightning return strokes and other large-scale activity. Small-scale breakdown, however, is bright in VHF between 30–300 MHz. This is the mechanism allowing the LMA to track lightning development on a smaller scale, though only utilizing a very narrow band of 60–66 MHz. As the name suggests, the broadband interferometer is sensitive to a much larger range of frequencies between 20–80 MHz waveforms and is digitized at 180 MHz, allowing high-resolution observations of the most intimate processes involved in breakdown. The INTF consists of three flat plate antennas (Figure 5.6) installed in a triangle with average baselines of 113 m (Figure 5.7). The antennas have similar design to those of the sferic sensors, but without the discharging RC circuit, allowing sensitivity to higher frequencies. The array's radiation pattern is symmetric about its central vertical axis and has an effective range of ∼50 km for high-quality observations covering all of TASD. At low elevation angles, however, the detection efficiency drops from ∼100% at 20 km range to ∼80% at the farthest Southwest side of TA at 36 km.

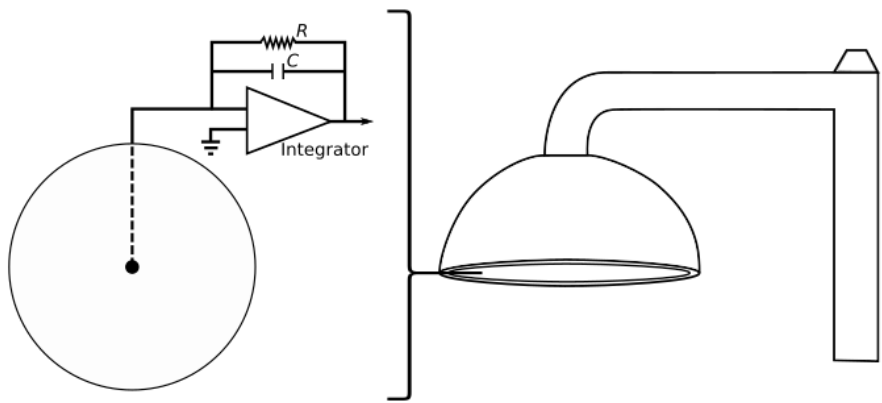

**Figure 5.4:** Simplified schematic of sferic sensors at Telescope Array. The sensitive, flat-plate antenna faces downward under the electronics shielding and is positioned ∼2.3 m above the ground. The sensor's position and accurate time are attained via a GPS antenna and recorded alongside data on two local 512 GB SD cards. The system runs on battery power charged by solar panel to ensure ∼100% uptime. Positions of slow antennas (SAs) are labeled on Figure 4.1 with a single fast antenna (FA) at the INTF site.

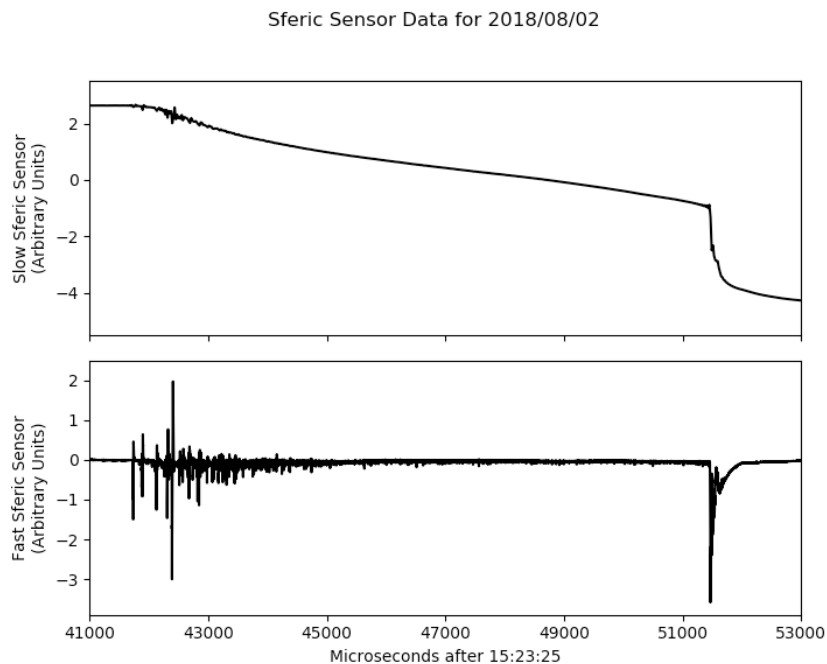

**Figure 5.5:** Sferic sensor data for the parent flash of TGF B. SA data in the upper panel records the relative change in ambient electric field during the entire flash, showing a slow discharge during leader progression and a sharp drop at the time of the –CG return stroke. FA data in the lower panel records impulsive variations, highlighting the extremely strong IBPs at the beginning of this flash which were comparable in strength to that of the return stroke. For a more quantitative analysis of this flash, see chapter 7.

Unlike the LMA and slow sferic sensors which observe and record lightning data in real time, the interferometer's three antennas record information faster than it can be processed and would quickly fill up data storage. Instead, the system only stores data for which the triggering criteria has been met. Triggering logic and data storage is handled by a local computer on site, seen as the building in Figure 5.7. As the data from antenna C is digitized into bins at 180 MHz, the change in signal strength between data points is monitored against a preset threshold — when the signal changes by more than this value over a set number of bins, the system declares a trigger and associated flash data is recorded for post-processing. At Telescope Array, this threshold is typically set to 8.8 mV (out of 1.5 V range) over 10 bin counts, but can be adjusted based on local conditions.

Once an event has been recorded, it is processed using the method of cross correlation with multiple baselines, a technique common in similar experiments (Sun et al. (2013)). This method can be performed with a minimum of three antennas (two pairs sharing one antenna, e.g. A–B and C–B) in order to determine 2D locations of a point source in spherical coordinates. For each pair, the angle $\alpha$ between the first baseline and the source direction is given by

$$cos(\alpha) = \left(\frac{\Delta\phi}{2\pi}\right)\frac{\lambda}{d} \tag{5.2}$$

where $\Delta\phi$ specifies the phase difference between the two signals, $\lambda$ is the radiation wavelength, and $d$ is the baseline length. The result is symmetric about the baseline's axis, resulting in a conical surface of possible sources. The same calculation can be performed for $\beta$, the angle between the second baseline and the arrival direction, producing a second conical surface with the same vertex (the shared antenna). For any valid solution, these two surfaces have exactly two linear intersections. The array's horizontal geometry ensures that only one of these intersections is above ground, defining it as the sole arrival direction. A much more detailed description of this technique can be found in the documentation of Stock et al. (2014); Stock (2014).

The uncertainty in point solutions is dependent on the values of $\alpha$ and $\beta$, but is typically on the order of $0.1°$ in angle and $<1\ \mu s$ in time after post-processing. Of course, these sources are only 2D arrival directions and carry no information about the range of lightning activity. Section 7.2 discusses the process of using the plan location of LMA data to recreate fully 3D positions, though with decreased accuracy (section 7.3). One solution to this issue is the implementation of stereo measurements as commonly used in cosmic ray fluorescence detection (section 4.3). A second, multi-baseline interferometer was temporarily installed at Telescope Array in summer of 2020 in order to test this method, but has not yet produced a stereo event (subsection 9.3.2).

## 5.4   National Lightning Detection Network

The U.S. National Lightning Detection Network (NLDN) is a commercial service offering lightning detection data since 1989. It has undergone significant upgrades over the years, currently consisting of a nationwide array of 114 IMPACT-ESP wideband detectors across the United States (Cummins & Murphy (2009)). NLDN datasets include latitude, longitude, time, current, and polarity of lightning flashes, with tags differentiating intracloud and cloud-to-ground activity. Although originally designed primarily for LF detection of CG flashes, later upgrades to VHF sensitivity allow the detection of intracloud strokes. In a study of rocket triggered lightning in Florida, the NLDN was found to be 100% efficient in detecting return strokes having currents of $\geq$20 kA and $\sim$80% for 5–20 kA (Nag et al. (2011)). In terms of location and timing, NLDN strokes were accurately reported within $\sim$300 m and $<$700 ns depending on range (Cummins et al. (2010)).

Data was originally obtained for all flashes within 25 km of Telescope Array before installation of the LMA, and its accuracy was good enough to link reconstructed burst events with individual IC lightning strokes (Abbasi et al. (2017)). Later upgrades of the LMA and INTF have provided much higher resolution lightning tracking, but NLDN data is still used for preliminary correlation and current measurements (chapter 7). Locations of all flashes between 2008–2013 as determined by NLDN are displayed in panel a of Figure 5.8, showing an approximately uniform distribution. Panel b shows the peak currents detected at TA for intracloud and cloud-to-ground flashes over the same period. Note that IC flashes have a nearly-symmetrical, bimodal distribution as a result of the symmetrical thundercloud charge structure discussed in chapter 2. The CG flash distribution, on the other hand, is dominated by negative polarity reflecting the rarity of positive cloud-to-ground strokes. TGFs at TA are typically associated with negative-polarity flashes having currents between –20 and –60 kA.

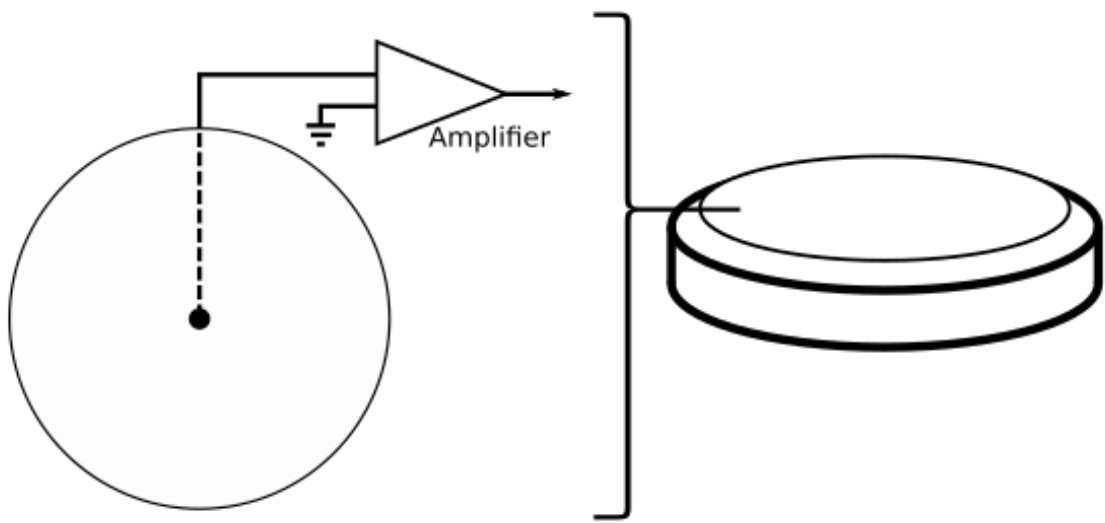

**Figure 5.6:** Simplified schematic of the three flat-plate antennas making up the broadband interferometer (INTF). The antenna is similar to that of the sferic sensors, but located close to the ground, facing upward, and sensitive to VHF radiation from 20–80 MHz. Voltage from each antenna station is amplified, digitized at 180 MHz, and sent to a nearby computer located in the building shown in Figure 5.7. When the triggering threshold is reached, data is stored on local hard drives for later post-processing.

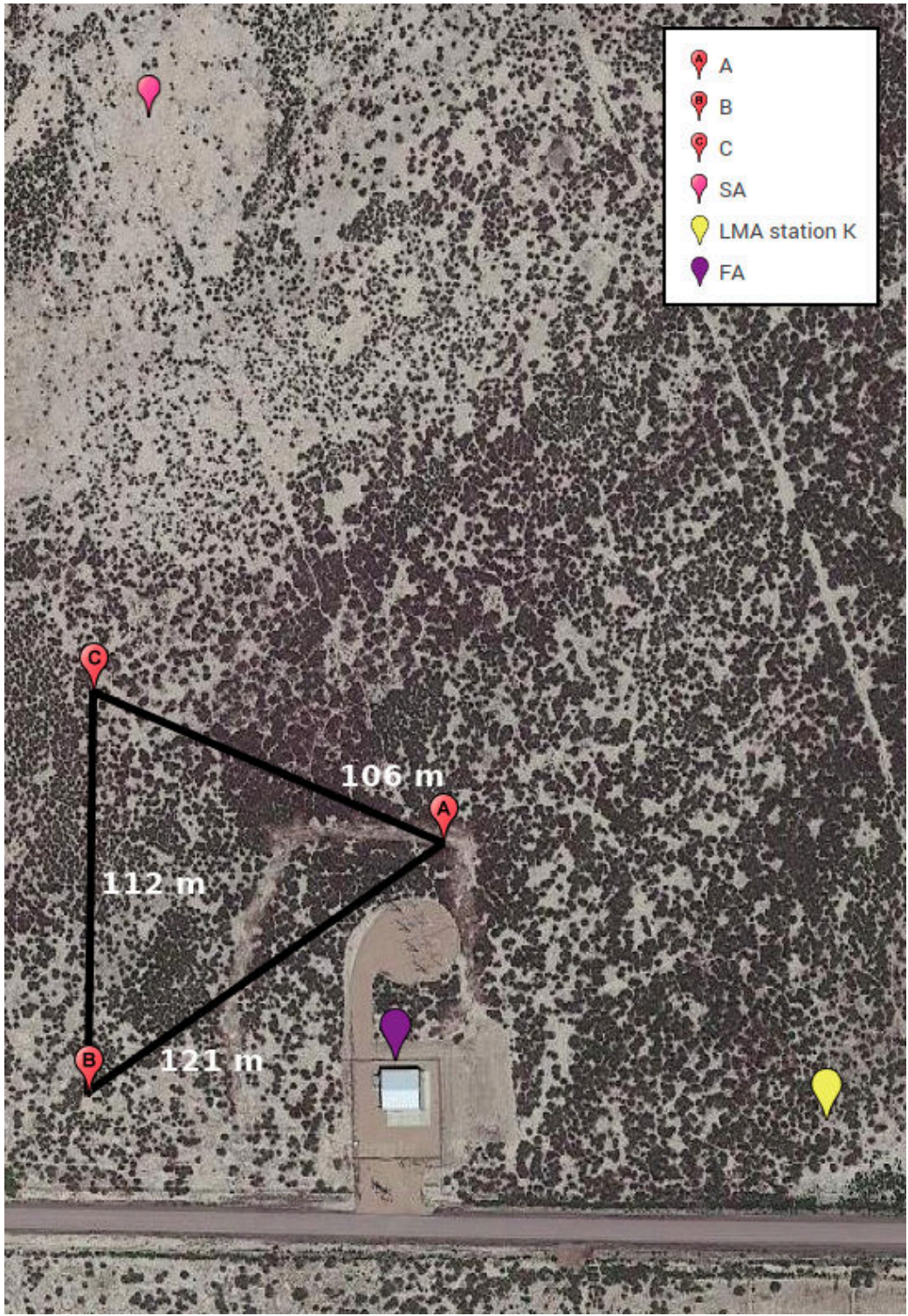

**Figure 5.7:** Geometry of the three flat-plate antennas (labeled A, B, and C) making up the broadband interferometer (INTF). Nearby LMA station and slow (SA)/fast (FA) sferic sensors are also labeled. The southern building houses a computer for INTF data storage and processing. INTF antenna baselines are given to represent scale.

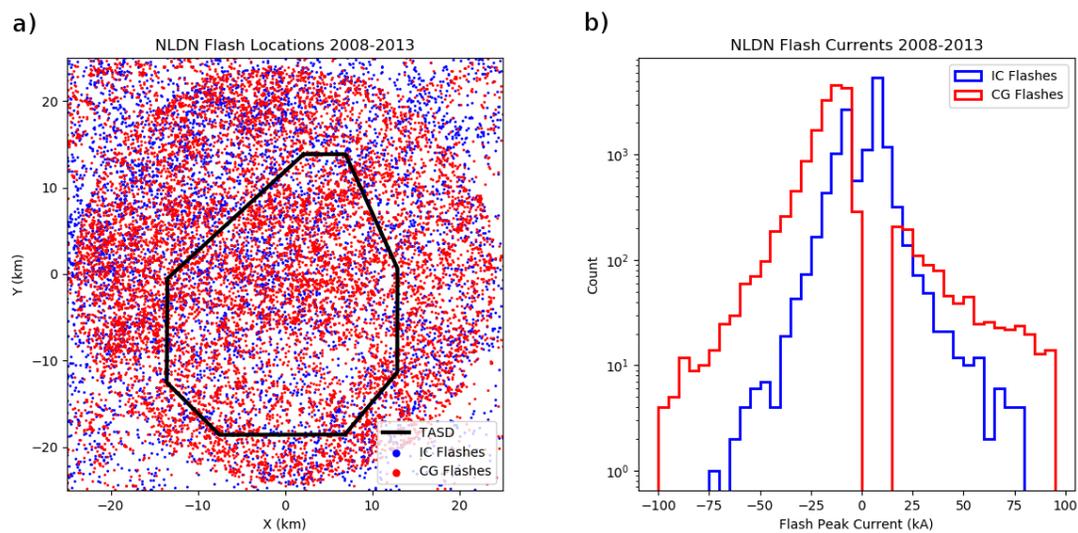

**Figure 5.8:** NLDN data of flashes occurring within 25 km of the TASD between 2008 and 2013. Cloud-to-ground flashes are given in red and cloud-to-cloud in blue. *a)* Approximately uniform distribution of the locations of lightning flashes near the TASD (black outline). Points outside the 25 km radius are from 2013, when further data was requested. *b)* Distribution of peak lightning flash currents as measured by the NLDN. IC flashes are similarly likely to be of either positive or negative polarity while CG flashes are dominated by negative polarity.

# CHAPTER 6

# EARLY TGF OBSERVATIONS AT
# TELESCOPE ARRAY

Early terrestrial gamma-ray flash detections at Telescope Array were limited based on available data. Just as the original discovery of TGFs by BATSE gamma-ray observatory was accidental, so were the detections by the TASD. For reference, this chapter splits data up into two sets based on available instrumentation: the initial 'burst' events rely solely on TASD detector responses correlated with lightning flash data from the National Lightning Detection Network (NLDN, section 5.4). The 'LMA' events include data from both the Lightning Mapping Array (LMA, section 5.1) and slow sferic sensors (SA, section 5.2) in addition to those detectors from the previous set. Each corresponding section describes the data and their implications at the time of their respective publications (Abbasi et al. (2017) and Abbasi et al. (2018)).

## 6.1  Burst Event Observations

The first set of events of Abbasi et al. (2017) was identified with a simple search in the TASD dataset for $\geq$3 triggers within 1 ms, dubbed a burst event, between May 2008 and May 2013. Of the ten events, five could be reconstructed using a slightly modified version of the standard cosmic ray analysis. Even though the locations of individual triggers were not considered in the search criteria, subsequent triggers within each burst were colocated within ~1 km of each other. Five of the bursts occurred within 1 ms of NLDN lightning flashes ('synchronized' events), and two within 200 ms ('related' events, shorter than the duration of a typical flash). All ten events took place during overhead thunderstorms. Of the five synchronized events, the correlated lightning flashes all occurred above the SD trigger locations within their respective footprints. All synchronized and related lightning flashes were negative polarity with currents ranging from $-12$–$-64$ kA.

Lightning data aside, it is clear that these events are not typical cosmic ray air showers due to their clustering in both space and time. In the example of Figure 6.1, the whole burst consisted of 5 triggers lasting ~450 $\mu$s and separated by 20–100 $\mu$s. Note that while the TASD processing software divided the activity into five triggers, there is clearly substructure (e.g., the three pulses in detector 2110 starting at 930,600 $\mu$s). The footprint plot of the first trigger in this burst is shown in Figure 6.2. For the burst events which could be reconstructed using modified TA analysis, the (presumed) cosmic ray primaries would have been between $10^{18}$–$10^{19}$ eV. The observed flux, however, is consistent with cosmic rays having energy of only $\sim 10^{13}$ eV (Abbasi et al. (2017)). The measured flux for cosmic rays between $10^{18}$–$10^{19}$ eV was $\sim$0.0001 Hz between 2008–2011 (Ivanov (2012)), whereas these events are triggering 3 or more times within 1 km$^2$ in less than 1 ms. The overall energy-independent trigger rate at TASD is $\lesssim$0.01 Hz and will be used as a more conservative estimate below.

Quantitatively, the expected rate of just two cosmic rays arriving within $\Delta t$ is approximately equal to the product of the fluxes:

$$R = \Gamma_1 \Gamma_2 \Delta t \tag{6.1}$$

where $\Gamma_1 = \Gamma_2 =$0.01 Hz and $\Delta t = 0.001$ s, giving a result of $\sim$3/year. Since all triggers of a burst arrive within 1 km of one another rather than randomly occurring across TA, we multiply by $\frac{\pi (1 \text{ km})^2}{680 \text{ km}^2}$ to obtain an expected coincidence rate of once per $\sim$72 years. Additionally, the bursts typically come in groups of 3–6 triggers and are usually separated by much less than 1 ms (on the order of 10–100 $\mu$s), reducing this rate by several orders of magnitude. Combined with the low average lightning flux in Western Utah of about 0.3/km$^2$/year, it was clear that these events were not cosmic ray-induced, but were instead associated with the overhead lightning.

### 6.1.1 Burst Event Analysis

At this point, the only data available were TASD waveforms and lightning flash times which, apart from the expected flux calculations above, restricted any detailed analysis. For TASD data, standard cosmic-ray shower reconstruction (Ivanov (2012)) struggled to make sense of the events; half of the bursts provided no meaningful results due to the difference from cosmic ray shower characteristics, e.g. signal shape and horizontal profile (Abbasi et al. (2017)). For the other five, a slightly modified TASD analysis produced the trajectory information presented in Table 6.1. Reconstructed event trajectories were then compared to the 3D position of lightning flashes recorded by the National Lightning Detection Network (section 5.4), with an example shown in Figure 6.3.

TASD data contains additional information in the raw waveforms themselves. While these plastic scintillators are not designed to identify individual particles (section 4.2), the signal shapes in each layer of scintillator give insight to the composition. For example, the upper panels of Figure 6.4 show a cosmic ray footprint alongside the waveform from the detector with most energy deposit. Note that the signals from upper and lower layers are effectively identical, which is typical of cosmic ray induced air showers. Both have a very steep rise and slower falloff, indicating that all particles are passing through the detector nearly simultaneously ($\sim$2 $\mu$s). The lower panels are from FL01 of the LMA event set and differ from cosmic rays in both shape and per-layer deposit. The signal continues over 10 $\mu$s and differs between layers. This event was chosen as a good example, but all detectors from both LMA and burst event sets exhibit the same characteristics.

Not only do the cosmic ray showers pass through each detector all at once, but they also arrive at adjacent detectors nearly simultaneously. The footprint plots (right panels of Figure 6.4) have circles representing deposit in each detector, colored by time. For the cosmic ray event, all but one of the detectors triggered within 0.5 $\mu$s, versus $\simeq$8 $\mu$s for the burst event. The implications of these differences is discussed in subsection 6.1.2.

### 6.1.2  Burst Event interpretation

The variation in trigger time between adjacent TASDs is well understood and regularly used in cosmic ray studies in the context of radius of curvature. For example, a high energy particle colliding with the upper atmosphere develops an extensive air shower hundred of kilometers above ground, meaning the path length traveled is nearly equal to reach adjacent surface detectors with 1.2 km separation. An extra-terrestrial gamma-ray primary penetrates much deeper into the atmosphere, causing longer delays in travel time. Following this logic, a TGF produced in lower storm clouds (3–4 km) would be significantly delayed in arriving at detectors up to 3 km away, an effect clearly seen in Figure 6.4. This method of curvature was used to calculate the altitudes in Table 6.1.

As discussed in section 4.1, muon detection is vital to studying cosmic rays. The two layers of scintillators in TASDs are designed to produce identical signals when muons pass through, while the steel enclosure and separating plate can filter out less penetrating particles. In the case of burst events, the variation between layer signals is indicative of electromagnetic incident particles (Figure 3.5). At high energies, photons have more opportunity to scatter electrons into the lower layer, resulting in a higher average energy deposition. On the contrary, electrons on the order of 10 MeV are less likely to penetrate the entire detector, depositing more energy in the upper layer on average. EM showers, then, are likely to produce the inconsistent signals seen in Figure 6.1. Unfortunately, the surface detectors at TA do not have capability to directly measure true particle energies (section 4.2), and particle energies were not examined further until the next set of events (Abbasi et al. (2018), subsection 6.2.2).

Data from the five reconstructable events are given alongside their correlated lightning strike data in Table 6.1. Notice that for the events with correlated lightning, IC events occur during or slightly before TASD triggers while CG events occur last, likely indicating the return stroke(s). Considering typical lightning development time of hundreds of milliseconds and accounting for propagation time, these measurements indicate that showers are produced during the initial stages of lightning flashes, possibly during strong in-cloud processes. For comparison, (Cummer et al. (2015)) showed TGF production within the first 2–3 ms of lightning leader progression in RHESSI events. The burst events at TASD were certainly emitted during early development, but more complete lightning data would be needed to confirm which stages they were associated with.

In terms of duration, TASD triggers tend to last up to 10 $\mu$s and are separated by 10–100 $\mu$s within bursts. The downward radiation reported by Dwyer et al. (2005) had similar timing characteristics, but arrived late in their respective flashes. Satellite observations, on the other hand, are typically more continuous over durations of hundreds of microseconds (Cummer et al. (2011)), though more recent studies suggest these are overestimations (Østgaard et al. (2019)). A rather simple explanation for the differences in duration is that the short, separated bursts at TASD may be smoothed out by atmospheric scattering into a seemingly-continuous signal seen by satellites. However, this has been ruled out as the sole culprit in more recent analyses (see chapter 8).

After the discovery of these events, the potential for lightning observations at Telescope Array prompted a collaboration with Langmuir Laboratory for Atmospheric Research for their expertise in studying lightning and TGFs. In order to get a better look at the (suspected) TGFs' relation to lightning initiation, the LMA (section 5.1) and slow sferic sensors (section 5.2) were implemented at TA in 2014. Together, these systems track the 3D lightning activity as it develops while measuring the electric field changes from ground level. The detectors were left running for about two years to collect data for the following investigation of Abbasi et al. (2018).

## 6.2 LMA Event Observations

An additional ten bursts, collectively called 'LMA' events and numbered FL01–FL10, were found in TASD data between 2014 and 2016 with the criteria of having at least two consecutive triggers and an NLDN flash within 1 ms (see supporting information of Abbasi et al. (2018) for full data). Intermittent LMA coverage during this period meant that only

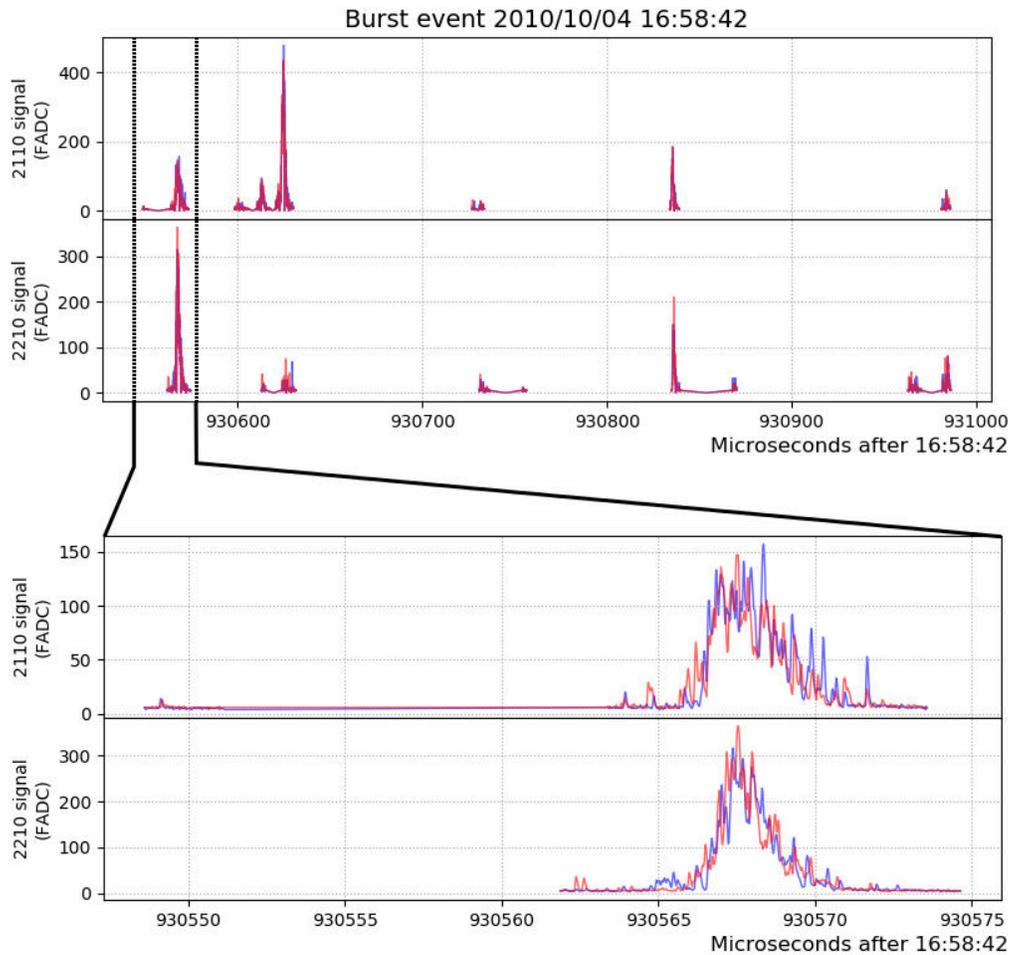

**Figure 6.1:** Responses from two adjacent surface detectors of a burst event of five triggers on 2010/10/04. Red (blue) represents the upper (lower) layer of scintillator. Signals during the first trigger are shown in more detail in the lower panel. The footprint of active detectors during this trigger are shown in Figure 6.2.

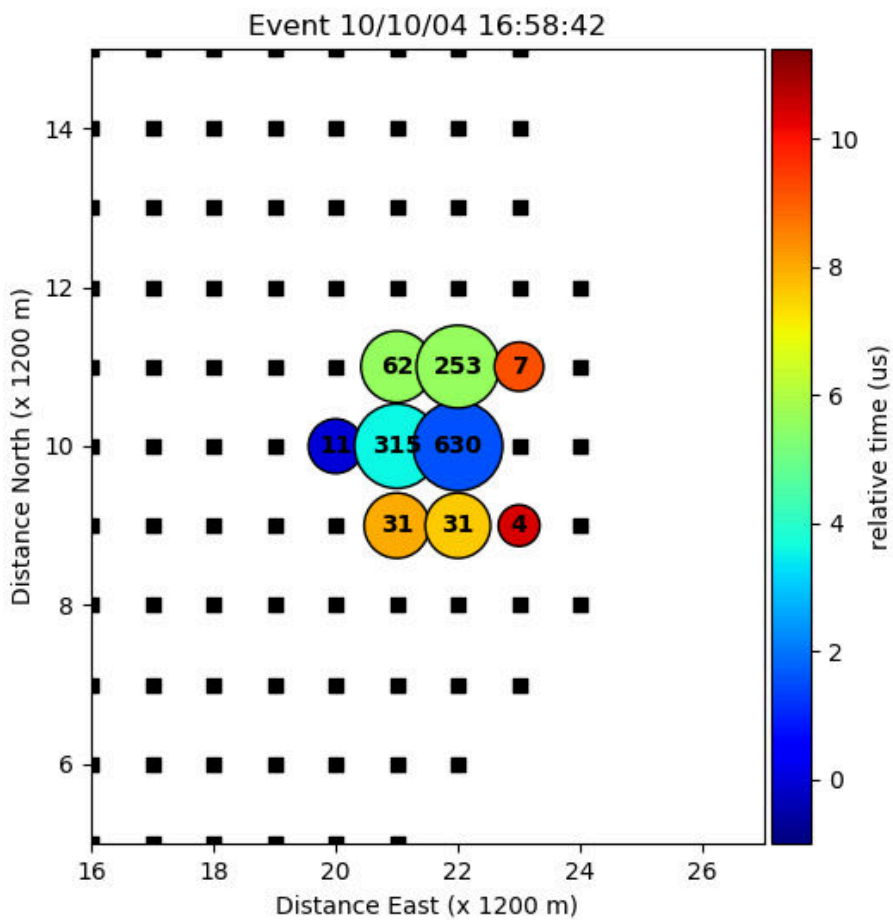

**Figure 6.2:** Footprint plot showing active detectors in the first trigger of the burst event of 2010/10/04 16:58:42. TASDs which did not trigger are shown as black squares, while those that did are represented by the numbered circles. Colors represent relative timing of individual level-0 triggers at each detector and sizes are proportional to the log of energy deposit (given in units of FADC count, see subsection 4.2.3).

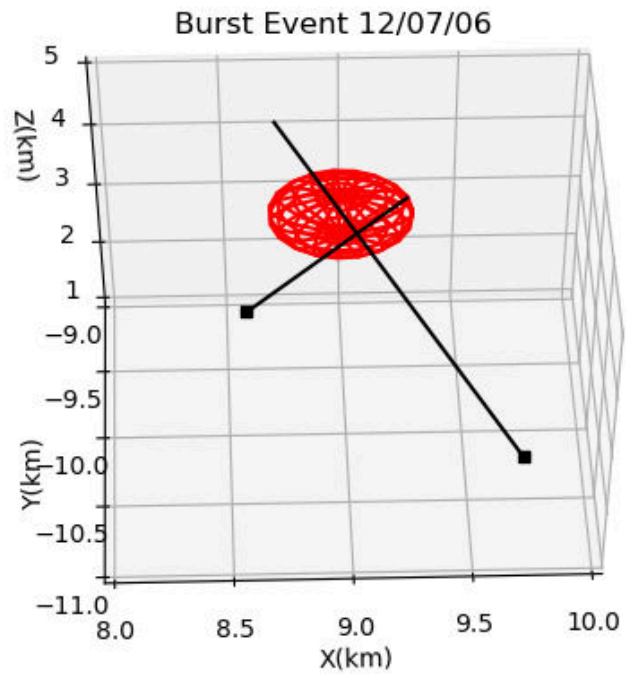

**Figure 6.3:** The two TASD triggers of burst event 2012/07/06 and their trajectories produced by modified cosmic ray analysis (Abbasi et al. (2017)). The two triggers are shown as black points at ground level (1.4 km) with plotted trajectories. The associated NLDN flash is shown at 3.8 km altitude within a red sphere having the typical position error of ∼300 m (Nag et al. (2011)).

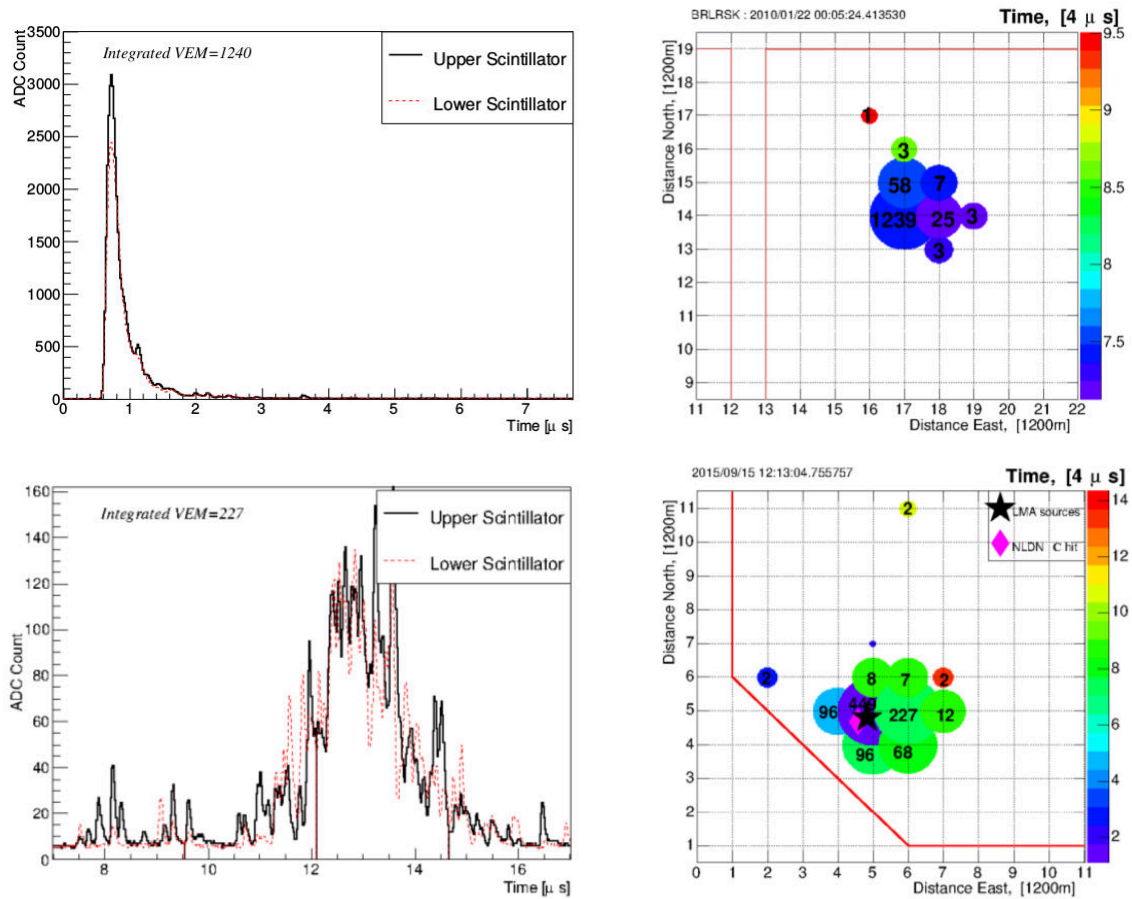

**Figure 6.4:** between a typical cosmic ray event (top) and TGF event FL01 from the 'LMA' dataset (bottom). Left panels are detector responses of the single TASD station with most energy deposit from the event. Right panels show the 'footprint' of the shower with energy deposit in each TASD given in units of VEM (section 4.2). Circle size is proportional to energy deposit while colors represent relative TASD trigger time in units of 4 $\mu$s.

**Table 6.1:** Quantitative burst event values

TASD data for burst events which were reconstructable (alongside NLDN data where available). In the first column, AS refers to air shower and LG to lightning. In the seventh column, zenith values refer to the reconstructed trajectories of AS events while current refers to NLDN data of lightning flashes. In the eighth column, alt refers to the reconstructed radius of curvature of AS events while LG type specifies the type of lightning flash detected by NLDN. X and Y values are relative to the center of the TASD.

| Event | | Date | Time | $\mu$sec | X [km] | Y [km] | Zenith [deg]/ Current [kA] | Alt [km]/ LG type |
|---|---|---|---|---|---|---|---|---|
| AS | | 10/10/04 | 16:58:42 | 930565 | 11.36 | -7.43 | 15.7 | 4.0 |
| AS | | 10/10/04 | 16:58:42 | 930612 | 10.48 | -7.37 | 13.1 | 4.4 |
| AS | | 10/10/04 | 16:58:42 | 930835 | 11.14 | -8.16 | 27.7 | 3.3 |
| | LG | 10/10/04 | 16:58:42 | 930608 | 12.5 | -5.1 | -63.5 | IC |
| | LG | 10/10/04 | 16:58:42 | 934058 | 10.6 | -8.1 | -35.8 | CG |
| AS | | 11/07/27 | 08:06:15 | 124319 | 3.45 | 1.95 | 5.3 | 4.1 |
| AS | | 11/07/27 | 08:06:15 | 124543 | 2.90 | 2.23 | 19.2 | 3.1 |
| | LG | 11/07/27 | 08:06:15 | 124303 | 3.7 | 2.3 | -35.6 | IC |
| | LG | 11/07/27 | 08:06:15 | 130887 | 3.1 | 2.0 | -28.0 | CG |
| AS | | 11/09/16 | 19:40:56 | 567481 | -3.21 | -9.29 | 38.8 | 3.3 |
| AS | | 11/09/16 | 19:40:56 | 567566 | -3.52 | -9.41 | 32.2 | 3.1 |
| AS | | 12/07/06 | 01:49:11 | 184219 | 9.85 | -10.70 | 24.2 | 3.8 |
| AS | | 12/07/06 | 01:49:11 | 184307 | 7.64 | -9.67 | 23.6 | 3.4 |
| | LG | 12/07/06 | 01:49:11 | 184122 | 9.0 | -9.7 | -36.3 | IC |
| AS | | 12/09/07 | 01:55:45 | 380684 | -8.64 | 1.25 | 14.9 | 4.5 |
| AS | | 12/09/07 | 01:55:45 | 380755 | -9.86 | -0.34 | 11.4 | 4.8 |
| AS | | 12/09/07 | 01:55:45 | 380881 | -9.45 | -0.96 | 31.7 | 3.4 |
| | LG | 12/09/07 | 01:55:45 | 380675 | -9.0 | 0.7 | -53.9 | IC |
| | LG | 12/09/07 | 01:55:45 | 390411 | -9.6 | -2.0 | -20.1 | CG |
| | LG | 12/09/07 | 01:55:45 | 409370 | -8.6 | -1.7 | -12.2 | CG |

three of these had fully-reconstructed LMA data during their bursts Figure 6.5. The other seven had full sferic data, some of which are shown in Figure 6.6. Unfortunately none of the events had both LMA and sferic data available together, however the LMA-correlated and sferic-correlated events produced essentially the same results (subsection 6.2.2).

As described in section 5.1, processed LMA data consists of data points with 6 fields: time of day, latitude, longitude, altitude, $\chi^2$ fit value, and power. These identify source points in 3D space, collectively showing the overall flash development (Figure 5.2). Slow sferic data (section 5.2) represents the relative electric field changes during a lightning flash. Together these detectors supply unique information about the overall flash structure and development which were unavailable for analysis of the previous burst event set.

In general, the TGF candidates looked similar to the previous set of burst events. As before, TASD signals had durations of tens of microseconds with spacing on the order of 100 $\mu$s, came in groups of 2–5 triggers, and deposited energy in the wide range of 50–26,418 VEM (e.g. Figure 6.7). NLDN data showed synchronized lightning flashes within 1 ms for all ten events. Most of these were again recorded as strong negative currents ($-20$–$-140$ kA), although some weak positive currents are recorded early in some of the flashes. Flashes with active LMA data, however, could be analyzed in more detail.

### 6.2.1   LMA Event Analysis

Three of the events in this set had correlated LMA activity and are shown in Figure 6.5. The left panels show each entire flash in windows of 400 ms, while the right panels are a zoomed-in view of the few milliseconds around each burst. The most obvious similarity in all events is that the TASD triggers (vertical dashed lines) occur in the first 1–2 ms of the flashes, during the most energetic LMA data (shown as the size of filled diamonds and stars). NLDN data for these events is also displayed along the abscissa. Note that highest-power LMA points and –IC NLDN events also occur in the first couple milliseconds. In the first event FL1, for example, each of the five TASD triggers are closely associated with individual –ICs, and the first two are also closely correlated with the two highest-power LMA points of the entire flash.

The other seven events had accompanying slow sferic data, but not LMA. Four of these are shown in Figure 6.6. These plots follow the same style of Figure 6.5; the left panels display the entire flash with zoomed-in views in the right panels. TASD triggers are again shown as red dashed lines, and NLDN points are displayed on the horizontal black dashed line and are accompanied by vertical dashed lines. The timing in these events is consistent with the results of the LMA-correlated events, with TASD triggers occurring in the first 1–2 ms during the leader stage as predicted during study of the burst event set. Perhaps more importantly, most of these are aligned with the very beginning of field change, meaning the events are closely associated with the leader initiation process itself.

Another interesting feature of these events is their duration and spacing. TASD signals of the first three events of this set (FL1–FL3) are shown in Figure 6.7. Energy deposit in the most-energetic detector for these bursts is plotted against time, showing pulse durations of 5–20 $\mu s$ separated by $\simeq$50–200 $\mu s$, clearly displaying that these are not continuous processes, but distinct EM showers produced sequentially. This reinforces the findings of the burst event set which differed from the many satellite-detected TGFs reporting continuous durations lasting hundreds of microseconds or more (see subsection 6.2.2).

From Figure 2.1, the typical symmetry of thundercloud charge regions allows flashes to develop either upward or downward. Just as upward TGFs are produced by upward +IC flashes between the upper regions, downward TGFs are expected to be produced during –IC and CG flashes between the lower regions. The three events with LMA data from this set suggest this is the case. In the left-hand panels of Figure 6.5, sources appear to be grouped at two altitudes, from which we infer that the mid-level negative region is at 5–6 km MSL (3.5–4.5 km AGL) and the lower-level positive region is at 3–4 km (1.5–2.5 km AGL). The right panels show that TASD triggers occur during LMA points between these regions (3–5 km MSL), but due to the sparse groupings and rapid altitude changes, we cannot pinpoint source altitudes better than within hundreds of meters or more. Better data would be needed to confirm these sources, leading to the installation of the broadband interferometer with a much higher sampling rate and resolution (section 5.3).

Other measurements of the source were more reliable. For example, see the footprint plot of FL1 (lower-right panel of Figure 6.4); LMA and NLDN points are directly above the gamma-ray detections, implying the source plan location is well understood, within $\sim$200 m, and that gamma production is mostly vertical, resulting in a opening half-angle of 23–37°. Therefore, the timing accuracy of the TGFs largely depends on the altitude; a 1 km range of possible production altitudes $\Delta z$ would result in a possible source timing window of $\frac{\Delta z}{c} \simeq 3 \ \mu s$. A similar method has been used for satellite observations (Cummer et al. (2011)), for which altitude uncertainties on the order of a few km translate to timing uncertainties of 10 $\mu s$ or more.

Thus far, the measurements above have shown that large-scale, ground-based arrays can efficiently and reliably study downward TGFs. Event sources can be located within the lightning development, specifically during the initial 1–2 ms of downward negative flashes.

In addition, the close proximity of measurements can identify TGF opening angles and approximate source locations within an overhead thundercloud, although the sparsity of LMA data limits precision. Differences between upward and downward observations also remain unanswered without more detailed information on lightning development, leaving plenty of room for improvement. After analysis and publication of these events, another upgrade was planned to increase measurement resolution and add additional information in order to identify the production mechanisms associated with TGFs in general.

### 6.2.2  LMA Event interpretation

After installation of the lightning mapping array and slow sferic sensors, the additional analysis lead to new, unique conclusions about downward TGFs and their relation to lightning initiation. Thus far, only a handful of downward TGFs had been successfully linked to natural lightning flashes due to the poor coverage of ground-based detectors. Two of these reported gamma-rays $\simeq$200 $\mu$s after the return stroke of negative CG lightning (Dwyer et al. (2012); Tran et al. (2015)), indicating that the production mechanisms may be different than those causing upward TGFs, which are assumed to be produced during the upward propagation of negative leaders in +IC flashes (Shao et al. (2010); Lu et al. (2010, 2011); Mailyan et al. (2016)). The typical thundercloud charge structure of Figure 2.1 consists of a central region of negative charge separated from positive charge regions above and below. This symmetry means that negative leaders can initiate both upward and downward flashes in a storm. It is expected, then, that the downward breakdown process should also produce TGFs well before the return stroke. Note from Figure 5.8 that the rate of positive and negative IC flashes are nearly equal. With the addition of CG flashes, downward lightning is much more common than upward, so the same should be true of TGFs if the production mechanisms are the same.

All three LMA-correlated events of this set (Figure 6.5) show gamma activity during the first couple milliseconds of downward leader development between the mid-level negative and lower positive charge regions. This suggests that these events are indeed the downward analog of satellite-detected TGFs, being produced during downward negative breakdown in the initial stages of –CG lightning. That said, the issue of duration still separates the two. As seen in Figure 6.7, overall bursts can span up to 500 $\mu$s, comparable to some upward TGFs, but the burst-like quality is unique to TGFs at TA and this study was unable to conclusively explain the difference. The durations do resemble the x-ray observations of Dwyer et al. (2012), in which lower-energy radiation was correlated with individual leader steps of a downward negative flash. As discussed below, however, energy analysis shows that photons of the TASD events have gamma-level energies of at least a couple MeV.

Section 3.2 describes the widely-accepted RREA spectrum of upward TGFs. From Celestin et al. (2015), both the energy spectrum and fluence are dependent on the potential difference available for accelerating electrons at the leader tip. Since there were no proper measurements of the potential difference for TA events, showers were simulated in GEANT4 to find how these compared to TGFs detected by satellite. However, since the TASDs cannot reconstruct individual particles, multiple simulations were run with different starting conditions to approximate the true observations. Only flashes 1–3 were compared to these simulations since only they had accompanying LMA data and therefore altitude estimates. Figure 6.8 shows the result of these simulations, for which a fluence of $10^{12}$–$10^{14}$ gamma-rays is consistent with the data from FL01 and FL02. FL03 had a relatively high energy density, but was clearly an outlier in its proximity to ground level, resulting in drastically less shower development and atmospheric attenuation (and thus an overabundance of low-energy photons), making it a poor comparison to the other events. The simulation studies presented in Celestin et al. (2015) show that showers with fluences of $10^{12}$–$10^{14}$ photons should be produced by potential drops of 10–50 MV in the leader tip region. Individual photons would then have maximum energies of 2–8 MeV.

Notice the significant differences between the simulated values presented here and those from the satellite observations of section 3.1; fluence estimates here are 1–4 orders of magnitude lower, and the upper limit of maximum photon energies at TA barely reaches the well-defined characteristic photon energy of RREA (subsection 3.2.1). Additionally, upward TGFs are typically interpreted as a fairly smooth flux of gamma-rays lasting hundreds of microseconds or more (Stanley et al. (2006); Cummer et al. (2015)), while the detections at Telescope Array come in discrete bursts of 5–20 $\mu s$ (Figure 6.7). These burst observations are more similar to the recent upward TGFs of Lyu et al. (2018) as well as previous analyses of GBM-Fermi events which appeared to consist of multiple, partially overlapping pulses possibly explained by atmospheric scattering and detector effects (Briggs et al. (2010); Fishman et al. (2011)). Simulations by Celestin & Pasko (2012) give similar conclusions, but suggest that the artificial extension of pulses due to atmospheric scattering cannot account for the discrepancy on its own — the issue was revisited in the later analyses of Belz et al. (2020) and is discussed in chapter 8.

There are still some conclusions to be drawn about particle energy, however, when comparing observed detector responses to simulation. First, individual particles can be distinguished in detector responses with low energy deposit. See the upper-left panel of Figure 6.9, in which individual particle signals are distinct. The integrated energy deposit of one such signal from the figure is ∼32 FADC counts, consistent with a VEM (as defined in subsection 4.2.3). Due to the well-understood physics of these detectors (simulations of Figure 3.5 and section 4.2), energetic electrons with energy above 2 MeV will not deposit more than ≃2 MeV in one layer of scintillator. Therefore, we can only place lower limits on the energies of incident particles. Assuming that these particles are indeed Compton-scattered electrons, the minimum-energy case requires that the photon perfectly back-scatters, giving the majority of its energy to the electron. In this case, a photon of energy ≥2.2 MeV would be required to produce an electron of 2 MeV. However, due to the higher probability of grazing coincidences, the initial photon's energy would likely need to be several times higher to produce the same electron measured by the TASD. The angular distribution for Compton scattering is described by the Klein-Nishina differential cross section (Klein & Nishina (1929)), showing the expected energy given to a scattered electron is close to 1/3 of the photons initial energy. Without more detailed simulations, though, we cannot make a definitive statement about the expected photon energies that produced various signals at Telescope Array. Instead, we merely set the lower limit that at least some photons arrive at ground level with at least 2.2 MeV. These claims were further expanded in the analysis of Belz et al. (2020) (chapter 8).

The distinction between x-rays and gamma-rays varies depending on whom you ask due to the history of their discoveries. Originally, gammas were defined as resulting from nuclear transitions while x-rays resulted from electron transitions. These rules are outdated since we now know that these rays are both electromagnetic radiation, and the border of these regimes is now taken to be between 100–1,000 keV, placing the photon detections at Telescope Array in the gamma regime, as opposed to the x-rays of Dwyer et al. (2012) having energies of hundreds of keV. Of course, any EM shower will be dominated by lower-energy particles, but photons and electrons with less than a few MeV deposit little to no energy in the TASD scintillators (Figure 3.5), suggesting the signals are actually dominated by much fewer high-energy particles. This is likely why the TASD responses are impulsive and less smooth than would be expected for a larger flux of x-rays (Figure 6.1). Accordingly, the value of 2.2 MeV above should be interpreted (confusingly) as a lower-limit estimate of the maximum photon energy of a given TGF. This is an important distinction when comparing these results with Celestin et al. (2015) to place a lower limit on available potential of 10 MV, consistent with Figure 6.8.

In an attempt to glean further insight into TGF production, additional instruments were prepared for a 2018 installation to the East of the TASD. The broadband interferometer (section 5.3) provides higher-resolution lightning tracking to give better estimates of source locations (specifically altitude). The fast sferic sensor works similarly to the slow sferic sensors (section 5.2), but on a shorter timescale. These have been used in other studies to identify and examine initial breakdown pulses (IBPs) and their relation to TGFs Marshall et al. (2013).

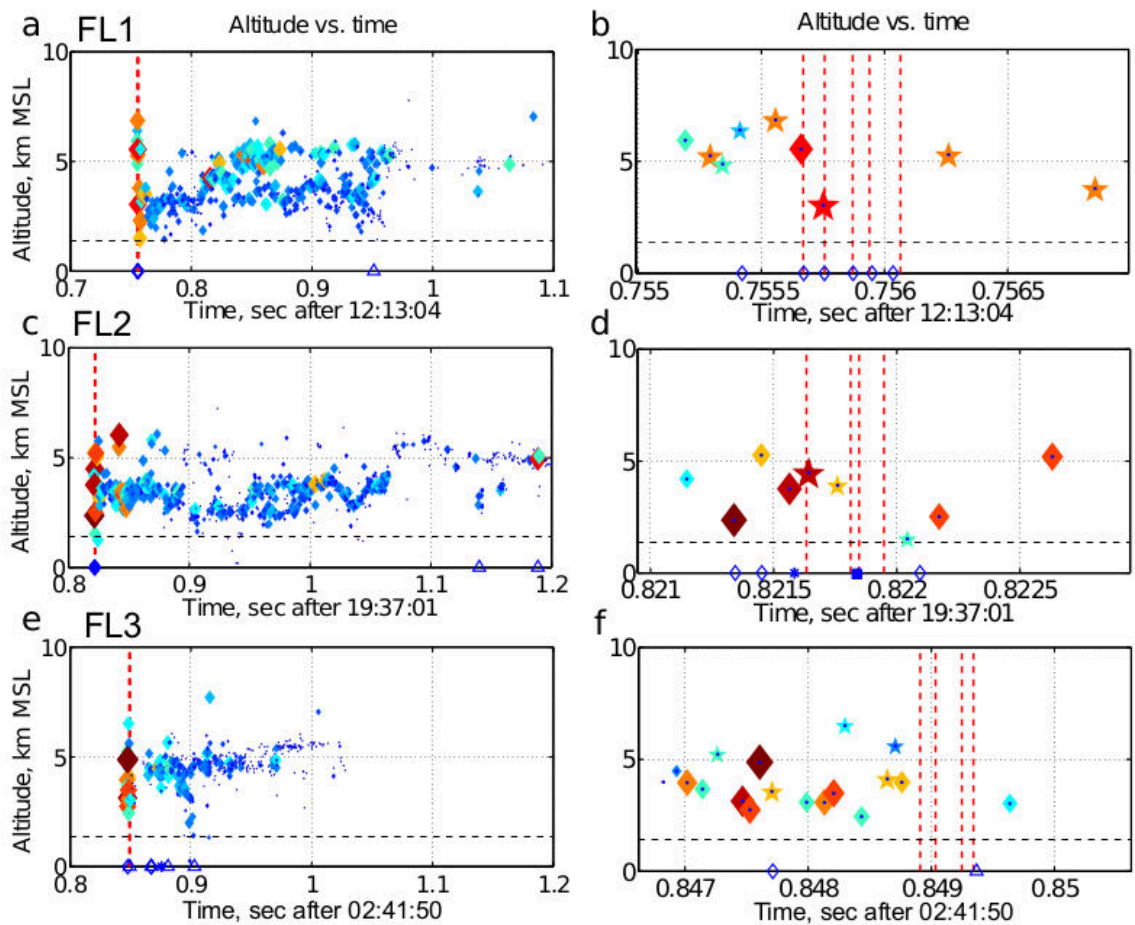

**Figure 6.5:** Data for FL1–FL3, showing altitude versus time of LMA sources (filled diamonds sized and colored based on their radiated power between –20 dBW and +25 dBW for blue through red). Filled stars are the same, except representing sources with large $\chi^2$ fit values. NLDN events are given on the abscissa, with diamonds (squares) representing –IC (+IC) flashes and triangles (hollow stars) representing –CG (+CG) flashes. TASD trigger times are also represented by red dashed lines. The horizontal dashed line indicates the ~1.4 km average ground-level altitude of the TASD. Left panels show development of the entire flash and right panels show only the first 2–3 ms of each.

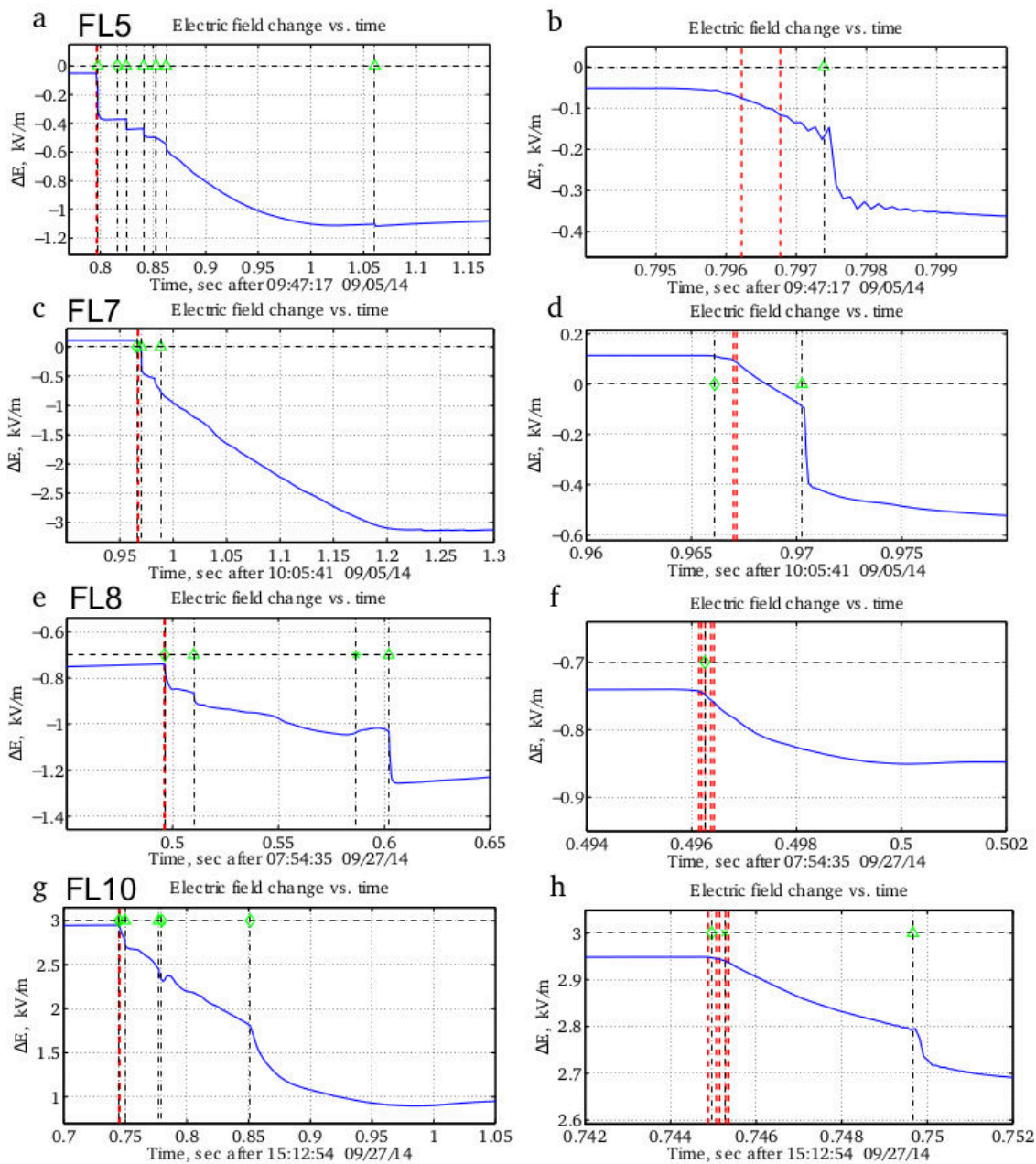

**Figure 6.6:** Slow Antenna Data for FL5, 7, 8, and 10. TASD trigger times are represented by red dashed lines and NLDN events are indicated by black dot-dashed lines with green markers. Diamonds (stars) represent –IC (+IC) flashes and triangles (crosses) represent –CG (+CG) flashes. Left panels show development of the entire flash and right panels show only the first 5–20 ms of the flash.

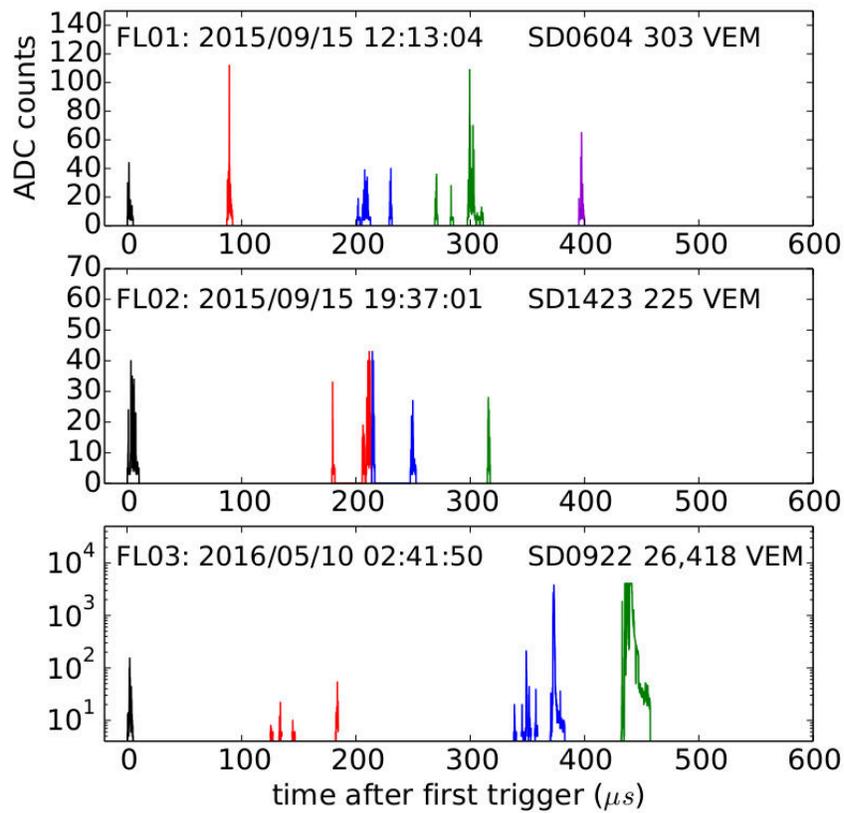

**Figure 6.7:** TASD signals of the detector with highest energy deposit from FL1–FL3. The waveforms are resolved into discrete showers, most of which last less than 10 $\mu s$ and occur in succession over hundreds of microseconds.

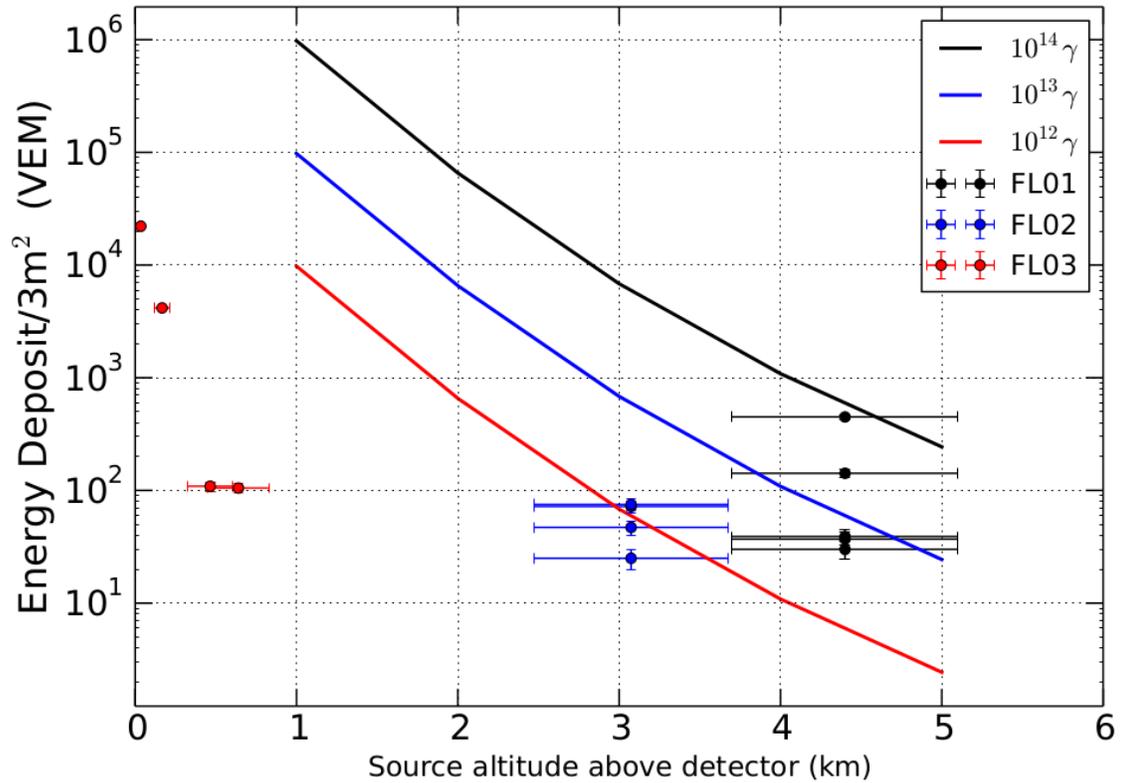

**Figure 6.8:** Results of energy deposited in TASDs by simulated photon showers with total fluences of $10^{12}$ (red line), $10^{13}$ (blue line), $10^{14}$ (black line) versus source altitude. Data from FL01–FL03 are superimposed based on approximate altitudes provided by LMA sources. The altitude and energy density of FL03 was an outlier among events and is a poor comparison. Flashes 01 and 02 are consistent with showers having $10^{12}$–$10^{14}$ photons.

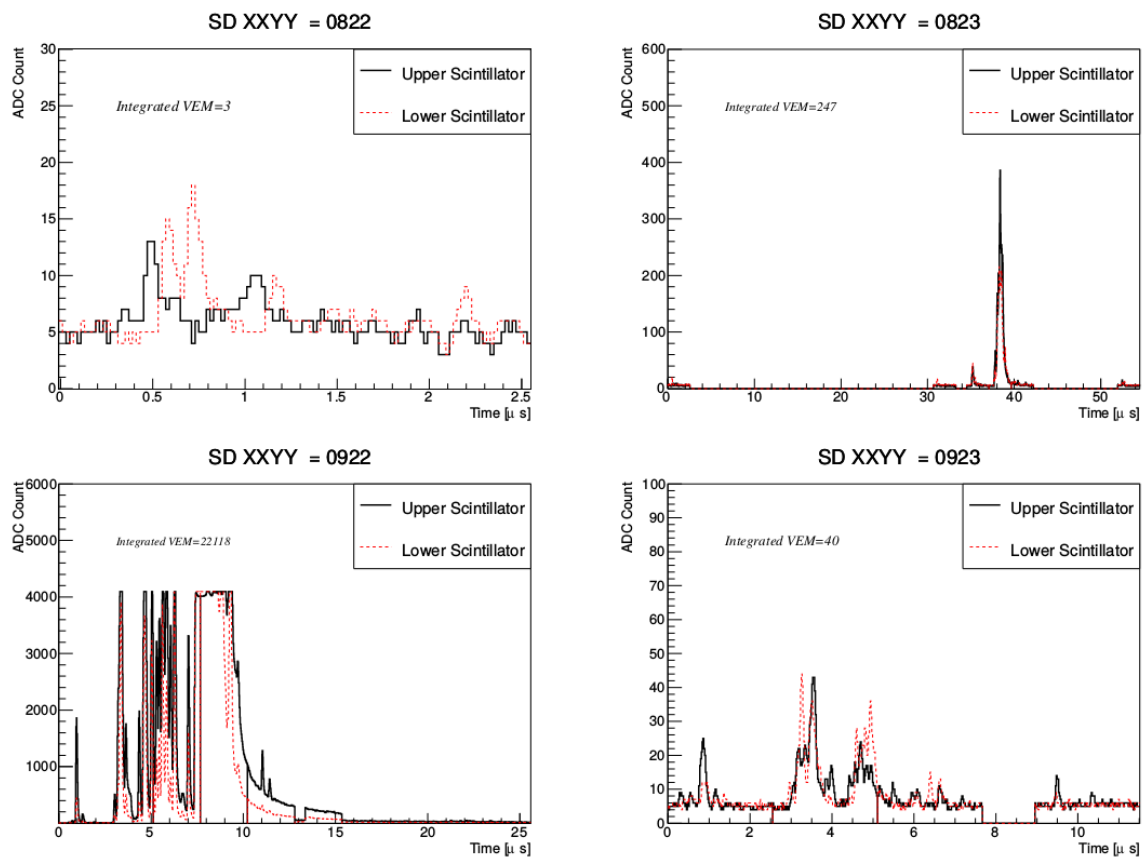

**Figure 6.9:** Responses from four TASDs of the fourth trigger of FL03, demonstrating the range of energy deposit in four adjacent detectors. The flash terminated in an extremely high-current –94.1 kA ground stroke only 78 m from TASD 0922 (bottom left) and saturated the detector. In contrast, TASD 0822 (upper left) recorded only ∼0.01% as much energy deposit 1.2 km away; the flux was low enough that individual particles can be distinguished in both layers of scintillator.

# CHAPTER 7

# INTERFEROMETER OBSERVATIONS
# AND ANALYSIS

The strength of this study is in the diversity of complimenting detectors. These were added and improved over >7 years, meaning that the methods of analysis have adapted as necessary. Therefore, the most recent set of observations from 2018 ('INTF' events) is the focus of this dissertation and is documented in detail here. Instrumentation for these events includes the Telescope Array Surface Detector (TASD, section 4.2), broadband interferometer (INTF, section 5.3), sferic sensors (FA, section 5.2), and Lightning Mapping Array (LMA, section 5.1). Information on earlier observations from the papers of Abbasi et al. (2017) and Abbasi et al. (2018) can be found in chapter 6.

## 7.1   INTF Event Observations

After the interferometer upgrade in summer of 2018, events came in quickly. Rather than the previous criteria of coming in bursts of 2+ TASD triggers in 1 ms, the search was simply for time correlations of INTF triggers with any TASD trigger. Position and activity were then examined more closely in order to identify TGF candidates. Three candidate events were recorded within three weeks, with a fourth occurring a couple months later (Table 7.1). All four had TASD triggers coincident with downward lightning activity collected by the LMA, INTF, and sferic sensors. The first three ended in negative cloud-to-ground flashes while the last was a negative intracloud flash. The first three events produced NLDN triggers, but the last did not, possibly due to the NLDN's poor efficiency in detecting IC flashes (Zhu et al. (2016)). Each of these instruments supply a unique piece of the puzzle that must be assembled during analysis (section 7.2) in order to form a coherent picture of the TGFs and their sources.

Because of the complexity of these different data types, several plots are given for each event to highlight specific aspects. These are shown at the end of their respective sections and grouped by format rather than by event. The left panels of Figure 7.1–Figure 7.4 show the 'footprints' of TASDs which triggered during each event, including energy and timing information, while right panels give examples of corresponding TASD signals. Figure 7.5–Figure 7.8 show the azimuth-elevation development of VHF sources detected by the INTF. Basic data from each of the other detectors is overlaid to give an idea of a TGF's relation to its parent flash. Figure 7.9–Figure 7.12 show the same VHF sources plotted alongside fast sferic signals as a function of time and with increasing levels of detail. TASD responses are also included after applying the geometric corrections of subsection 7.2.1. Additional varied information for each event is given in Figure 7.13–Figure 7.16.

TGF A occurred during a storm on August 2, 2018 relatively close (17.0 km) to the INTF station in the first millisecond of a –38.3 kA –CG flash. The associated TASD data consisted of two triggers, labeled $a$ and $b$, the first of which consisted of a total of 561 VEM at ground level over 9 detectors and covering ∼12 km$^2$. The second trigger occurred ∼100 $\mu$s later and contained (apparently) two distinct, weaker showers totaling 192 VEM over 8 detectors and covering ∼10 km$^2$. These footprints are shown in Figure 7.1 alongside the detector response with the highest energy deposit for each trigger. Unfortunately, both were observed by the edge of the surface detector array, meaning that the energies and footprints were likely larger than recorded. Note from Table 7.1 that trigger $a$ was also associated with a –36.7 kA NLDN current only 13 $\mu$s earlier, comparable in strength to the return stroke itself. Further analysis would identify this as a strong IBP (subsection 2.4.1). VHF sources detected by the INTF are not listed in this table due to their high sample rate, but can be seen in Figure 7.5. Figure 7.9 gives the flash's same INTF data points versus time with the fast sferic sensor waveform overlaid, showing that overall leader development lasted ≃8 ms with an average vertical speed of 5×10$^5$ m/s. TGF A's parent flash was also the only of these four to be initiated by a narrow bipolar event (Figure 7.17), which traveled upward 150 m in 11 $\mu$s (1.3×10$^7$ m/s). The significance of this difference in speeds is discussed in chapter 8.

The same storm produced two additional TGFs (B and C) about an hour later. The former only triggered the TASD once in the first millisecond of a –26.5 kA CG flash 16.6 km from the INTF station. The total energy deposit was fairly low at 112 VEM in 10 TASDs with an area of ∼13 km$^2$ (Figure 7.2). The sole TASD trigger occurred at the same time as a sferic pulse of current –30.1 kA, even stronger than the return stroke. INTF data for TGF B is shown in Figure 7.6 and Figure 7.10, in which the leader proceeded to ground in ∼10 ms with an average vertical speed of 4×10$^5$ m/s. Surprisingly, unlike the other events, TASD 1421 in the center of the footprint area did not measure energy deposit above background levels. Examination of this detector revealed that it was not malfunctioning during this time, and that stray muons were detected before and after the storm as expected during normal operation, indicating that this hole likely reflects the true nature of the TGF's resulting shower — this idea is explored in chapter 8.

TGF C was detected nearby just 2.5 minutes later during a –26.8 kA CG flash 16.0 km from the INTF station. It contained two gamma bursts separated by ∼120 $\mu$s, similar to TGF A. In this event, the second trigger was the more energetic and was associated within 5 $\mu$s of a –21.7 kA NLDN current. The first trigger measured only 35 VEM in 4 detectors covering ∼5 km$^2$, while the second totaled 212 VEM over 9 detectors and ∼11 km$^2$ (Figure 7.3). INTF data of Figure 7.7 and Figure 7.11 show the near-vertical early leader development before branching toward ground over ∼11 ms at a slightly slower average vertical speed of $3.4 \times 10^5$ m/s.

The fourth and final TGF of this set took place during a late season storm in October and was associated with a low-altitude, downward –IC flash. As stated above, this flash did not produce any NLDN current data. The event consisted of two TASD triggers separated by ∼140 $\mu$s. The first trigger contained two gamma-ray showers separated by about 11 $\mu$s, as seen in Figure 7.4, and totaled 100 VEM energy deposit in 9 detectors covering ∼11 km$^2$. As with TGF A, the first trigger was detected by the edge of the array, likely underestimating the showers' sizes and intensities. The second, more-energetic trigger was captured in its entirety slightly north of the first. 440 VEM of energy deposit was captured within a large footprint area of 12 detectors covering ∼14 km$^2$. The northward shift in ground detection is explained by the flash's largely horizontal leader development as seen by the interferometer (Figure 7.8 and Figure 7.12), in which the leader propagated between charge regions in the cloud for ∼25 ms. Due to the horizontal nature of this intracloud flash, as well as the lack of a well-defined return stroke, it is meaningless to compare average vertical leader development speed. Instead, small-scale breakdown speed during the TGF will be discussed in chapter 8 and compared to the other flashes.

The upgraded lightning detectors make it much easier to identify lightning-correlated events without relying on NLDN data or sequential bursts of TASD triggers. As such, the events of this data set have different characteristics to those of previous observations. For example, TGF B of this set consisted of only a single trigger, but was undoubtedly produced by its associated lightning activity, same as the other events which came in groups of two triggers each. As with the previous event sets, all TASD triggers occurred in the first 1–2 ms of lightning activity. With the new fast sferic sensors colocated with the INTF station, strong IBPs are also clearly seen at the beginning of each flash, but their relationship to TASD data was convoluted due to the vast separation between detectors. Figure 7.9–Figure 7.12 show this correlation, but these plots were only possible after the thorough analysis of section 7.2.

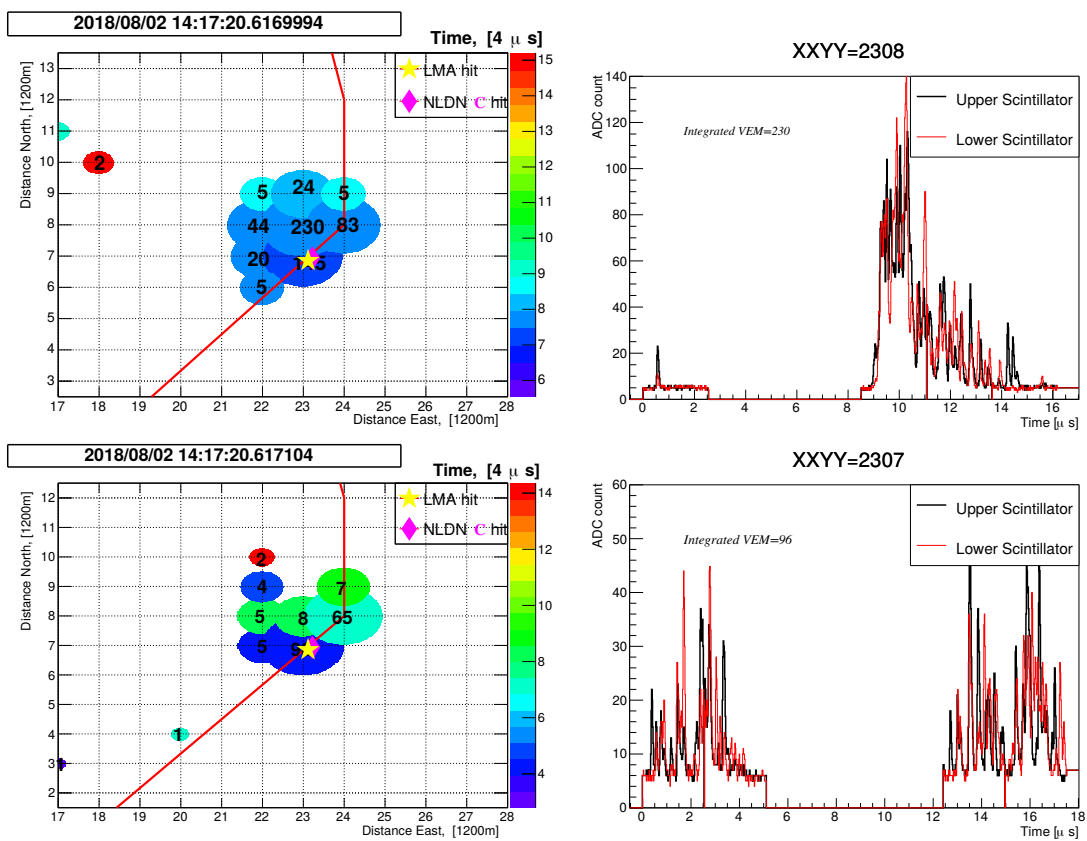

**Figure 7.1:** Footprints (left panels) for the two triggers of TGF A and the responses of each trigger's TASD with highest energy deposit (right panels). Numbers in footprint circles indicate the detectors' integrated energy deposit in VEM while colors represent relative onset times in units of 4 $\mu$s. The yellow star indicates the median plan location of LMA sources within $\pm 1$ ms of the gamma burst. The purple diamond indicates the NLDN event closest in time prior to the TGF (Table 7.1), but is hard to see due to being closely colocated with the LMA median point. Both LMA and NLDN points occurred beyond the edge of the Southeastern TASD boundary (red line).

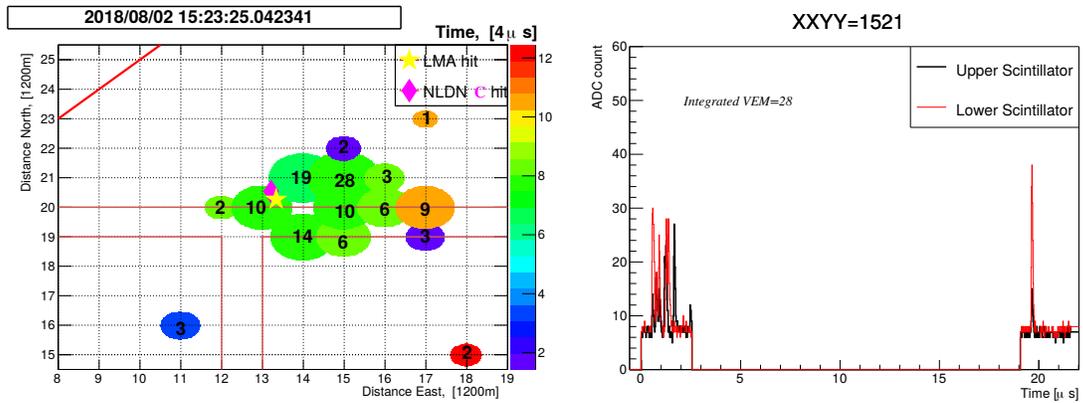

**Figure 7.2:** Footprint (left panel) for the single trigger of TGF B and the response of the TASD with highest energy deposit (right panel). Numbers in footprint circles indicate the detectors' total energy deposit in VEM while colors represent relative onset times in units of 4 $\mu$s. The yellow star indicates the median plan location of LMA sources within $\pm1$ ms of the gamma burst. The purple diamond indicates the NLDN event closest in time prior to the trigger (Table 7.1). TASD 1420 did not record a trigger and corresponds to the central hole in the footprint plot. The detector did not report any errors and appeared to be in perfect working condition during the TGF (more details in chapter 8).

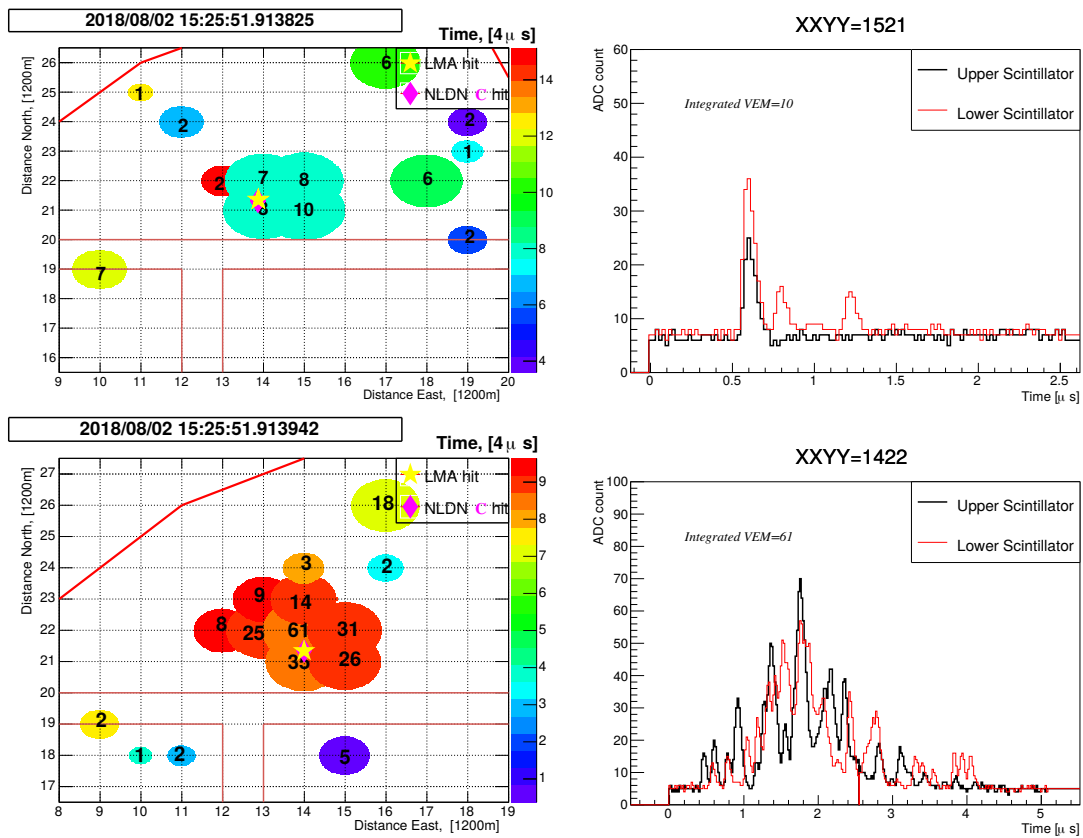

**Figure 7.3:** Footprints (left panels) for the two triggers of TGF C and the responses of individual TASDs (right panels). Numbers in footprint circles indicate the detectors' total energy deposit in VEM while colors represent relative onset times in units of 4 $\mu$s. The yellow star indicates the median plan location of LMA sources within $\pm 1$ ms of the gamma burst. The purple diamond indicates the NLDN event closest in time prior to the trigger (Table 7.1), but is hard to see due to being closely colocated with the LMA median point. The signal of TASD 1521 (top right) during the first gamma burst illustrates the fact that the SDs are capable of distinguishing individual Compton electrons in low-fluence detections. More discussion on this waveform can be found in chapter 8.

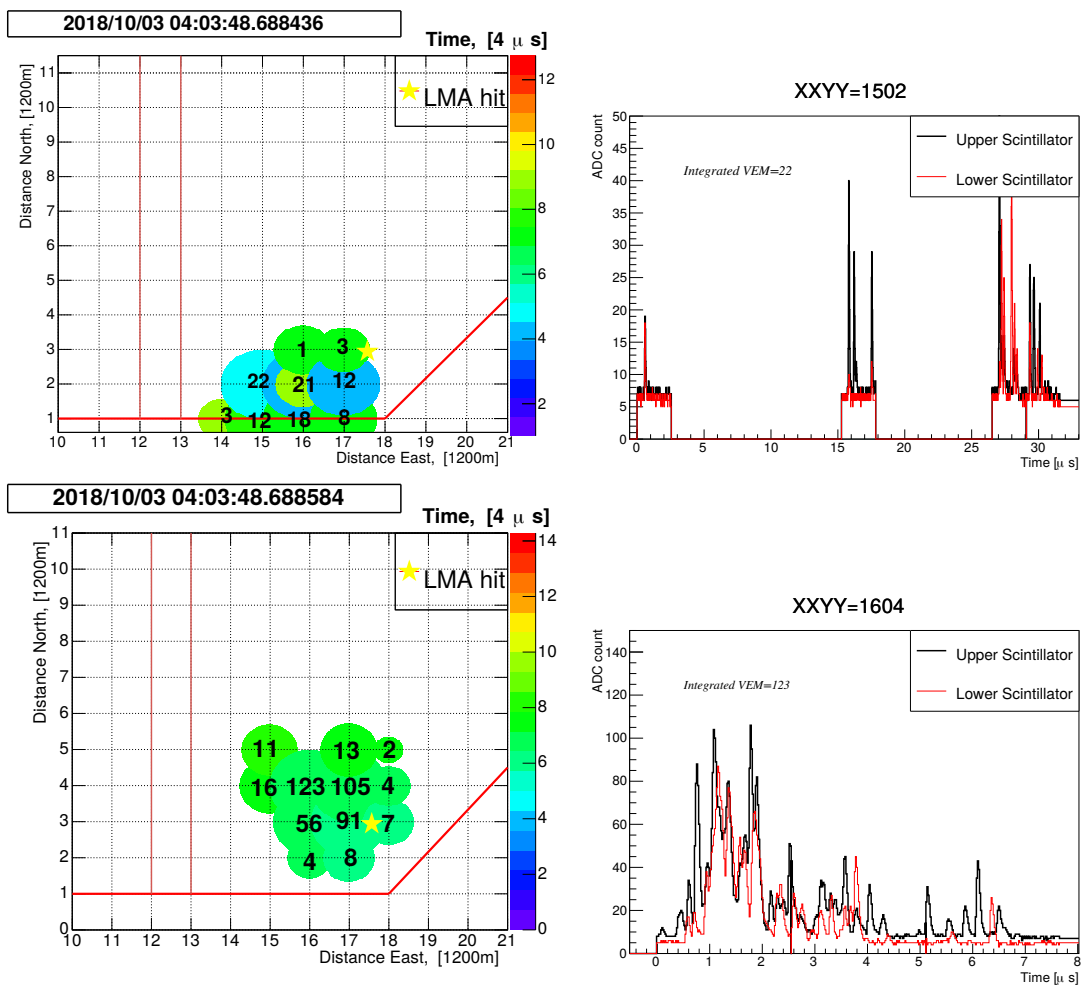

**Figure 7.4:** Footprints (left panels) for the two triggers of TGF D and the responses of each trigger's TASD with highest energy deposit (right panels). Numbers in footprint circles indicate the detectors' total energy deposit in VEM while colors represent relative onset times in units of 4 $\mu$s. The yellow star indicates the median plan location of LMA sources within ±1 ms of the gamma burst. The first event occurred on the edge of the TASD boundary (red line) while the second, main trigger appears to have been recorded in its entirety. The parent flash of TGF D was the only intracloud flash of this dataset and was not detected by the NLDN.

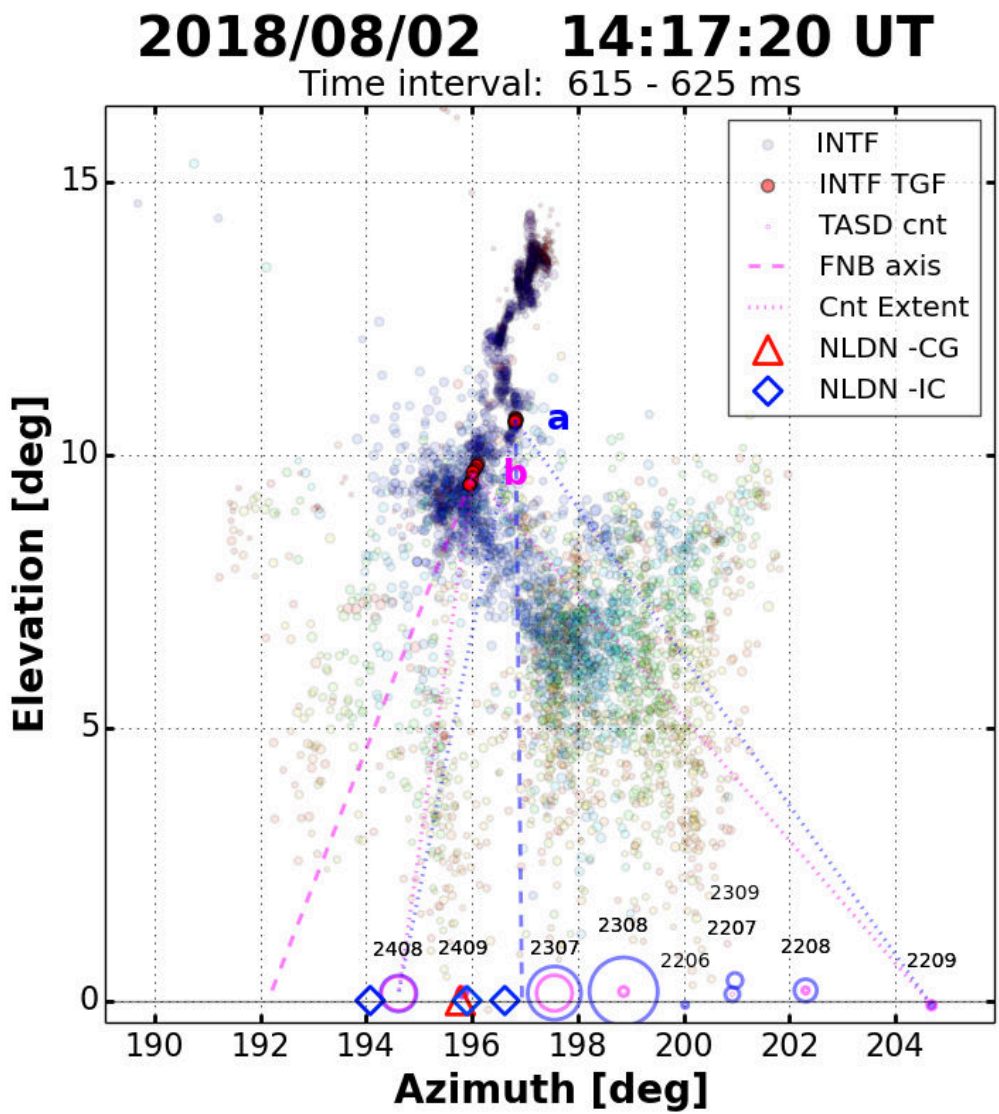

**Figure 7.5:** Azimuth-Elevation plots showing downward development of TGF A's parent flash. Red highlighted points and a, b labels indicate the TGFs' sources as calculated by the analysis of section 7.2. Dashed lines show the FNB axes associated with the TGF and extended to ground level. Finely dotted lines represent the angular extent of particle detections at the ground as detected by TASD stations. Baseline symbols represent TASD particle detections and NLDN flash data.

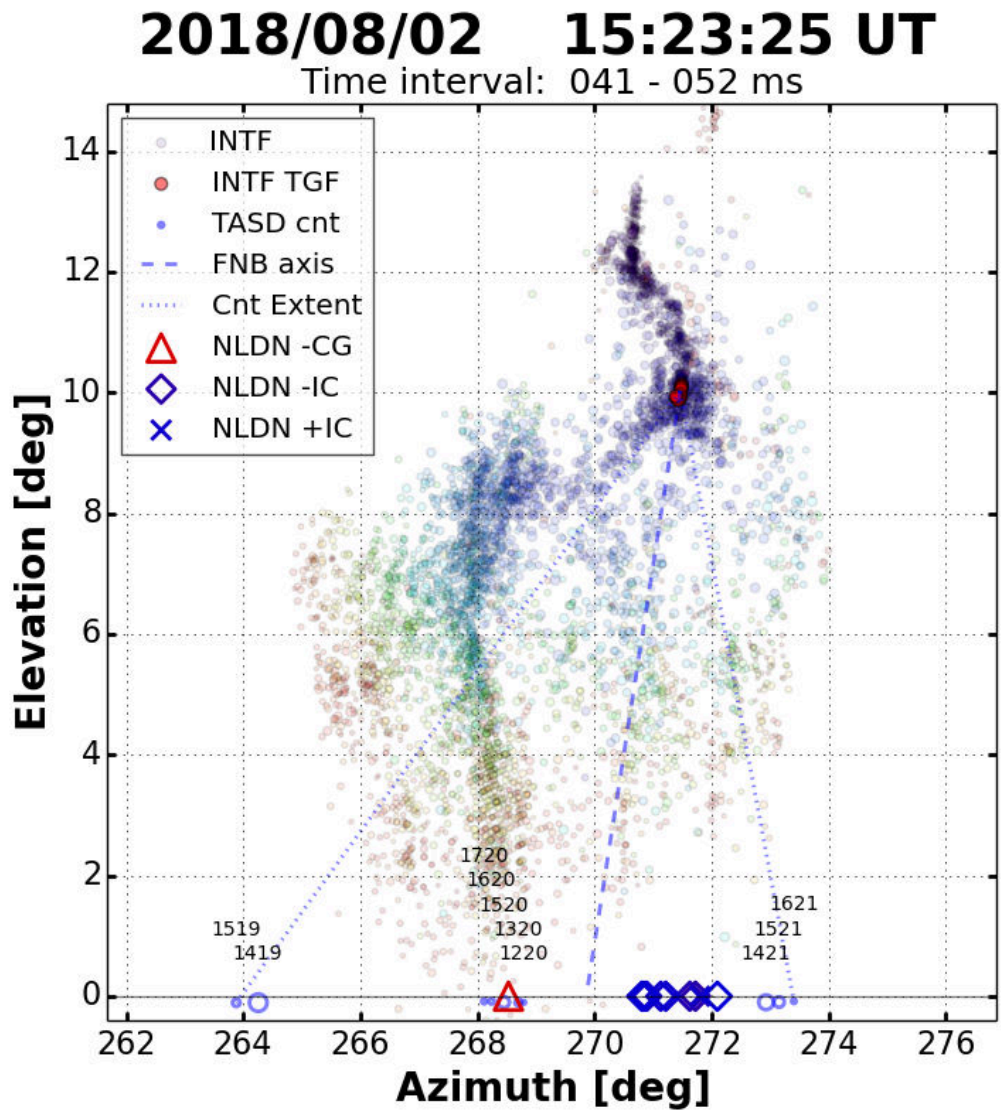

**Figure 7.6:** Azimuth-Elevation plots showing downward development of TGF B's parent flash. Red highlighted points indicate the TGF's source as calculated by the analysis of section 7.2. the dashed line shows the FNB axis associated with the TGF and is extended to ground level. Finely dotted lines represent the angular extent of particle detections at the ground as detected by TASD stations. Baseline symbols represent TASD particle detections and NLDN flash data.

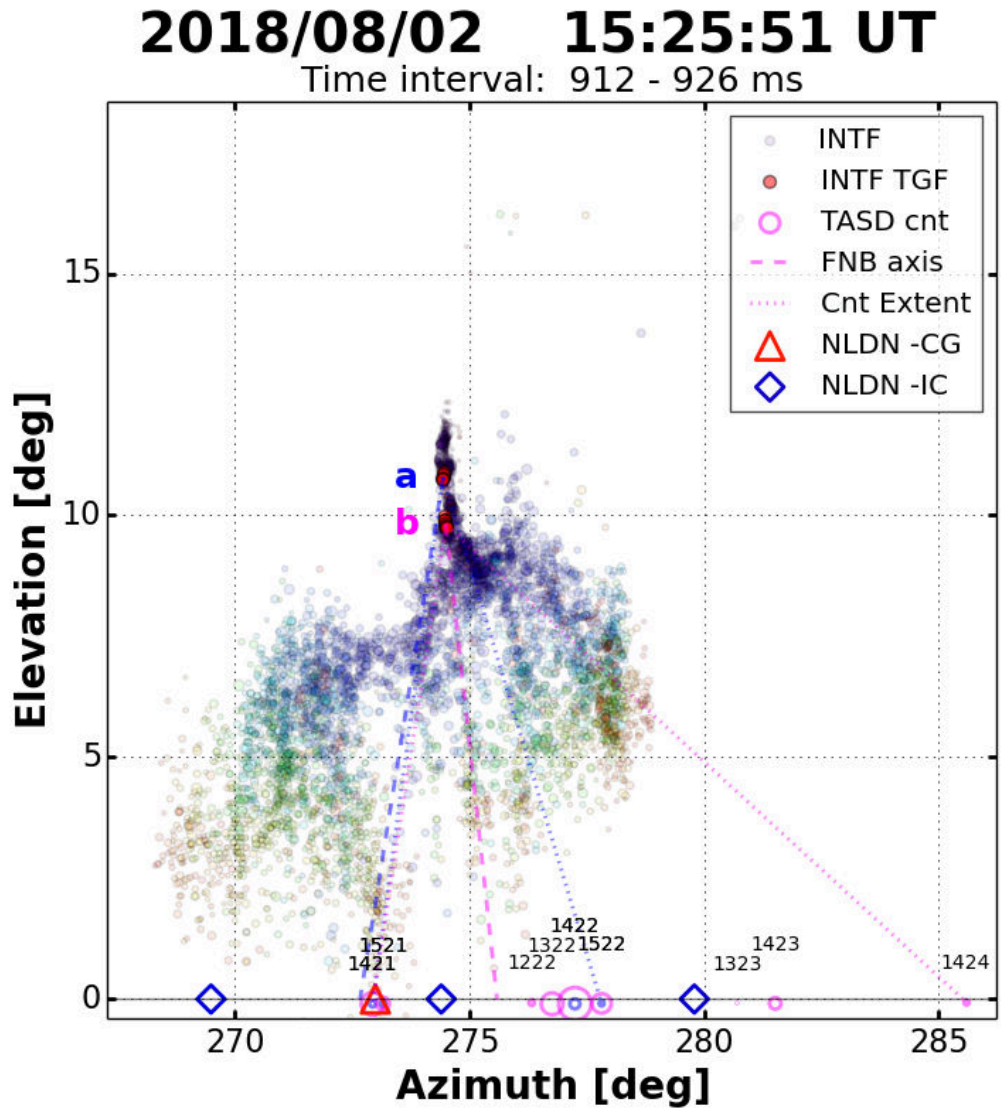

**Figure 7.7:** Azimuth-Elevation plots showing downward development of TGF C's parent flash. Red highlighted points and a, b labels indicate the TGFs' sources as calculated by the analysis of section 7.2. Dashed lines show the FNB axes associated with the TGF and extended to ground level. Finely dotted lines represent the angular extent of particle detections at the ground as detected by TASD stations. Baseline symbols represent TASD particle detections and NLDN flash data.

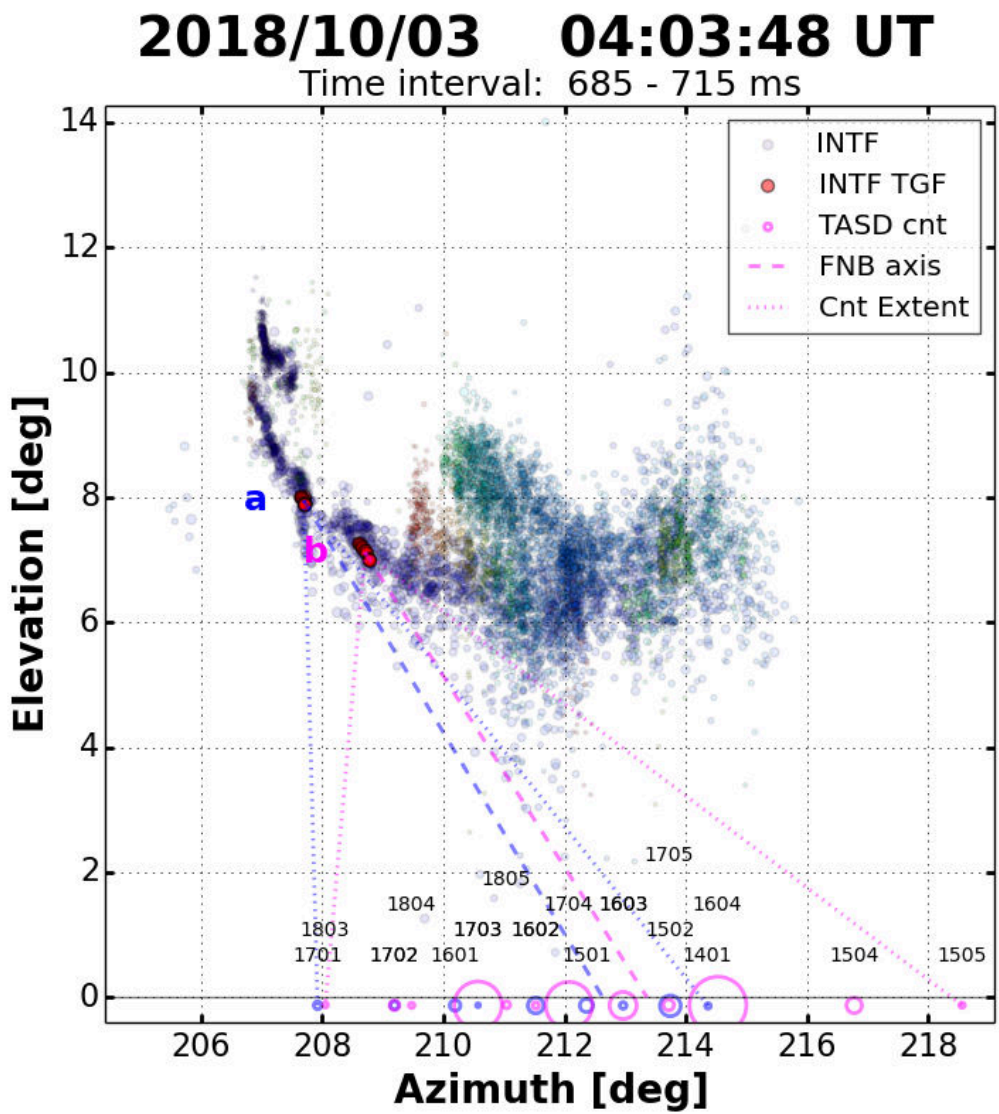

**Figure 7.8:** Azimuth-Elevation plots showing downward development of TGF D's parent flash. Red highlighted points and a, b labels indicate the TGFs' sources as calculated by the analysis of section 7.2. Dashed lines show the FNB axes associated with the TGF and extended to ground level. Finely dotted lines represent the angular extent of particle detections at the ground as detected by TASD stations. Baseline symbols represent TASD particle detections and NLDN flash data.

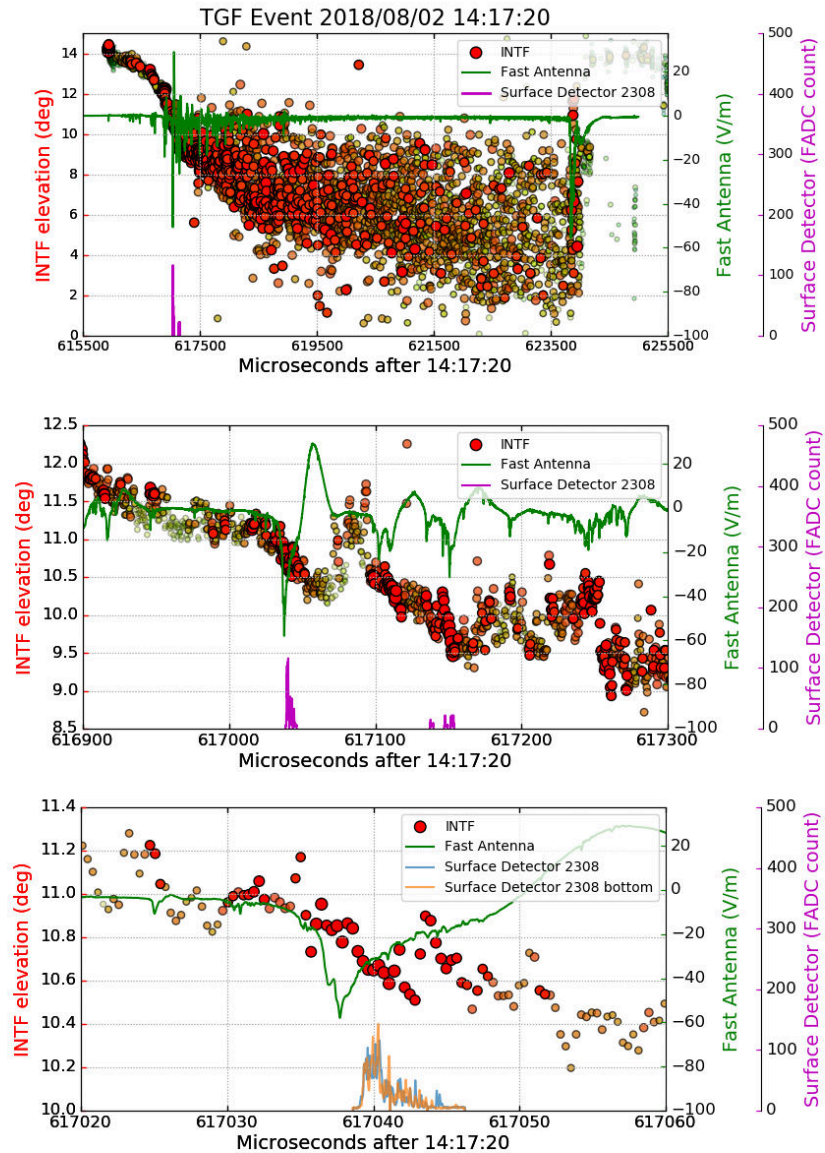

**Figure 7.9:** Combined data of TGF A on 2018/08/02 at 14:17:20 UT. Panels show INTF elevation data points versus time (circles), fast sferic waveform (green curve), and TASD gamma detections (purple curve). **Top:** Observations over the course of the full flash ending in a –38.3 kA cloud-to-ground stroke. Initial TGF detection occurred during the strongest (–36.7 kA) sferic pulse 326 $\mu$s after flash start. **Middle:** 400 $\mu$s of observations centered on the two triggers, showing their correlation with the two strongest sferic pulses. **Bottom:** Expanded 40 $\mu$s view of TASD 2308's upper and lower scintillator responses of the first (and strongest) TGF relative to the IBP sferic and downward FNB (see chapter 8). Other detector responses are shown in Figure 7.13.

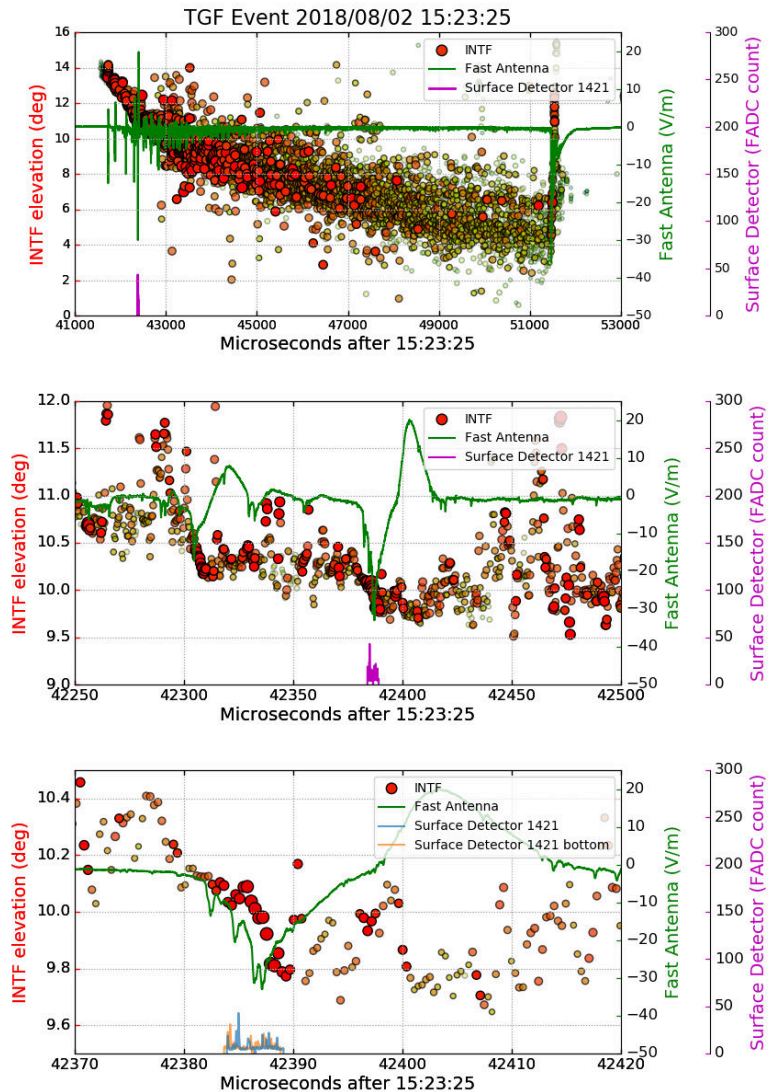

**Figure 7.10:** Combined data of TGF B on 2018/08/02 at 15:23:25 UT. Panels show INTF elevation data points versus time (circles), fast sferic waveform (green curve), and TASD gamma detections (purple curve). **Top:** Observations over the course of the full flash ending in a –26.5 kA cloud-to-ground stroke. TGF detections occurred during the strongest (–30.1 kA) sferic pulse 341 μs after flash start. **Middle:** 250 μs of observations centered on the TASD trigger, showing its correlation with the strongest IBP. **Bottom:** Expanded 50 μs view of TASD 1421's upper and lower scintillator responses of the TGF relative to the IBP sferic and downward FNB. Other detector responses are shown in Figure 7.14.

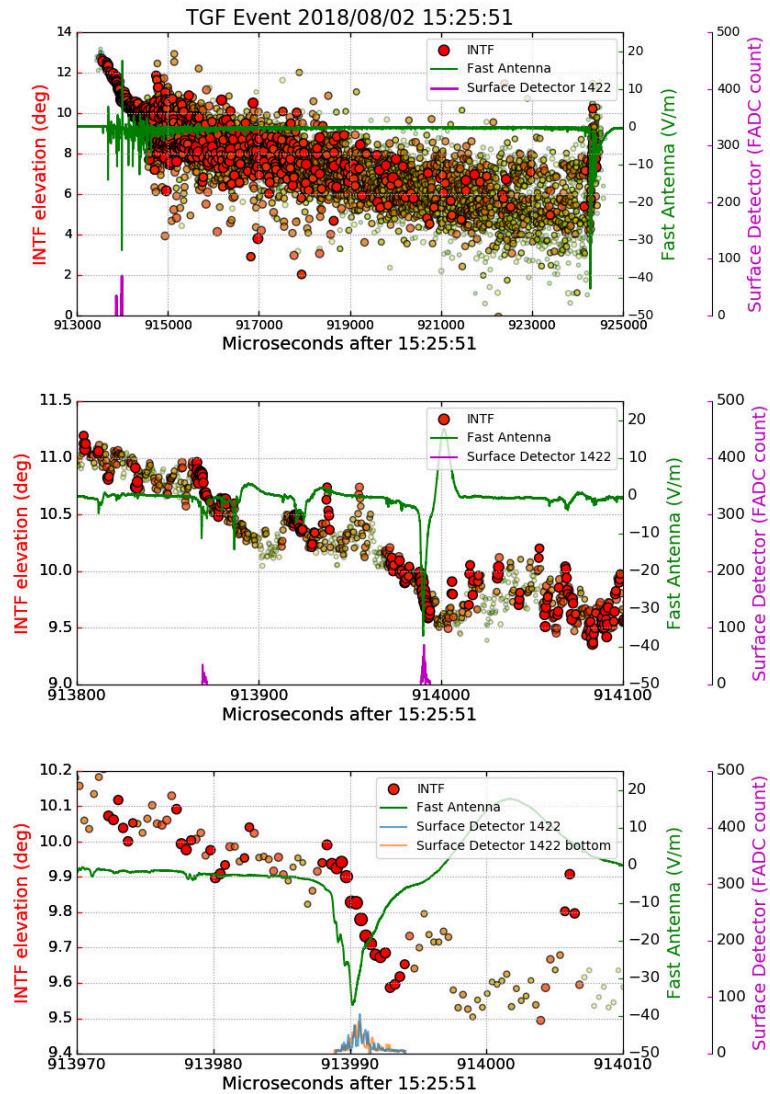

**Figure 7.11:** Combined data of TGF C on 2018/08/02 at 15:25:51 UT. Panels show INTF elevation data points versus time (circles), fast sferic waveform (green curve), and TASD gamma detections (purple curve). **Top:** Observations over the course of the full flash ending in a –26.8 kA cloud-to-ground stroke. In this case, the flash produced two gamma bursts 825 and 942 $\mu$s after flash start. The first was associated with a weaker sferic, but still during an episode of downward FNB. The main, second burst occurred during the strongest sferic pulse (–21.7 kA). **Middle:** 300 $\mu$s of observations centered on the two TASD triggers, showing their correlation with strong sferic pulses. **Bottom:** Expanded 40 $\mu$s view of TASD 1422's upper and lower scintillator responses of the second TASD trigger relative to its associated IBP sferic and downward FNB. Other detector responses are shown in Figure 7.15.

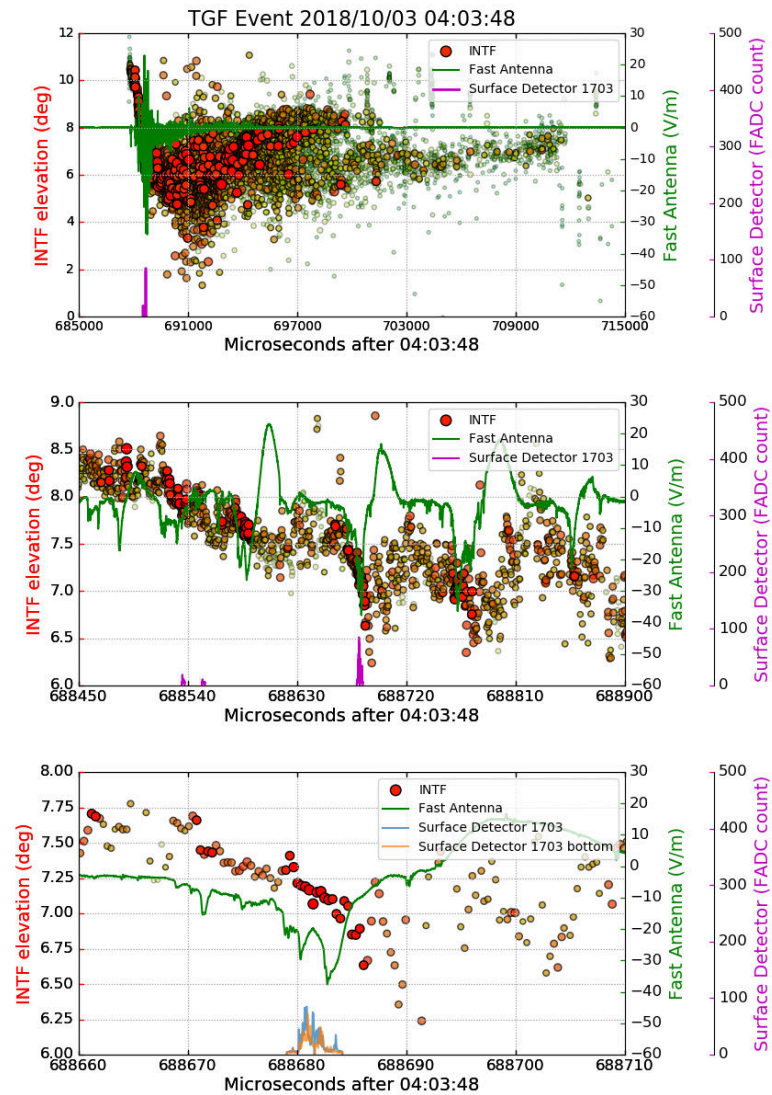

**Figure 7.12:** Combined data of TGF D on 2018/10/03 at 04:03:48 UT. Panels show INTF elevation data points versus time (circles), fast sferic waveform (green curve), and TASD gamma detections (purple curve). **Top:** Observations over the course of the full low-altitude intracloud flash. The flash produced two gamma bursts ∼600 and 740 μs after flash start. The first was associated with relatively weak sferic pulses while the second occurred during the flash's strongest IBP. **Middle:** 450 μs of observations centered on the two triggers, showing their correlation with IBP sferics. **Bottom:** Expanded 50 μs view of TASD 1703's upper and lower scintillator responses of the second, stronger TGF relative to the IBP sferic and downward FNB. Other detector responses are shown in Figure 7.16.

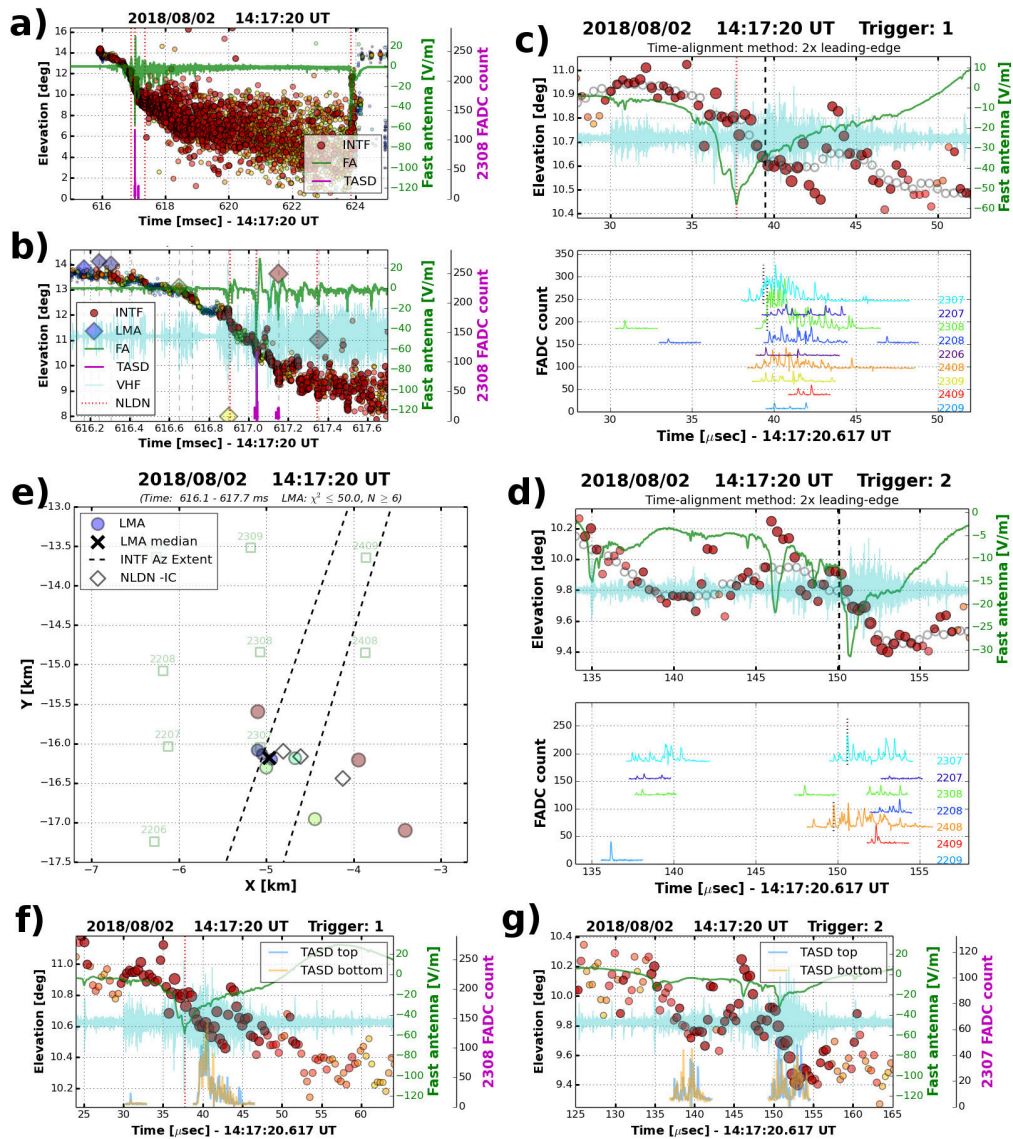

**Figure 7.13:** More complete summary results for TGF A, including observations for the two triggers (note from panel d that the second trigger appeared to contain data from two gamma bursts separated by 12 $\mu$s). **a,b:** Overview plots similar to Figure 7.9 but with NLDN detections (red dotted vertical lines), LMA data points (diamonds), and raw VHF waveform (cyan trace). **c,d:** Correlation results from the 'stepping' method of section 7.2, showing the mean onset time given by the two TASD stations of highest energy deposit (vertical dashed lines) compared to the signals of all active TASDs. **e:** Overhead (plan) view of LMA sources within $\pm0.8$ ms of the TGF (the median of these points is marked by the black 'x'). Dashed lines indicate the middle 90% of INTF azimuth values within the same $\pm0.8$ ms window. Light green squares represent TASDs that were triggered by the TGF. **f,g:** Detailed view of the observations of panels a and b showing responses from both upper and lower scintillators. Note the correlation with FNB even during the weaker IBPs and gamma bursts of the second trigger.

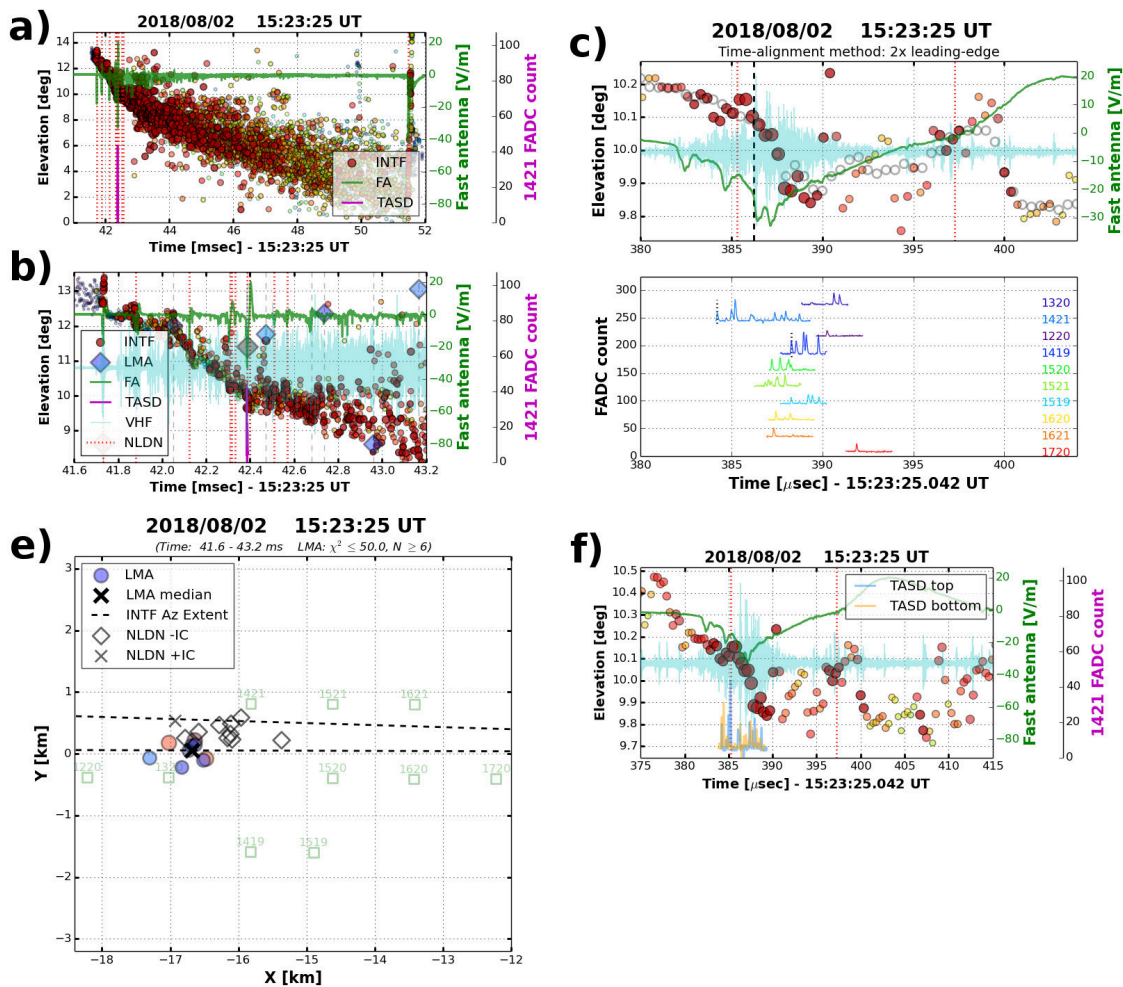

**Figure 7.14:** Same as Figure 7.13, but for the single gamma burst of TGF B. Unfilled gray circles in panel c indicate the 0.5 μs averages from higher-resolution INTF data used in the stepping method of section 7.2. TASD signals are sorted top to bottom according to increasing range from the TGF source. Note the correlation in panel f of the close TASD 1421 waveform with the sferic sub-pulses and sequence of upward FPB followed by downward FNB which was not detected by any other TASDs for at least 3 μs, reinforcing the interpretation that the gamma burst consisted of more than a single, synchronized onset (see chapter 8). Also note the tight clustering of NLDN and LMA points in panel e which lead to the superior resolution of the analysis for this event (details in section 7.3).

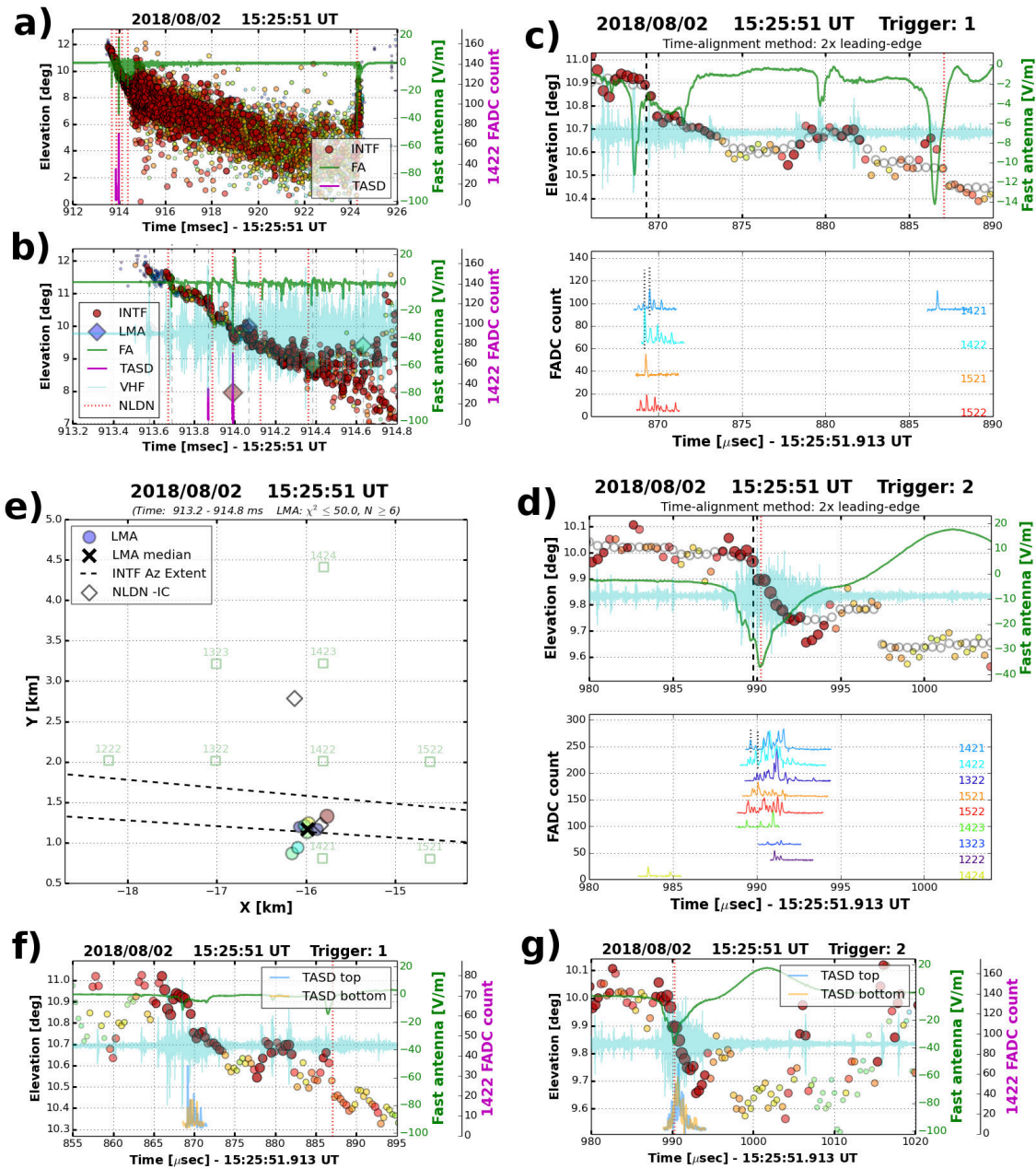

**Figure 7.15:** Same as Figure 7.13, except for the two triggers of TGF C. Again, the NLDN correctly identified many of the strongest IBPs as seen in panel b. One of these was noticeably mislocated from the otherwise tight grouping of NLDN and LMA points (panel e). The main gamma burst occurred during the second trigger associated with a simple, classic IBP with leading sub-pulses and smooth FNB (panel d). Most TASDs of the first trigger were correlated with a single, weaker IBP and FNB, but TASD 1421 also registered some low energy deposit in the following IBP 17 μs later as seen in panel c.

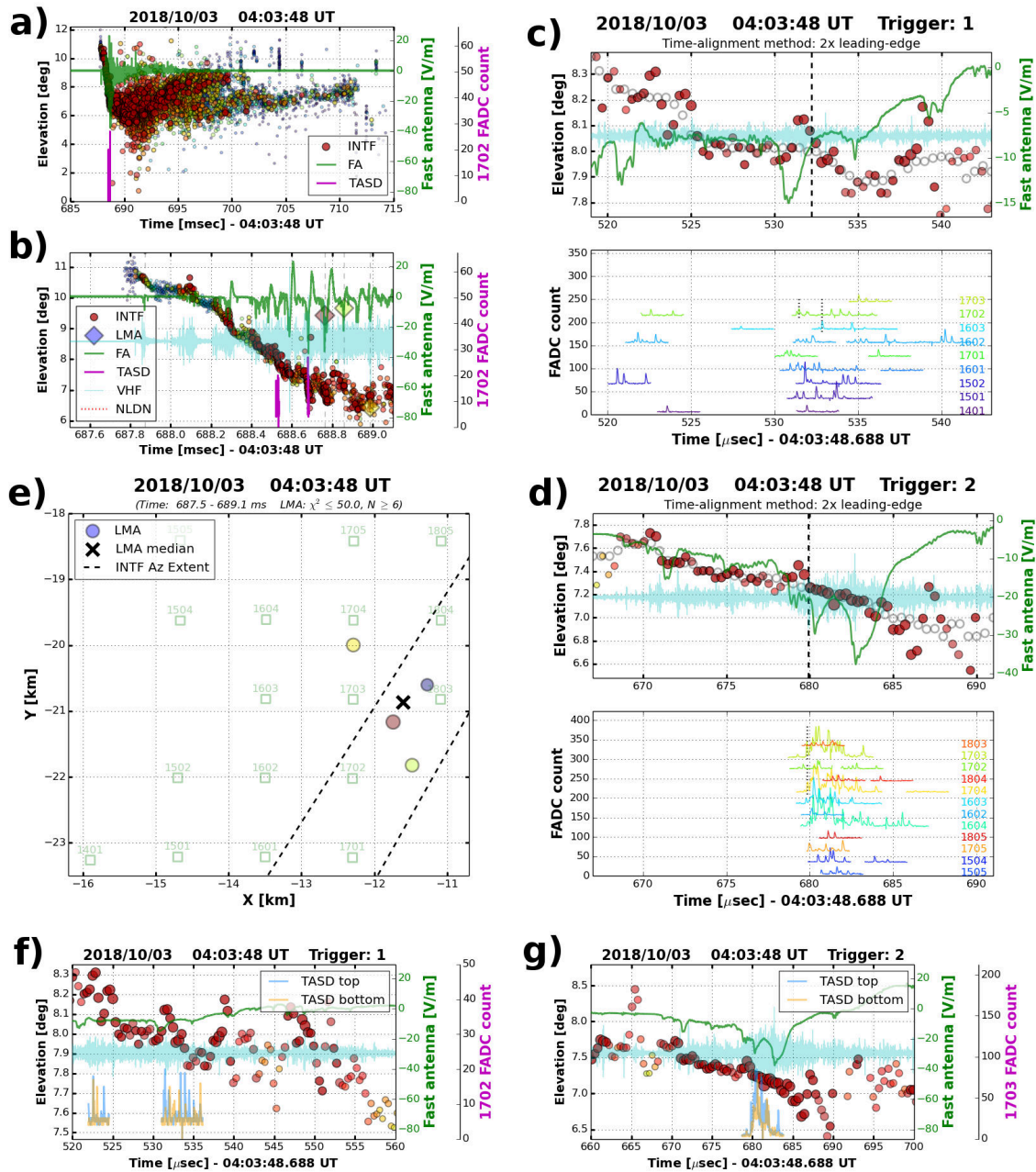

**Figure 7.16:** Same as Figure 7.13, except for the two triggers of TGF D. This –IC flash did not continue to ground and was not detected by the NLDN, but its initial breakdown was otherwise similar to those of TGFs A–C. The second trigger was much more energetic and was associated with the strongest, rather complex IBP of the flash (panel d). The onset itself was further correlated with a strong sub-pulse before the peak during a long period of FNB. The first trigger again contained data from two gamma bursts separated by $\simeq 10 \ \mu$s, both occurring during weaker IBPs and periods of noisy FNB. The poor grouping of associated LMA points during the main trigger (panel e) resulted in the largest uncertainty in the analysis of this event (section 7.3).

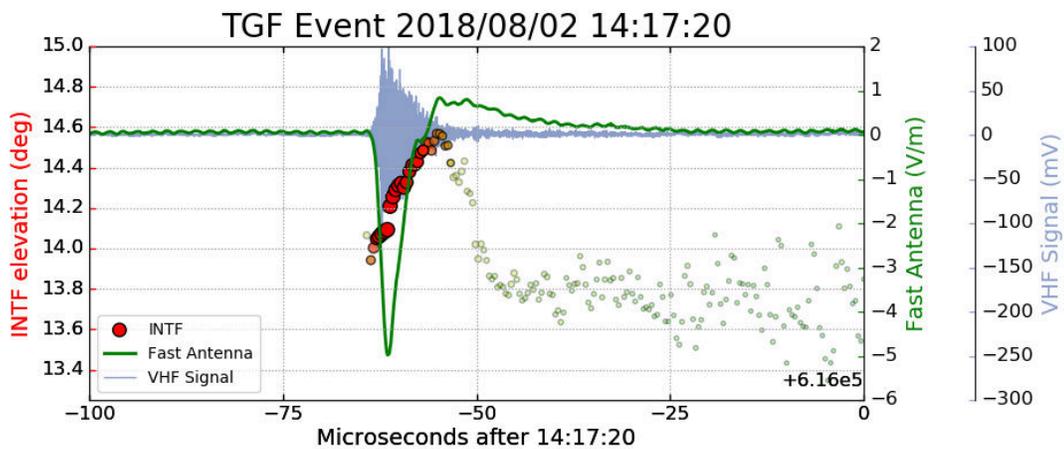

**Figure 7.17:** The NBE which initiated the parent flash of TGF A. This is a classic example of a narrow bipolar event produced by fast positive breakdown (subsection 2.4.2). The FPB propagated upward over 150 m in 11 $\mu$s, corresponding to an average speed of $1.3\times10^7$ m/s. While the peak VHF source powers were essentially equal to those of the TGF (27.7 dBW vs. 27.6 dBW), the TGF's IBP sferic was 24 times stronger (58 V/m vs. 2.4 V/m). As a result, the NLDN measured a peak current of –36.7 kA for the IBP sferic but did not detect the initiating NBE's sferic, whose peak current would have been only ∼2 kA.

**Table 7.1:** Quantitative INTF event values

Observations of the LMA, NLDN, and TASD observations for the flashes of TGFs A–D of the INTF event set, during the initial 1–2 ms of the flashes. LMA source powers are given in units of dBW, NLDN currents are given in kA, and total energy deposit per TASD is given in both VEM and approximate MeV units. Times of these events are given in microseconds after the second — times corresponding to TASD triggers are given as the time of detection at ground level, delayed by propagation from the source. LMA and NLDN times correspond to the time of the source itself as determined by standard processing.

| Date | Time | $\mu$sec | LMA dBW | NLDN $I_{pk}$ | TASD energy sum VEM/MeV | # of TASDs |
|---|---|---|---|---|---|---|
| 2018/08/02 | 14:17:20 | 616655 | 20.3 | | | |
| | | 848 | | -10.0 kA C | | |
| | | 851 | 21.5 | | | |
| | | 981 | | -36.7 kA C | | |
| | | 987 | 27.7 | | | |
| | | 994 | | | 561/1150 | 9 |
| | | 617094 | 27.2 | | | |
| | | 104 | | | 192/393 | 8 |
| | | 191 | 20.3 | | | |
| | | 288 | | -12.0 kA C | | |
| | | 294 | 27.6 | | | |
| | | 444 | 21.8 | | | |
| 2018/08/02 | 15:23:25 | 042253 | 25.4 | | | |
| | | 255 | | -10.9 kA C | | |
| | | 259 | | +12.0 kA C | | |
| | | 279 | | -3.1 kA C | | |
| | | 329 | | -26.9 kA C | | |
| | | 332 | 27.2 | | | |
| | | 341 | | -30.1 kA C | | |
| | | 341 | | | 112/229 | 12 |
| | | 416 | 18.1 | | | |
| | | 459 | | -7.6 kA C | | |

**Table 7.1:** (continued)

| Date | Time | μsec | LMA dBW | NLDN $\mathbf{I}_{pk}$ | TASD energy sum VEM/MeV | # of TASDs |
|---|---|---|---|---|---|---|
| 2018/08/02 | 15:25:51 | 913524 | 4.9 | | | |
| | | 615 | | -6.4 kA C | | |
| | | 633 | 15.3 | | | |
| | | 816 | 20.1 | | | |
| | | 825 | | | 35/72 | 5 |
| | | 832 | | -5.0 kA C | | |
| | | 937 | | -21.7 kA C | | |
| | | 938 | 26.7 | | | |
| | | 942 | | | 212/434 | 8 |
| | | 914016 | 14.4 | | | |
| | | 076 | | -5.9 kA C | | |
| 2018/10/03 | 04:03:48 | 687336 | 22.2 | | | |
| | | 696 | 23.2 | | | |
| | | 688177 | 26.1 | | | |
| | | 200 | 22.0 | | | |
| | | 346 | 24.6 | | | |
| | | 436 | | | 100/205 | 9 |
| | | 508 | 31.8 | | | |
| | | 584 | | | 440/902 | 12 |
| | | 599 | 30.5 | | | |

## 7.2  INTF Event Analysis

The concoction of detectors of this study observe different characteristics of lightning phenomena and have different locations throughout the array, making synchronized detection difficult. Signal propagation time between these stations can be up to 100 $\mu s$, assuming a lightning initiation of $\sim$3 km (as observed in previous events of Abbasi et al. (2017, 2018)). The upgrade, however, was intended to increase resolution to the order of a microsecond, in which details of lightning activity at the time of TGF production could be distinguished and examined more closely. Detailed methods were developed for the purpose of correcting the array's geometrical effects as accurately as possible, and are documented here in great detail. For each of these steps, data from TGF A is provided in-text as an example. The complete set of values used in the analysis and its results are given at the end of subsection 7.2.1 in Table 7.2 and Table 7.3.

This section also includes analyses on the TGF-initiated showers themselves based on energy and timing data provided by the TASDs. This begins with a further examination of individual particle energies, explored briefly in subsection 6.2.2, followed by the effects of Compton scattering on detection times and shower composition.

### 7.2.1  Geometry Corrections

The broadband interferometer's high sample rate and reconstruction accuracy support the resolution goals of this study (section 5.3), but the instrumental differences still remain a problem. The measurements are not trivial to compare to one another when they do not detect the same lightning impulses — i.e., the broadband interferometer is tuned for VHF radiation between 20–80 MHz, SDs record incident particle energy deposition, and sferic sensors measure relative electric field change. Therefore, a new method was needed to align the high-quality measurements and make sense of their interactions. In order to maximize resolution, propagation times from each source to each detector needed to be precisely accounted for. Of course, this requires accurate 3D positions both sources and detectors. Panel a of Figure 7.18 shows a simplified coordinate system centered on a TGF source ($a$) at reference altitude 1.4 km with propagation vectors to active TASDs ($b$) and to the INTF station ($c$). Unfortunately the INTF processes only 2D field-of-view sources using zenith and azimuth. Plan location, then, must be given by another measurement in order to convert an elevation measurement ($\theta$) into a true altitude ($z_a$). Originally, it was assumed that the plan location should be directly over the center of SD energy deposit (shower core, or the average TASD location weighted by energy deposit), but this led to poorly-aligned signals. Figure 7.8 of TGF D, for example, shows that a slanted trajectory during a TGF can lead to significant lateral offset between its source position and detection at ground level. Sferic sensors don't have the capability to reliably locate lightning impulses, so the remaining option is to use the GPS data of LMA points.

The LMA's slower data rate of $\sim$100 $\mu$s does not work well for tracking leader propagation on these small scales, but can work as a starting point for a single-point estimation of a source plan location. Given this horizontal location, and the relative flatness of the Telescope Array, INTF elevation data can be converted to altitude using simple trigonometry: $z_a = D \times tan(\theta)$, where $D$ is the horizontal distance between the INTF station and the estimated source location. Another issue of this method is that the TASDs are the only detectors recording the TGF particle showers themselves, and do not have a direct connection to the LMA or interferometer. Thus, despite the INTF's frequent and accurate elevation data, there is no obvious way to know which VHF sources correspond to the TGF itself except by shifting the timing of TASD triggers ($t_b$) into a frame relative to the INTF location ($t_c$) by removing all effects due to propagation time.

In summary, the 4-dimensional TGF source determinations (3 spatial dimensions with time) are defined by the minimum 4 measurements: TASDs supply the onset time ($t_a$), the LMA gives its plan location ($x_a$, $y_a$), and the INTF defines its altitude ($z_a$). Before starting calculations, we define the following quantities as being fixed (uncertainties in these values are provided in the following section): TASD locations ($x_b$, $y_b$, and $z_b$) are known via long-term GPS measurements and trigger times are have a fixed value at each detector for each event based on local triggering software (section 4.2). The INTF location is also fixed at the well-defined $x_c$, $y_c$, and $z_c$. Other measurement coordinates are determined from these data as well as LMA sources as outlined below.

The idea is simple enough in its description, but the execution is less than straightforward. To begin, all LMA point positions within $\pm$1 ms of the TASD trigger are averaged and used as the TGF source plan location. For TGF A, the LMA mean plan location was ($39.193273$, $-112.754214$). As described in (section 5.1), non-impulsive and VHF-noisy sources can lead to mislocations by LMA, resulting in poor fit parameters in most cases (large $\chi^2$). This is avoided by excluding all LMA points with $\chi^2$ >50. For TGF A, 4 such LMA points were cut from the original 15 with $\chi^2$ values between 86–300. In rarer cases, however, these sources can produce points with passing $\chi^2$ values, but with large offsets misrepresenting the true position. Using the median of these points could protect against outliers, but medians can be more biased than means with small sample sizes like these. Therefore, for these groups of $\leq$15 points, any outliers ($> 2.5\sigma$) within the 2 ms window were also excluded before taking the average. For TGF A, one such outlier was excluded, leaving 10 LMA points which passed the cuts and produced the gps coordinates given above. In addition, averages allow for the use and propagation of standard error $\sigma_{\bar{x}}$ (error in the mean) to quantify these measurements and their limitations. The outlier and $\chi^2$ cutoff values here were chosen to maximize precision of the plan location; too high and the outliers would not be excluded and would skew data. If too low, too many points would be excluded and the sample size would be too small to represent the true error. The cuts of TGF A reduced the latitude (longitude) standard error from $0.00253°$ $(0.00246°)$ to $0.00172°$ $(0.00213°)$. More details on the error analyses are given in section 7.3.

At this point, the plan location of the source has been identified, but timing and altitude are still unknown; unfortunately the two are irrevocably tied to one another. Figure 7.9 shows data from TGF A — note that the time of TGF production cannot be determined from the INTF points alone. The elevation of lightning activity is a function of time ($z_a(t_c)$, with $t_c$ relative to the INTF position), so a reliable TGF arrival time is needed to determine the source altitude $z_a$. As we have established, timing information is known by active TASDs alone, and must be back-propagated to the source (at an altitude of $z_a$) before accounting for the TGF–INTF propagation distance $R$. In this way, INTF detection time is a function of altitude ($t_c(z_a)$). Now the issue begins to reveal itself: time depends on source altitude and vice-versa ($z_a(t_c(z_a))$).

This interdependency cannot be analytically inverted, but numerical methods can provide a solution without too much fuss. In the early stages of analysis, two basic methods were developed independently to connect INTF and TASD data: scanning and iterating over INTF points to determine the most likely altitude at which the TGFs were initiated. The former iterative method was used for solutions in Table 7.2 and started by assuming an altitude estimate to begin the process. In previous observations, TGF altitudes were observed at ∼3 km, which turned out to be a good approximation for these four events as well (see Table 7.3). With this initial value of $z_a$, we can move on to solving backward for source time $t_a$ using propagation time from point a to b (panel a of Figure 7.18). Mathematically,

$$t_a = t_b - \frac{r_b}{c} \tag{7.1}$$

for which $r_b$ is the slant range between the TGF source and a given SD (note that this has a unique solution for each active SD). For TASD 2208 of TGF A, for example, $r_b = 3.55$ km and $t_b = 616994.0$ $\mu$s after the second, resulting in $t_a = 616981.7$ $\mu$s.

With a source time, we can now determine the time the signal should have reached the interferometer ($t_c$) using an analog of Equation 7.1:

$$t_c = t_a + \frac{R}{c} \tag{7.2}$$

where $R$ is the slant range between the source and the INTF station. For the same SD 2208 of TGF A, $R = 17.26$ km, resulting in $t_c =$ 617039.2 $\mu$s (net difference of $\Delta t_b = t_c - t_b = 45.2$ $\mu$s).

Now with the INTF time, the circular dependence is satisfied and the elevation $\theta$ at time $t_c$ can be used to find $z_a = D \times tan(\theta(t_c))$. Due to some noise in the INTF elevation measurements, a mean is taken of INTF points within $\pm 4$ $\mu$s of $t_c$ to determine the approximate elevation and equivalent altitude. Similar to LMA point solutions, the interferometer can produce outliers, but these are removed during processing as outlined in section 5.3 and do not contribute to the mean. Remember that these calculations are done for each active SD, producing a list of solutions at each step as seen in Table 7.2. Again for SD 2208 of TGF A, $D = 16.96$ km and the average $\theta(t_c) = 10.73°$, giving an altitude of $z_a = 3.21$ km.

Of course, these resulting altitudes were determined assuming a rough estimate and are not perfect solutions. The preceding process can be iterated through again (see Figure 7.19) using the new altitude value as the initial estimate until the estimate and resulting values are consistent, which happens after only a few iterations for these events. The final altitudes are indeed close to the assumed 3 km, but the initial estimate can be varied and has no significant effect on the final converged value. Because this process is carried through for each participating surface detector, there are many, slightly varied solutions to the iterative calculations; after the iteration process converged, a median of these solutions was then taken to best represent all coordinates of the source (bold values in Table 7.2). The resulting uncertainties were calculated alongside each step in the process and are shown in the same table (see section 7.3 for more details).

The alternative stepping method used a strategy of smoothing out the noisy INTF points into averaged, 0.5 $\mu$s steps (visualized in panel b of Figure 7.18), each of which was taken in turn as the assumed altitude and time ($z_a$ and $t_c$). These values were then shifted to account for propagation time to predict the onset time of the TGF's arrival at participating TASD stations using inverse forms of Equation 7.1 and Equation 7.2:

$$t_b = t_c - \frac{R}{c} + \frac{r_b}{c} = t_c - \frac{R - r_b}{c} \tag{7.3}$$

In this method, rather than making this calculation for each participating TASD individually, this value of $t_b$ was compared to only the two most energetic detectors. This way, the difference between the predicted onset time and the true measured TASD trigger time can be minimized and interpolated to find the most likely corresponding source altitude. As confirmation of both methods' accuracies, the two were developed independently and produced consistent results within the reported standard error. For the example of TGF A in panel b of Figure 7.18, the stepping method resulted in a $t_c$ of 617039.6 $\mu$s after the second, while the iterative method resulted in 617039.2 $\mu$s, within the 0.7 $\mu$s uncertainty. Correspondingly, the stepping method's solution of $z_a$ is ∼3.18 km and is within the 30 m uncertainty of the iterative method's solution of 3.21 km. Although the two methods are consistent and are a good test of accuracy, the iterative method's results are those reported in data tables and provided figures because it takes all active TASDs into account and supports a more thorough error analysis.

After the source coordinates have been determined, it is now possible to put all data types into the same time frame, effectively removing the effects of propagation. All plots given here display data relative to the INTF station ($t_c$) because it is the most well-defined frame of reference and is colocated with the fast sferic sensor. TASD data is shifted to this frame using a form of Equation 7.3:

$$\Delta t = t_c - t_b = \frac{R - r_b}{c} \tag{7.4}$$

such that $\Delta t$ is the effective time delay between the two frames of reference. Again, this value is unique to each participating TASD, and the set of solutions for each TGF is given in Table 7.2. After shifting, TASD detector responses can be directly related to INTF data and sferic signals, as in the the plots of Figure 7.9–Figure 7.12. For these figures, TASD data is shown only from the detector with most energy deposit for simplicity. Figure 7.20– Figure 7.23 show full-page versions of the zoomed-in view of each event with all participating TASD data as well as the raw VHF signal detected by the INTF. In these plots, the solid black vertical and horizontal lines represent the median TGF onset time ($t_c$ in the INTF frame) and elevation ($\theta_a$), respectively, as determined by the iterative analysis method (bold values in Table 7.2). The intersection of these lines, then, indicates the onset of TGF-producing activity in the lightning flash.

### 7.2.2 Particle Energy

Despite the analysis of subsection 6.2.2 (Abbasi et al. (2018)), the energies of particles detected at Telescope Array have been called into question. Typical TGF studies use calorimeter data to measure individual particle energy due to the low flux at relatively small, distant satellites (section 3.1). The scintillator detectors of the TASD (section 4.2), on the other hand, are optimized for timing resolution and do not track or entirely capture incident particles. In addition, the study of Dwyer et al. (2005) presented results of x-ray emissions up to a few hundred keV during the stepped leader stage of negative lightning. The similarity of these observations may suggest that the events at TASD also consist of photons in the x-ray regime. This appears to be the case when ignoring energy limitations at TASD, but the analysis of Abbasi et al. (2018) presented evidence of at least some gamma photons having minimum energies of 2.2 MeV, fitting within the range of energy reported by satellite measurement (Nemiroff et al. (1997)).

These lower limits are obtained by examining TASD responses in detectors with low energy deposit, in which individual particle responses can be distinguished without significant signal pileup. From extensive simulations of the surface detectors, visualized in Figure 3.5, individual photons below ∼1 MeV (and their Compton scattered electrons) deposit negligible amounts of energy. Of course, in detectors observing high flux, these can sum up to considerable energy. In those which experience low flux, however, detectors can resolve singular incident particles. From the top right panel of Figure 7.3 from TGF C, three distinct particles were detected depositing energy consistent with a single VEM as defined in subsection 4.2.3. The simulation figure suggests that these particles, if electrons, were most likely on the order of 10 MeV or higher. In an effort to place more definitive lower limits, however, we look at the minimum-energy case. Following the example of subsection 6.2.2, energy deposit of 1 VEM corresponds to ∼2 MeV. The two latter signals of the TGF C example deposited ∼2.4 MeV each in the lower layer of scintillator. Assuming that 100% of electron energy was absorbed and that photons were perfectly back-scattered, the electrons could not have been produced by photons of energy less than 2.6 MeV. The earlier, larger signal of the same figure represents the penetrating case in which the particle deposits energy in both layers. The detector response is obviously larger, but still consistent with the high-energy tail of a single VEM (Figure 4.5). Accounting for both layers of scintillator and an expected 1.4 MeV energy loss in the separating steel plate, total energy deposit amounts to 6.2 MeV and must have been produced by a generating photon of at least 6.4 MeV.

Again, these values should be interpreted as minimums since the high likelihood of grazing angles suggests that electrons at these energies were scattered by even higher-energy photons. From Abbasi et al. (2018), a generating photon is expected to transfer about 1/3 of its energy to a Compton electron, raising the photon energies above to 7.2 and 18.6 MeV for the single-scintillator and penetrating cases respectively (note that the latter corresponds to the upper-end of spectra measured by RHESSI from Nemiroff et al. (1997) while the former is equal to the typical average energy in RREA models like Dwyer (2004)). Since we cannot make definitive statements about expected energies, we settle for the ironclad lower limits placing these photons well into the gamma regime, much closer to satellite observations than to the x-ray observations of Dwyer et al. (2004, 2005, 2012), which only reach up to a few hundred keV. Rather than alienating these observations from those two distinct cases, this may suggest that all of these types of events are in fact manifestations of the same phenomenon with differing characteristics. This idea is explored further in section 8.5.

### 7.2.3  Compton Scattering Effects

As discussed in chapter 3, the basic models attributed to TGF production are based on the strong ambient and local electric fields in thunderstorms which accelerate electrons into energies high enough to allow for avalanching, or the process in which particles gain energy faster than they lose it due to scattering interactions. In this way, the term 'terrestrial gamma-ray flash' is actually a misnomer, instead referring to the general electromagnetic showers produced by these runaway electrons. In fact, when subjected to sufficiently strong electric fields, EM showers can become dominated by electrons rather than photons (Skeltved et al. (2014), see subsection 3.2.4). Once a shower propagates outside of the ambient field, it develops into a standard EM shower dominated by photons primarily due to their longer interaction length. The more a particle interacts with the medium, the more it is delayed from the overall shower.

After the time-shifting analysis of subsection 7.2.1, TGF are shown to occur during periods of fast negative breakdown, and the signals persist until the FNB dies out and returns to normal leader behavior, after which TASD responses return to background levels and VHF power decreases (see Figure 7.20–Figure 7.23. We can safely say that the duration of signals detected at ground level accurately portray the duration of electron acceleration at the source after verifying the time delay due to EM shower interactions is negligible, as seen in Figure 7.24. In these simulations, 95% of particles seeded at 3 km above ground arrive at the detector plane within 60 ns, an order of magnitude less than the uncertainty of the timing analysis. The remaining 5% of particles are delayed due to atmospheric scattering, by which they lose energy and are therefore less likely to deposit energy in scintillators as seen in Figure 3.5. Additionally, lower energy particles are more likely to scatter again, further reducing their potential to deposit any detectable energy. This effect is also seen in simulations of Celestin & Pasko (2012) which were performed for upward TGFs.

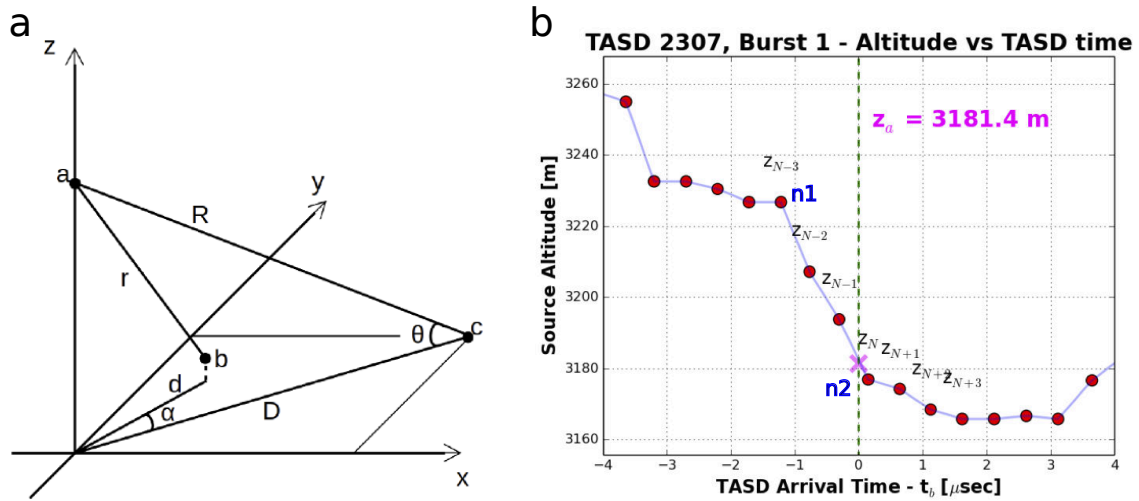

**Figure 7.18:** *(a):* Simplified coordinate system for temporal correlations. The TGF source is at point *a* with its plan *x, y* location serving as the coordinate origin. A TASD is located at point *b* and the INTF station is at the more distant point *c*. *(b):* INTF data in 0.5 $\mu$s time steps used in the stepping method for determining source altitude and time (TGF A in this case), showing the correlation of downward FNB leading up to the TGF onset (red 'x').

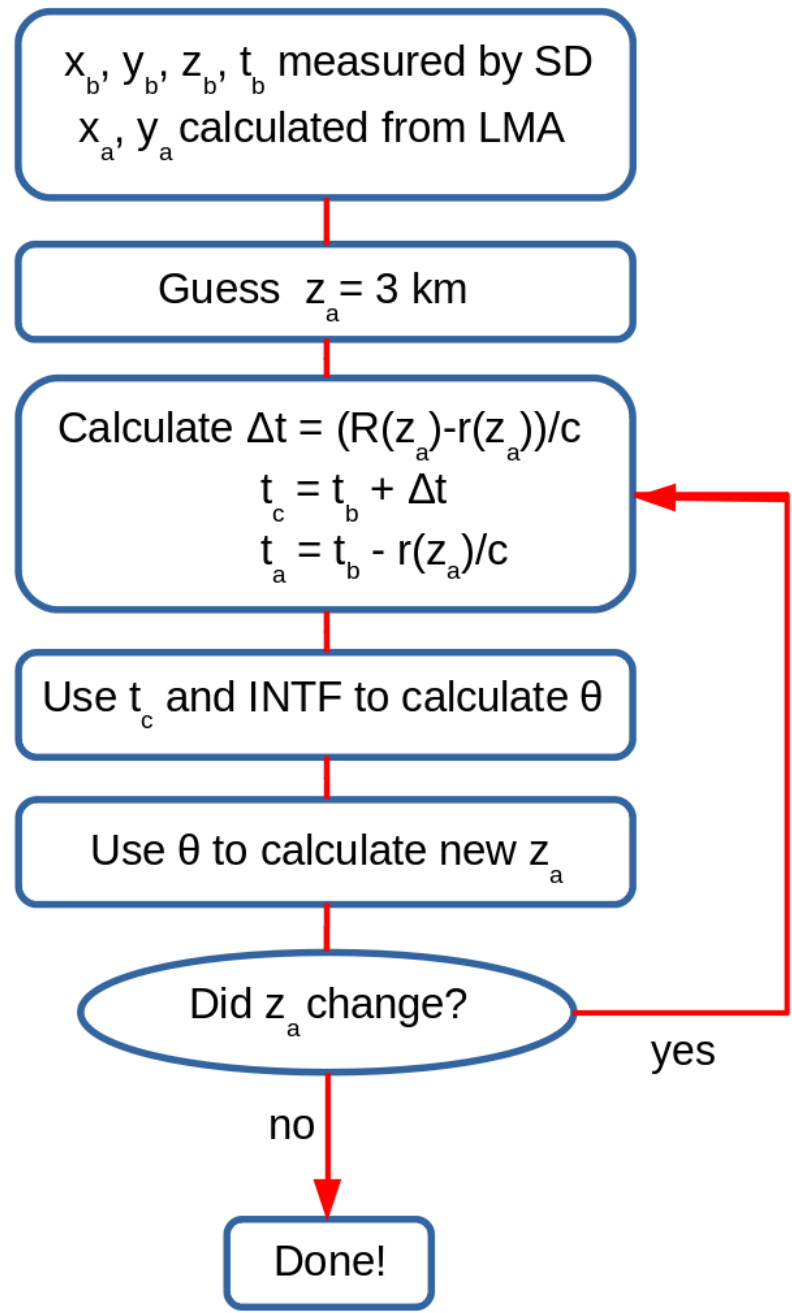

**Figure 7.19:** Iterative procedure for determining altitude ($z_a$) and onset time ($t_a$) of the TGF source. This method is performed individually for each participating TASD station and converges after only a few iterations for these events.

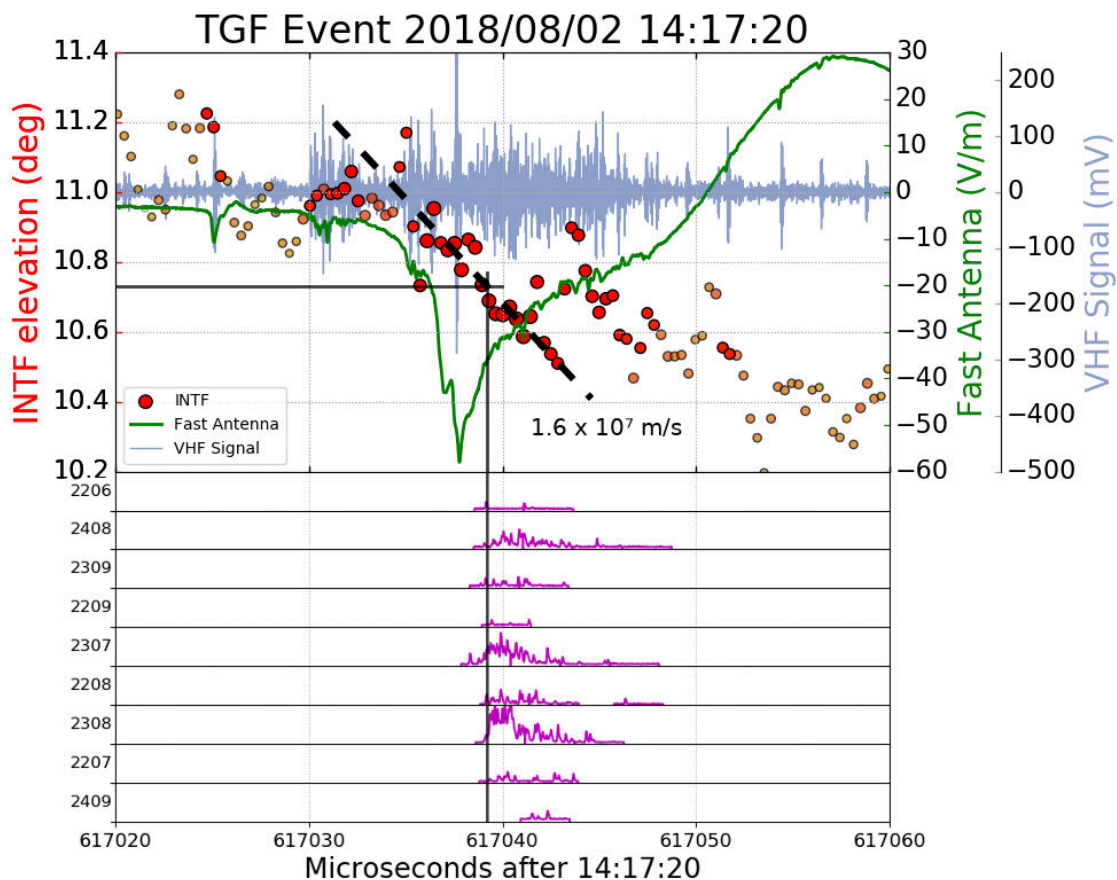

**Figure 7.20:** Enlarged view of the bottom panel of Figure 7.9 showing signals from each participating TASD. Black vertical and horizontal bars mark the solutions to the iterative method and point to the onset of TGF-producing activity in the INTF data. The light blue trace shows the raw VHF data recorded by the INTF, and circles represent radiation sources with color and size representing relative power. Purple traces in the lower axes are surface detector responses with station numbers XXYY identifying their easterly (XX) and northerly (YY) locations within the array in 1.2 km grid spacing units. The detection times are in good agreement with one another as well as with the median, highlighting the correlation between the TGF onset time and the flash's strongest IBP during a period of downward FNB.

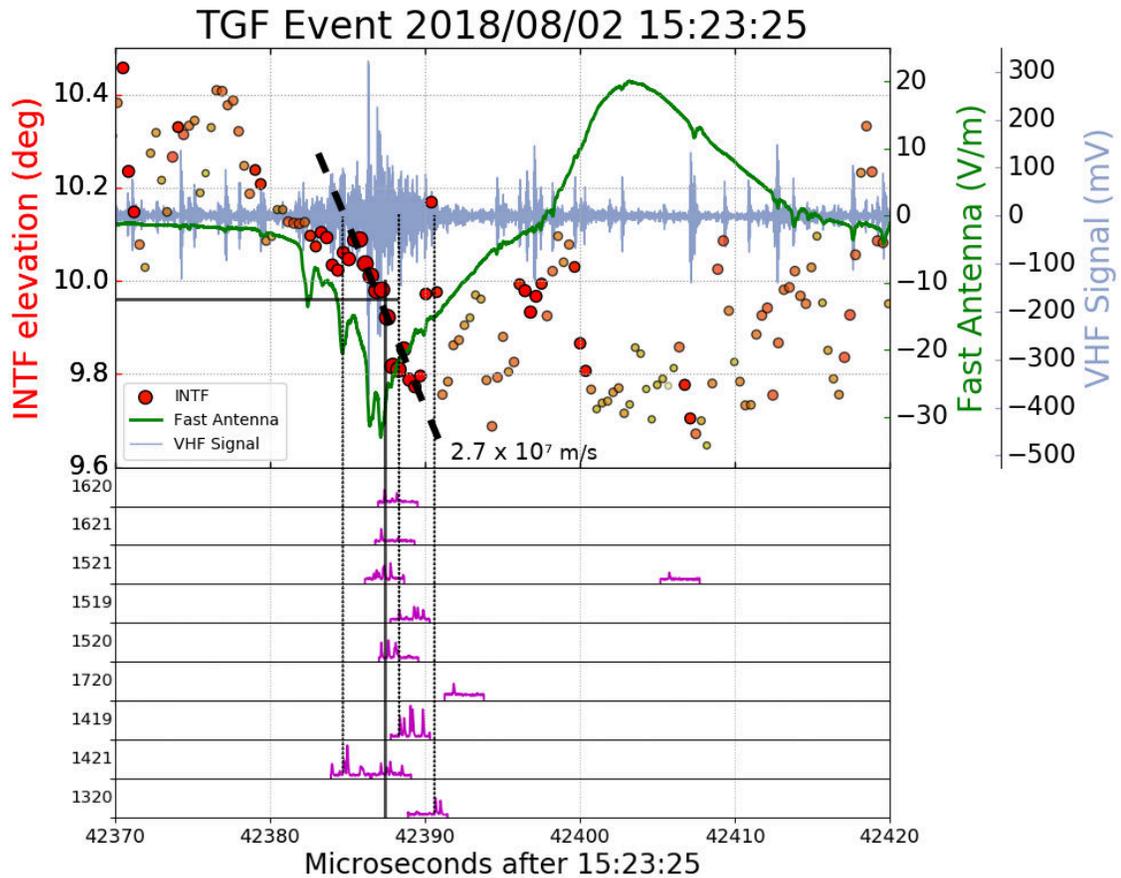

**Figure 7.21:** Same as Figure 7.20, except for the bottom panel of Figure 7.10. In contrast to the other TGFs, the TASD onset times are not all consistent with the median; TASD 1421 had a noticeably early onset time associated with the second strong sub-pulse, while the median onset was associated with the peak of the IBP and with a step-discontinuity in the downward FNB. Slightly delayed TASD signals suggest additional onsets as discussed in chapter 8 and are visualized by vertical dotted lines. The late detection by TASD 1521 is consistent with the energy deposit of a single stray muon, and may not be associated with lightning activity.

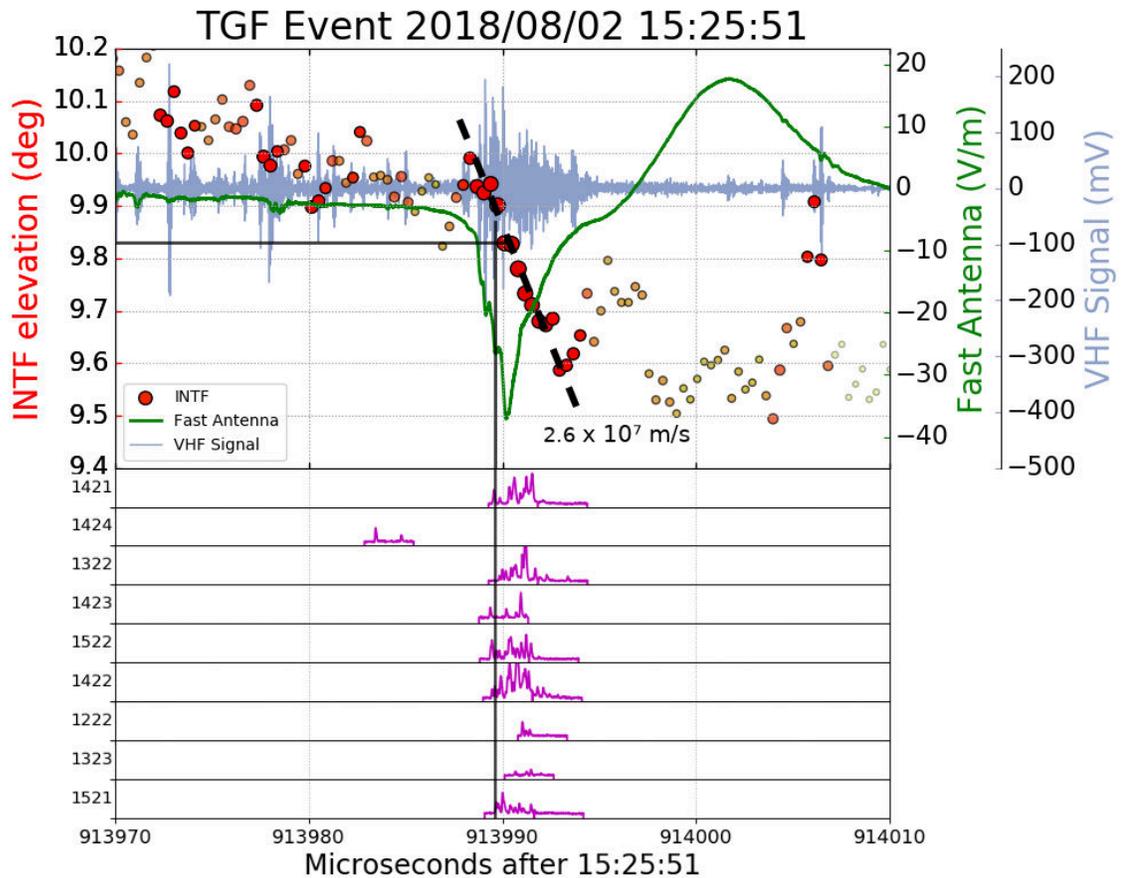

**Figure 7.22:** Same as Figure 7.20, except for the bottom panel of Figure 7.11. This event is an example of a classic, simple IBP. A brief period of upward FPB precedes the IBP and its associated downward FNB. The TASD onsets are mostly in good agreement with one another and with the median except for a single low-energy detection by TASD 1424. The median is also closely correlated with a sub-pulse and step-discontinuity in the VHF radiation development.

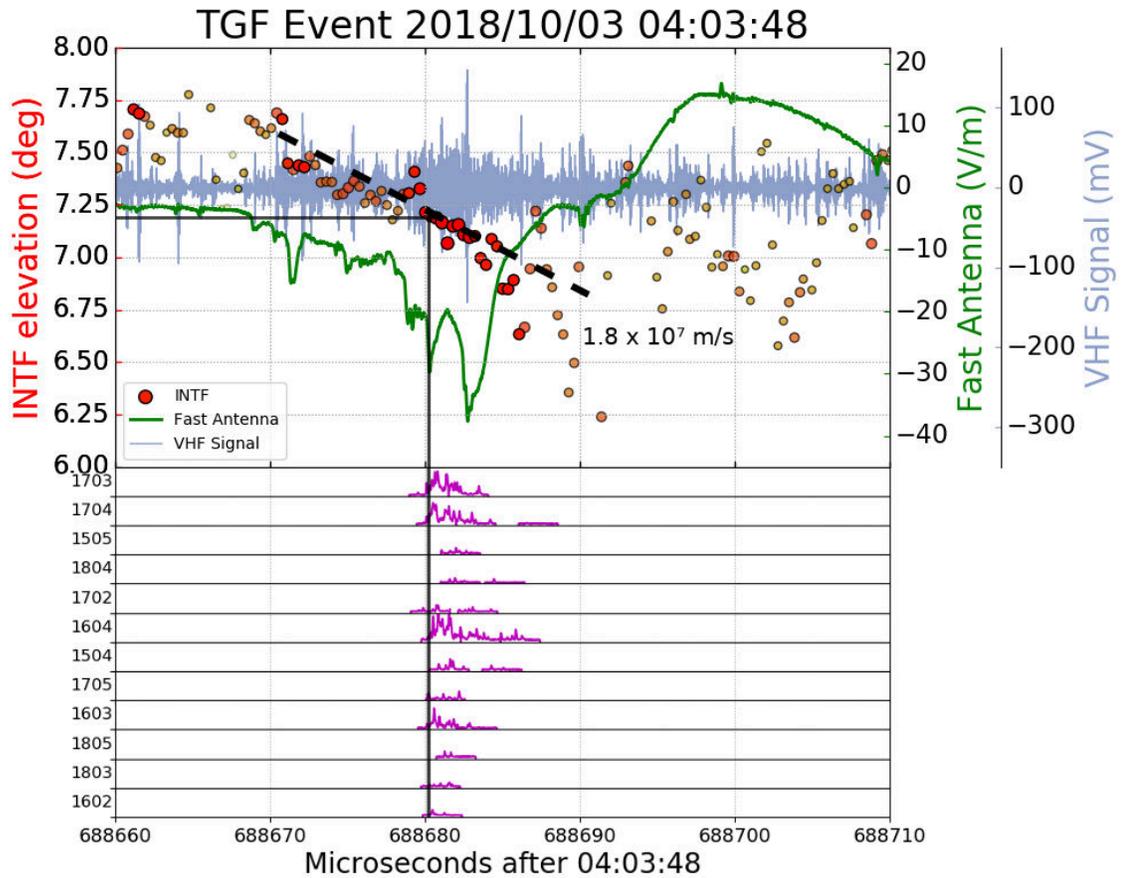

**Figure 7.23:** Same as Figure 7.20, except for the bottom panel of Figure 7.12. The sferic of this event is similar to that of TGF B, with slower build-up and multiple strong, embedded sub-pulses. The TASD onsets are closely correlated with one another as well as with a strong sub-pulse before the IBP's main peak. The associated downward FNB had a long duration and extent (Table 7.3) with a step-discontinuity occurring immediately before the TGF's onset.

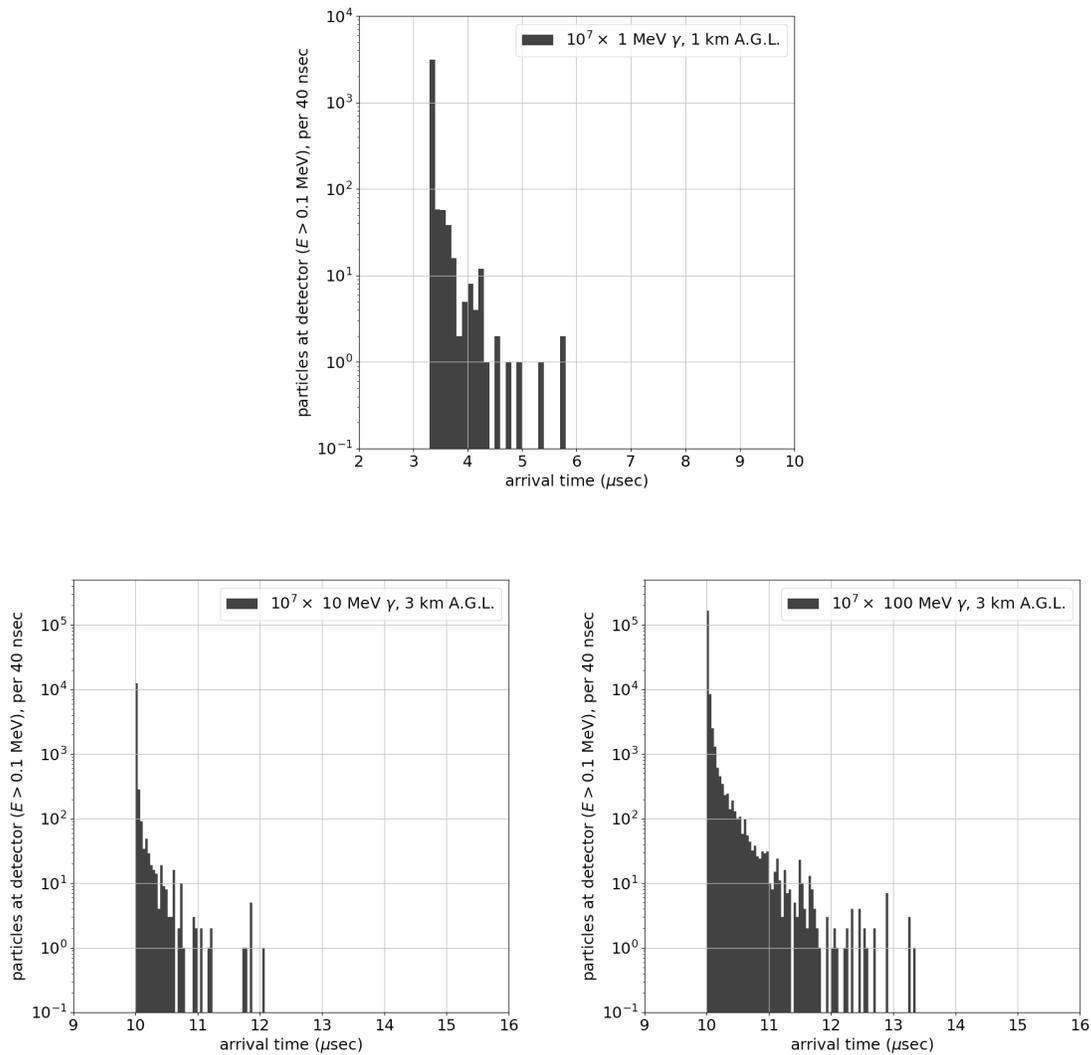

**Figure 7.24:** Simulations of atmospheric scattering in EM showers and the effect on arrival time. In the lower left (right) panel, a shower of $10^7$ photons having energy of 10 MeV (100 MeV) are generated downward from a height of 3 km, representative of the source altitudes at Telescope Array. In the upper panel, a shower is generated at only 1 km, since there was no resulting signal for 1 MeV photons from 3 km. In all cases, $\geq 95\%$ of particles reach the detector plane within 60 ns of the first. Note that these plots show particle count, not simulated energy deposit; any EM particle below a few MeV is unlikely to deposit energy above background levels. Because of this, any particles which arrive after the first bin have been delayed due to scattering, lost energy to atmospheric particles, and are further suppressed from depositing energy in a TASD. Therefore, signals produced by EM showers from 3 km altitude are not significantly delayed or extended due to atmospheric effects.

**Table 7.2:** Iterative analysis method values

Calculated values and associated uncertainties of the TGF source for each active TASD station. SD is the surface detector station number XXYY identifying their easterly (XX) and northerly (YY) locations within the array in 1.2 km spacing units. $\theta_c$ is the elevation angle corresponding to the altitude $z_a$ (given in km above the reference plane of 1.4 km). $t_c$ is the calculated time of TGF signal arrival at the INTF station and $\Delta t$ is the relative timing difference between the INTF and TASD signals ($t_c - t_b$). Values given in bold are the medians for that column and are most representative of the TGF source.

| Event | SD | $\theta_c$ (deg) | $z_a$ (km) | $t_c$ ($\mu$s) | $\Delta t_b$ ($\mu$s) |
|-------|------|------------------|----------------|--------------------------|------------------|
| TGF A | 2208 | **10.73 ± 0.03** | **3.21 ± 0.03** | **617,039.2 ± 0.7** | **45.2 ± 0.4** |
|       | 2206 | 10.73 ± 0.03 | 3.21 ± 0.03 | 617,039.2 ± 0.9 | 45.1 ± 0.7 |
|       | 2308 | 10.73 ± 0.03 | 3.21 ± 0.03 | 617,039.1 ± 0.6 | 45.9 ± 0.3 |
|       | 2209 | 10.73 ± 0.03 | 3.21 ± 0.03 | 617,039.3 ± 0.7 | 42.6 ± 0.3 |
|       | 2307 | 10.79 ± 0.03 | 3.23 ± 0.03 | 617,038.3 ± 0.6 | 46.9 ± 0.4 |
|       | 2409 | 10.71 ± 0.02 | 3.21 ± 0.03 | 617,041.1 ± 0.6 | 43.6 ± 0.1 |
|       | 2207 | 10.73 ± 0.03 | 3.21 ± 0.03 | 617,039.8 ± 0.8 | 45.9 ± 0.5 |
|       | 2408 | 10.76 ± 0.03 | 3.22 ± 0.03 | 617,038.7 ± 0.6 | 45.7 ± 0.2 |
|       | 2309 | 10.78 ± 0.03 | 3.23 ± 0.03 | 617,038.6 ± 0.7 | 43.5 ± 0.2 |
| TGF B | 1419 | 9.93 ± 0.03 | 2.91 ± 0.02 | 42,388.4 ± 0.4 | 45.0 ± 0.2 |
|       | 1421 | 10.06 ± 0.02 | 2.95 ± 0.01 | 42,384.0 ± 0.3 | 45.7 ± 0.2 |
|       | 1520 | **9.96 ± 0.03** | **2.92 ± 0.02** | 42,387.2 ± 0.3 | 44.35 ± 0.09 |
|       | 1320 | 9.87 ± 0.02 | 2.90 ± 0.02 | 42,391.0 ± 0.4 | 46.5 ± 0.3 |
|       | 1519 | 9.92 ± 0.02 | 2.91 ± 0.02 | 42,389.3 ± 0.4 | 43.8 ± 0.2 |
|       | 1620 | 9.96 ± 0.03 | 2.92 ± 0.02 | **42,387.4 ± 0.4** | 41.72 ± 0.05 |
|       | 1521 | 9.98 ± 0.02 | 2.93 ± 0.02 | 42,386.4 ± 0.3 | **44.1 ± 0.1** |
|       | 1621 | 9.96 ± 0.03 | 2.92 ± 0.02 | 42,387.2 ± 0.4 | 41.48 ± 0.07 |
|       | 1720 | 9.87 ± 0.02 | 2.89 ± 0.02 | 42,391.4 ± 0.4 | 38.58 ± 0.03 |

**Table 7.2:** (continued)

| Event | SD | $\theta_c$ (deg) | $z_a$ (km) | $t_c$ ($\mu$s) | $\Delta t_b$ ($\mu$s) |
|---|---|---|---|---|---|
| TGF C | 1424 | $9.94 \pm 0.01$ | $2.80 \pm 0.01$ | $913{,}983.5 \pm 0.3$ | $39.9 \pm 0.2$ |
|  | 1521 | $\mathbf{9.83 \pm 0.03}$ | $\mathbf{2.77 \pm 0.01}$ | $913{,}989.4 \pm 0.2$ | $43.70 \pm 0.05$ |
|  | 1522 | $9.83 \pm 0.03$ | $2.77 \pm 0.01$ | $913{,}989.4 \pm 0.2$ | $\mathbf{43.45 \pm 0.07}$ |
|  | 1421 | $9.83 \pm 0.03$ | $2.77 \pm 0.01$ | $\mathbf{913{,}989.6 \pm 0.2}$ | $44.7 \pm 0.1$ |
|  | 1423 | $9.85 \pm 0.02$ | $2.78 \pm 0.01$ | $913{,}988.9 \pm 0.3$ | $42.6 \pm 0.2$ |
|  | 1422 | $9.83 \pm 0.03$ | $2.77 \pm 0.01$ | $913{,}989.6 \pm 0.2$ | $44.4 \pm 0.1$ |
|  | 1323 | $9.77 \pm 0.03$ | $2.75 \pm 0.01$ | $913{,}991.3 \pm 0.3$ | $42.2 \pm 0.2$ |
|  | 1322 | $9.81 \pm 0.03$ | $2.76 \pm 0.01$ | $913{,}990.0 \pm 0.2$ | $43.9 \pm 0.2$ |
|  | 1222 | $9.78 \pm 0.03$ | $2.76 \pm 0.01$ | $913{,}991.0 \pm 0.3$ | $42.0 \pm 0.2$ |
| TGF D | 1704 | $7.20 \pm 0.02$ | $3.02 \pm 0.04$ | $688{,}680.1 \pm 1.4$ | $69.3 \pm 0.8$ |
|  | 1603 | $7.20 \pm 0.02$ | $3.02 \pm 0.04$ | $688{,}680.1 \pm 1.5$ | $68.7 \pm 1.1$ |
|  | 1705 | $\mathbf{7.19 \pm 0.02}$ | $\mathbf{3.02 \pm 0.04}$ | $\mathbf{688{,}680.2 \pm 1.4}$ | $67.3 \pm 0.6$ |
|  | 1702 | $7.21 \pm 0.02$ | $3.03 \pm 0.04$ | $688{,}679.9 \pm 1.7$ | $69.5 \pm 1.3$ |
|  | 1805 | $7.16 \pm 0.03$ | $3.01 \pm 0.04$ | $688{,}681.0 \pm 1.3$ | $67.3 \pm 0.3$ |
|  | 1703 | $7.20 \pm 0.02$ | $3.02 \pm 0.04$ | $688{,}680.1 \pm 1.4$ | $70.2 \pm 1.0$ |
|  | 1803 | $7.17 \pm 0.02$ | $3.01 \pm 0.04$ | $688{,}681.0 \pm 1.3$ | $70.2 \pm 0.8$ |
|  | 1504 | $7.17 \pm 0.02$ | $3.01 \pm 0.04$ | $688{,}680.9 \pm 1.6$ | $65.7 \pm 1.1$ |
|  | 1804 | $7.14 \pm 0.03$ | $3.00 \pm 0.04$ | $688{,}681.7 \pm 1.3$ | $69.4 \pm 0.5$ |
|  | 1602 | $7.19 \pm 0.02$ | $3.02 \pm 0.04$ | $688{,}680.1 \pm 1.7$ | $\mathbf{68.1 \pm 1.4}$ |
|  | 1604 | $7.21 \pm 0.02$ | $3.03 \pm 0.04$ | $688{,}679.9 \pm 1.5$ | $67.9 \pm 1.0$ |
|  | 1505 | $7.16 \pm 0.03$ | $3.01 \pm 0.04$ | $688{,}681.1 \pm 1.6$ | $64.1 \pm 1.1$ |

**Table 7.3:** TGF source and breakdown characteristics

Values of the TGF source calculated by the iterative analysis alongside the extents and durations of FNB associated with the most energetic event of each TGF. $D$ is the horizontal distance between the event and the INTF station. $z_a$ ($t_a$) is the TGF's median altitude (onset time) of the iterative analysis performed on each active TASD (all values given in Table 7.2). Uncertainty in these values is included as calculated by the error analysis of section 7.3. $z_{FNB}$ and $t_{FNB}$ are the observed vertical extents and durations of downward FNB associated with each respective TGF, with the average speed given as $v_{FNB}$. No uncertainties are given for the final three columns as they are simply approximated based on VHF source development (e.g. Figure 7.20–Figure 7.23).

| Event | D (km) | $z_a$ (km) | $t_a$ ($\mu$s) | $z_{FNB}$ (m) | $t_{FNB}$ ($\mu$s) | $v_{FNB}$ (m/s) |
|---|---|---|---|---|---|---|
| TGF A | 16.96 ± 0.15 | 3.21 ± 0.03 | 616,981.7 ± 0.6 | 150 | 10.0 | $1.5 \times 10^7$ |
| TGF B | 16.64 ± 0.08 | 2.92 ± 0.02 | 42,331.7 ± 0.3 | 100 | 3.7 | $2.7 \times 10^7$ |
| TGF C | 15.98 ± 0.04 | 2.77 ± 0.01 | 913,935.1 ± 0.2 | 120 | 4.7 | $2.6 \times 10^7$ |
| TGF D | 23.9 ± 0.3 | 3.02 ± 0.04 | 688,600.1 ± 1.4 | 240 | 13.4 | $1.8 \times 10^7$ |

## 7.3   Measurement Uncertainties

Until the installation of the INTF, uncertainties for this study were simply reported as each detector's own systematic error, since the independent data types were reported separately. In the burst event set, for example, the independent NLDN flash times were compared to TASD trigger times looking for temporal coincidences. TASD data is constantly monitored and well understood with a trigger time uncertainty of $\pm 20$ ns corresponding to the bin size of energy deposit data (section 4.2). NLDN timing uncertainties are typically known within $<700$ ns depending on range (Cummins et al. (2010)). Lightning flashes last up to hundreds of milliseconds, however, meaning that NLDN data points are enough to identify a lightning flash. At the time, we didn't know the relationship between downward TGFs' production within the flash itself, and therefore without a more robust lightning detection method, the 1 ms search window for 'synchronized' events was sufficient to correlate downward TGFs with their parent flashes. Analysis of the burst data set did not include NLDN locations as part of the search criteria, yet all seven events with synchronized lightning were recorded within a few hundred meters of their respective flashes. For reference, NLDN flash locations at the Telescope Array are known to within $\sim 300$ m (Nag et al. (2011)).

After the INTF installation, however, analysis methods changed drastically (section 7.2). The uncertainty in reported source time and spatial coordinates now depends on the interaction between participating instruments. For this analysis method, INTF data requires LMA data for calibration, and its systematic error is therefore tied to it (section 5.3). Typical angular uncertainty in elevation for a single INTF point is $\sim 0.1°$ which results in a standard error in the mean of $0.02°$–$0.03°$ for these TGFs depending on how many points lie in the $\pm 4$ $\mu$s window (Table 7.2). This applies to uncertainty in azimuth measurements as well, which are used in calibration of the interferometer, but are not directly included in the geometry correction methods.

These errors are then carried through the complete calculations of section 7.2 using the general form of the equation for standard error propagation:

$$\delta f = \sqrt{\left(\frac{\partial f}{\partial x_1}\delta x_1\right)^2 + ... + \left(\frac{\partial f}{\partial x_n}\delta x_n\right)^2} \tag{7.5}$$

where $f = f(x_1, ..., x_n)$. All detector locations are known on the order of meters and have negligible contributions. Similarly, individual INTF point and TASD trigger times are known on the order of the sampling rate (10s of ns for both). These are all taken into account, but have very little effect on final uncertainties. Primary error sources, then, come from the two instances of taking averages — the TGF plan location is taken as the mean GPS location of LMA sources within 1 ms of particle detections, and its uncertainty is the standard error:

$$\sigma_{\bar{x}} = \frac{\sigma}{\sqrt{n}} \tag{7.6}$$

where $\sigma$ is the standard deviation and $n$ is the number of data points. TGF source elevations are done the same way — a mean is taken of all INTF elevation points within 4 $\mu$s of the TGFs inferred arrival at the interferometer (from Equation 7.2), and its uncertainty is taken as the standard error. In TGF A for example, standard error of the average LMA latitude and longitude are 0.00143° and 0.00213°, respectively, corresponding to a radial distance of ∼235 m. Standard error for the INTF elevation data of the same TGF is 0.03°, corresponding to an altitude uncertainty of ∼30 m at a distance of ∼17 km.

All subsequent calculations of section 7.2 can then be shadowed by their error counterparts using Equation 7.5. Since the iteration process is repeated for each participating TASD station, these errors are calculated individually and shown in Table 7.2 before a median is taken to best represent the TGF source. Typically, altitude measurements are much less precise than plan locations for TGF studies (see Abbasi et al. (2018) and the examples of section 3.1), but here the altitude determination comes from the higher-sampled INTF data whereas plan location data is supplied by only a few sparse LMA points. As a result, altitude uncertainties are as low as 10–40 m for these events, compared to horizontal location errors of 40–300 m. Source onset times are known as well as 0.2–1.4$\mu$s. TGF D has the highest error for all of these measurements.

Notice that the range of altitude errors is relatively small, but poor grouping of LMA data results in a much larger uncertainty in the plan location and source time. For example, TFG D has the poorest statistics in LMA data (visualized in panel e of Figure 7.16) — as these errors are propagated through each calculation, values for TGF D are consistently the highest in the given range for each measurement. This shows that low LMA sampling rate and possible mislocations during fast breakdown (described in section 5.1) are the main contributors to the reported errors. The error contributions of the four variables for TASD 2208 of TGF A are given in Table 7.4. The effect of 1$\sigma$ variations on solutions $z_a$ and $t_a$ are shown for each variable. As expected, LMA locations ($x_a$, $y_a$) are the main sources of error, which is typical of all events to varying degrees.

**Table 7.4:** Measurement errors for TASD 2208

The four measurements which define the four TGF source coordinates in the iterative method of subsection 7.2.1. These values are calculated for TGF A using the trigger time of TASD 2208 ($t_b$). The uncertainty in this trigger time is conservatively defined as two FADC bins, or 40 ns. LMA position ($x_a, y_a$) is given relative to the INTF station, and its errors are simply the errors in the mean of LMA points within 1 ms of the trigger time. The same is true of the elevation ($\theta$), except for all INTF points within 4 $\mu$s. Note that there is no change in $z_a$ if $t_b$ is shifted by 40 ns — this is because all of the same INTF points are still within the 4 $\mu$s window. Column 2 (3) indicates the value (error) used in the analysis. Column 4 (5) indicates the absolute change in $z_a$ ($t_a$) corresponding to a change of 1$\sigma$, alongside the percentage of that change relative to the total reported error from Table 7.3. In the first row, for example, changing $x_a$ by 0.19 km results in a 0.01 km change in the resulting altitude, or 33% of the reported error of 0.03 km (note that these percentages are not intended to add up to 100%). LMA plan location measurements are clearly the largest source of error in this analysis.

| Measurement | Value | Error | 1$\sigma$ effect on $z_a$ | 1$\sigma$ effect on $t_a$ |
|---|---|---|---|---|
| $x_a$ (km) | -4.6 | 0.187 | 0.010 km (33%) | 0.22 $\mu$s (37%) |
| $y_a$ (km) | -16.3 | 0.143 | 0.026 km (87%) | 0.24 $\mu$s (39%) |
| $t_b$ ($\mu$s) | 616,987.25 | 0.04 | 0 km (0%) | 0.04 $\mu$s (6.7%) |
| $\theta$ (deg) | 10.73 | 0.026 | 0.002 km (6.7%) | 0.01 $\mu$s (1.6%) |

# CHAPTER 8

# INTERPRETATION

This chapter shall discuss the interpretation of analyses of the 'INTF' data set of chapter 7 and how they affect our understanding of lightning in general. It begins by closely comparing the analysis results of each event to determine what lightning activity and characteristics are associated with TGF production. This is followed by further examination of data from each detector and what they can tell us about lightning processes and TGF shower characteristics. Measurements are then compared to those of other relevant studies, both historical and contemporary. This is followed by a reflection on the current understanding of TGF development and lightning initiation in general and how this study fits in.

## 8.1 Temporal Correlations

The results of the time-shifting analysis of subsection 7.2.1 form the basis of interpreting these events as they clearly show the relationship between ground-level TGF detections and lightning processes identified by the interferometer, sferic sensors, and other detectors. As seen in Figure 7.9–Figure 7.12, each TGF is clearly associated with the strongest IBPs in the initial stages of downward, negative-polarity lightning breakdown. For the events in which there were multiple gamma bursts (TGFs A, C, and D), the strongest TASD signal was always associated with the strongest IBP, with weaker bursts occurring before or after the main peak during weaker IBPs. The sample size is small, but the strong sferic pulses are clearly related to the production of TGFs. Also note that these strongest IBPs from each flash correspond to the transition from linear leader development to branching (best seen in Figure 7.5–Figure 7.8), perhaps indicating a 'maturing' of the leader development and allowing for TGF generation (see section 8.5).

Further, each TGF-associated IBP sferic is accompanied by isolated periods of high-power, rapid VHF source development (larger red points in Figure 7.20–Figure 7.23. These short intervals are embedded in the larger-scale leader development and are always associated with sferic pulses, further suggesting they are distinct from later leader propagation.

Tilles et al. (2019) recently identified this process as fast negative breakdown (subsection 2.4.2), the negative-polarity analog of fast positive breakdown (FPB) that is considered to be the cause of narrow bipolar events (NBEs) at the beginning of some lightning flashes (Rison et al. (2016)). The parent flash of TGF A was the only of this dataset to be initiated by an NBE, seen in Figure 7.17. INTF data for this event exhibits the same characteristics as IBP-related FNB except for propagating upward. The negative sferic pulse coupled with upward VHF propagation indicates its positive polarity, followed by weaker, downward propagation during a positive overshoot in the sferic pulse. The same activity (with opposite direction) is seen during the later IBPs, in which the strong downward FNB is associated with a negative sferic pulse followed by a large positive overshoot. Both polarities of fast breakdown propagate at speeds on the order of $10^7$ m/s, much faster than the average leader speeds of $\sim$$10^5$–$10^6$ m/s (Berger (1967); Chen et al. (1999)). In addition, modeling by Attanasio et al. (2019) shows that both polarities amplify the ambient electric field ahead of the advancing breakdown, thereby accelerating electrons into the regime of TGF production.

A common characteristic of "classic" IBPs is the presence of sharp, leading-edge sub-pulses (Weidman & Krider (1979); Nag et al. (2009); Karunarathne et al. (2014)), which are seen in all of Figure 7.20–Figure 7.23, but most prominent in TGFs B and D which have slower development. The subtlety of sub-pulses in TGFs A and C is likely due to their fast development times of <4 $\mu$s. Unlike FNB, sub-pulses are not always seen during the determined TGF onset times. In TGF A, for example, the IBP's sharp leading edge has at least one prominent sub-pulse, but the TGF is not initiated until after the main peak during a step-discontinuity in the VHF development. Higher-resolution processing of the INTF data revealed that the discontinuity was in fact due to a short period of faster development in which the streamer system propagated at $\sim$$3\times10^7$ m/s, nearly twice as fast as the overall FNB speed. This is seen in panel b of Figure 7.18, where the vertical dashed line represents the TGF's onset time as determined by the stepping method of section 7.2, placing the TGF onset at the end of this enhanced-speed interval. The later gamma bursts of TGF A were also associated with periods of FNB, but during weaker and noisier IBPs (Figure 7.13).

TGF B was unique in its poor temporal alignment of individual TASD responses. In Figure 7.21, the solid vertical line represents the median TGF onset as determined by the iterative method, which is again associated with a step discontinuity during the overall FNB development. In this case, the onset is also consistent with the main IBP peak. Additional vertical dashed lines mark the onset times of individual TASDs which were significantly separated from the median and from one another. Of course, these offsets could simply be a result of the stochastic nature of detections in stations with 1.2 km separation, but the delays of 1–3 $\mu$s between the dashed lines are quantitatively significant given the individual uncertainties in $t_c$ of 0.3–0.4 $\mu$s (Table 7.2). Since the initiating processes of TGFs remain unconfirmed, we can only postulate about what causes this behavior which is absent from the other events. TGF-producing activity could be occurring separately several times during the FNB, or perhaps a singular process that continues for a longer period during the sharp direction change in leader development at the time of the TGF as seen in Figure 7.6. The issue of TGF production is discussed in more detail in section 8.5. TASD 1421's early onset time was not shared with any participating detectors, but is directly correlated with a strong sub-pulse in the IBP's leading edge along with a short interval of upward VHF development. Similar upward breakdown is apparent in all TGFs from this set during an increase in VHF power (cyan waveform) and preceding or initiating the periods of FNB. TGF B, however, is the only event to record TASD activity during this upward development.

TGF C's second gamma burst and associated IBP represents a more classic example while still exhibiting many of the same characteristics as other events (Figure 7.22). The slower leader propagation transitions into fast negative breakdown beginning with a short interval of upward VHF development. Even though the FNB and associated IBP are relatively fast at $2.6 \times 10^7$ m/s, subpulses are visible along its leading edge. The last of these sub-pulses before the IBP peak is closely aligned with the median TGF onset as determined by both the iterative and stepping methods (panel d of Figure 7.15), as well as with a step-discontinuity in the VHF progression. In this case, the discontinuity was not much faster than the already quickly-propagating FNB. An early detection by TASD 1424 6 $\mu$s before the IBP appears to be unrelated to the following activity, and there is not enough information to determine its source. At least one gamma burst was also recorded 220 $\mu$s earlier during weaker IBPs as seen in panels c and f of Figure 7.15. The first of these was produced just after the IBP's sharp peak during an episode of slower FNB and even a step discontinuity in the VHF data. Detector 1421 recorded another short pulse during a similar IBP $\sim$17 $\mu$s later. Since there is no associated high-power FNB during the second pulse and there was only a small amount of energy deposit in a single detector ($<$5 VEM), it is unclear if this event was produced by lightning activity, despite its close coincidence with an IBP.

TGF D was composed of three gamma bursts detected during a negative intracloud flash in early October of the same year. The two weaker, earlier bursts were correlated with concurrent weak IBPs as seen in panels c and f of Figure 7.16. As with TGF C, the last burst was much stronger and was associated with the flash's strongest IBP (panels d and g). This IBP was unique from the other events in its long duration ($>12$ $\mu$s) and complex series of sub-pulses. Its VHF progression was of similar duration and very linear compared to the breakdown during other gamma bursts. The VHF sources had high power leading up to the IBP's peak, but gamma activity was not produced until after a short period of upward propagation and subsequent step-discontinuity before returning to its linear trend as seen in the detailed view of Figure 7.23. The individual TASD onset times were aligned very well with one another after the time-shifting analysis of section 7.2, suggesting a singular TGF source onset. The resulting median time coincided with a very strong sub-pulse in the IBP's leading edge, as well as with the end of the step-discontinuity during FNB development. Similar correlations were seen in the earlier gamma bursts, though with less energy deposit at ground level, weaker IBPs, and less well-defined FNB.

## 8.2    Shower Characteristics

In addition to lightning and TGF production, these observations can also tell us much about the resulting air shower attributes. For example, a common technique of cosmic ray analysis is to examine the lateral energy distribution, or the amount of energy deposit in a given TASD versus the distance of that station from the shower axis (the vector pointing from the shower source to the center of energy deposit at ground level). Figure 8.1 shows the lateral energy distribution for the first trigger of TGF A, in which 90% of energy deposit is contained within ~2.2 km of the core. With a source height of 3.2 km, this equates to a 35° half-opening angle from the shower axis. Figure 7.5–Figure 7.8 gives 2-dimensional visualizations of these angles for each TGF from the interferometer's point of view (finely dotted lines). These measurements are hard to compare with upward TGF observations which cannot detect an entire footprint, but simulations of Carlson et al. (2007); Mailyan et al. (2019) have used comparable estimates of 30°–45°.

TGF tilt angle (relative to the local zenith) is similarly challenging to measure via satellite and does not give the full picture. Again, because satellites are only sampling a single point within upward TGFs, it is difficult to determine at face value whether the sampling location is toward the shower's axis or its perimeter, though analysis and simulation by Xu et al. (2019) found that tilted geometries of 15–45° had a significant effect on the temporal profile of TGFs observed by FERMI's GBM. At Telescope Array, tilt angle is a straightforward measurement after determining the TGF source location, from which a line can be drawn along the shower's axis to its core location at ground level. In addition, using VHF source locations from the broadband interferometer, the axis of linear fast negative breakdown during each TGF can be extended from its source to ground level (dashed lines of Figure 7.5–Figure 7.8).

In terms of the showers' sources, thousands of TGFs and various related phenomena have been recorded and sorted according to their characteristics, but Tran et al. (2015) was the first to define the following criteria of terrestrial gamma-ray flashes: "(a) no sign of pile-up, characteristic of x-rays associated with leaders near ground, is seen in the recorded pulses, (b) the duration of the recorded pulse sequence is less than 1 ms, and (c) energy values for the largest pulses corresponding to individual photons exceed 1 MeV." We have certainly shown that TGF events at Telescope Array pass the latter two of these criteria, but not yet the first, i.e. that the signals are not overwhelmed by high numbers of low-energy particles masquerading as singular high-energy gamma-rays.

While detectors with low energy deposit are best for distinguishing individual particles as discussed in subsection 7.2.2, the others reveal important information about large-scale shower composition. Surface detector simulations show that individual photons below $\sim$1 MeV (and their Compton scattered electrons) deposit negligible amounts of energy (Figure 3.5). When enough of these particles arrive together however, the many small deposits can sum up to significant energy. Unfortunately, the TASD scintillators cannot quantitatively provide a spectrum of all detected particles, but the qualitative signal shapes suggest that this is not the case for the downward TGFs of this study. The right-hand panels of Figure 7.1–Figure 7.4 all lack the smooth profile that would be expected from a dominating amount of low-energy x-rays, even in the detectors which recorded the most energy deposit. In fact, none of the active detectors during this or previous data sets exhibit this behavior. Of course, most of these overlapping peaks are much larger than the expected energy deposit of any single Compton-scattered electrons, but their sharp, impulsive nature implies the showers consist of a countable number of overlapping, high-energy gamma-ray photons. The upper and lower-right panels of Figure 7.3 clearly show both low and high-energy cases. In the former, signals from three individual particles can be easily defined, while the latter shows the overlapping peaks expected from a gamma-ray-dominated shower depositing six times as much energy. By this interpretation, the detections at Telescope Array pass all of the criteria above and meet their definition of TGFs.

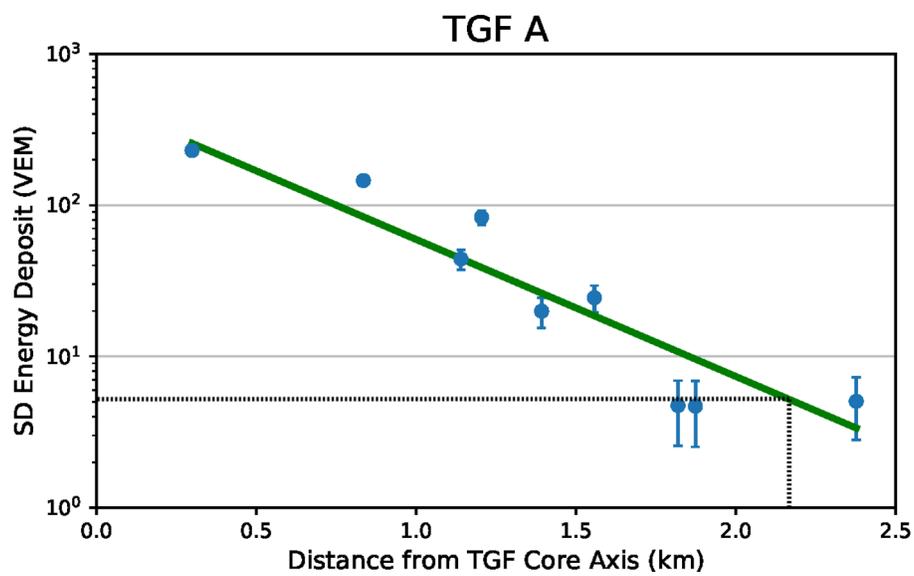

**Figure 8.1:** Lateral energy distribution for the first trigger of TGF A. Energy deposit is participating TASDs is given as a function of distance from the shower axis. Dotted lines mark the point in which 90% of energy deposit is contained. For TGF A, this is 2.2 km at ground level, or within 35° of the shower axis.

## 8.3   Comparison to other studies

The study of energetic radiation from lightning has grown significantly since its discovery in 1994. The most energetic of these phenomena are categorized as TGFs, but that does not mean they all share the same characteristics. For example, we have shown that the events of Telescope Array pass the TGF criteria of Tran et al. (2015), but they still vary in many ways from other observations. The same can be said of TGFs detected by different satellites; RHESSI has recorded photons of energy from 100 keV up to ∼20 MeV (Gjesteland et al. (2012); Grefenstette et al. (2009); Smith et al. (2005)), Fermi's GBM up to ∼30 MeV, and AGILE typically up to ≃40 MeV, though gamma-rays as high as 100 MeV were detected in a few cases (Marisaldi et al. (2010); Tavani et al. (2011)). Typically, however, measurements from all of these instruments are well-fit by a power law or RREA spectrum with average energy of ∼7.2 MeV (Dwyer & Smith (2005); Berge & Celestin (2019)).

Ground observations of TGFs are relatively limited, primarily consisting of the events by TETRA (Ringuette et al. (2013); Pleshinger et al. (2019)). Since this study is focused on the initiation mechanisms of natural lightning, in-depth comparisons to rocket- and laser-triggered lightning are generally avoided. The observations by TETRA and TETRA-II report a collection of downward TGFs with photon energies between 50 keV–6 MeV. Additional detections by Tran et al. (2015); Wada et al. (2019) recorded TGFs with photon energies ranging up to 13–20 MeV.

While the Telescope array cannot give upper energy limits, subsection 7.2.2 gives the lower limits of distinct particles detected by Telescope Array. Again, keep in mind that the values of 2.6 and 6.4 MeV are (confusingly) lower limits on the maximum photon energies measured by surface detectors. The energies are more likely to be higher due to the high likelihood of glancing collisions, but either way these values are consistent with TGFs recorded by other projects. One phenomenon that is distinct in this regard is x-ray emission, consisting of bursts of photons with energies up to several hundred keV, though signal pile-up made this determination difficult in some cases (Moore et al. (2001); Dwyer et al. (2004, 2005)). These observations all took place many years before the study by Tran et al. (2015), but plainly fail all of its TGF criteria and were categorized as a separate class of events.

The study of Dwyer et al. (2004) concerned emission from rocket triggered lightning, while those of Moore et al. (2001); Dwyer et al. (2005) examined x-ray observations from nearby natural cloud-to-ground flashes. These were each detected 1–2 ms before the first return stroke in natural and rocket-triggered lightning flashes, sometimes continuing for a short time before dying out. This alone is enough to distinguish them from traditional upward TGFs, which are have been understood to occur in the initial 1–3 milliseconds after lightning initiation (Stanley et al. (2006); Cummer et al. (2015); Lyu et al. (2016)). This same early production is seen clearly in all Telescope Array events with available lightning data. That said, ground detections by Tran et al. (2015); Wada et al. (2019) were categorized as TGFs despite occurring after the flashes' respective return strokes. The similar observation of Dwyer et al. (2012) was more conservatively identified as 'TGF-like,' matching the energies and durations of traditional TGFs, but occurring 191 $\mu$s after its associated return stroke.

Even though the TA events of this study have been identified as TGFs, their durations and burst-like behavior are more closely related to that of x-ray emissions. From chapter 6, The original defining characteristic of TASD burst events was their grouping of 3 or more SD triggers within 1 ms. In subsequent studies, this requirement has been removed in lieu of more accurate identification via lightning instrumentation, e.g. the single gamma burst of TGF B. Regardless, the events typically consist of multiple, distinct bursts of gamma-rays (Figure 6.1 and Figure 8.2), consistent with individual leader steps in early, negative-polarity breakdown. In these examples, the individual bursts typically last less than 10 $\mu$s, are separated by 10–140 $\mu$s, and can last for hundreds of microseconds depending on how many total bursts occur in succession. The x-ray observations of Dwyer et al. (2005) share similar attributes, though on slightly shorter time scales, consisting of distinct bursts of x-ray detections individually lasting only a couple microseconds, separated by 2–20 $\mu$s, and with overall durations of up to a few hundred microseconds. They were also tied to sequential leader steps, but in the final stages leading up to the return stroke rather than the early stepping during lightning initiation. The detectors involved in Moore et al. (2001)'s study of similar events have poorer time resolution, but the gaps between large spikes in the scintillator responses indicate the same burst-like nature of both x-ray emissions and of TGFs detected at Telescope Array. These events lasted multiple milliseconds overall, an order of magnitude longer than the others, perhaps due to the resolution and/or sensitivity to lower-energy photons.

In contrast, satellite observations have reported TGF durations of hundreds of microseconds. Early observations by BATSE even revealed TGFs lasting up to a millisecond or more (Fishman et al. (1994)), but further studies found that these were significantly affected by dead time losses and that the source durations were closer to $\sim$250 $\mu$s (Grefenstette et al. (2008); Gjesteland et al. (2010)). Fermi GBM reported a wide range of 50–1,000 $\mu$s, but with median durations of 240 $\mu$s (Fishman et al. (2011)), though in-depth simulations of TGF geometries by Mailyan et al. (2019) conclude that these events can be fully explained by average source durations closer to $\sim$200 $\mu$s. More recent observations by ASIM show a similar minimum of $\sim$40 $\mu$s, and ranging up to 500 $\mu$s, but with shorter typical durations of $\sim$100 $\mu$s (Østgaard et al. (2019)). Note that as instrumentation improves, TGFs are reported as having shorter durations. However the difference is still significant, and if both upward and downward TGFs indeed share the same production mechanism, there must be some other factors at play — this relationship is explored in section 8.4.

Another defining characteristic of TGFs is not a direct measurement, but rather the model-dependent metric of fluence, or the number of produced photons. This comes in large part from the modeling study of Celestin et al. (2015), in which electron acceleration is calculated during the corona flash stage during negative leader steps. Acceleration is driven by the local, enhanced electric field, which is in turn a result of the potential difference between the leader tip and the space ahead (subsection 3.2.3). From the available potential difference, there is a hard cutoff of maximum energy in the accelerated electrons and their produced bremsstrahlung photons. These results give a strong connection between detected maximum photon energy and available potential difference. Additionally, the total number of electrons and photons (fluence) is recorded from the Monte Carlo simulations and provided alongside corresponding values of potential difference. This study is incredibly useful in correlating measured energies with fluence and potential difference values, which are both difficult to measure directly, if not impossible. Since the simulated particle spectra have hard maximums, the results allow experiments to estimate the available source potential based on the highest-energy photon detected by instrumentation. The typical upper-limit photon energies from satellite-detected TGFs of 20–40 MeV imply potential differences of 160–300 MV, with the extreme cases of AGILE requiring even higher values. These correspond to total fluences in the range of $10^{16}$–$10^{18}$ photons.

Simplified GEANT4 simulations of RREA-initiated showers from Abbasi et al. (2018) suggest fluences on the order of $10^{12}$–$10^{14}$ photons based on measured energy deposit (Figure 6.8), several orders of magnitude lower than the satellite-inferred values, and corresponding to potential drops of 10–60 MV. Lower-limit photon energy estimates of 2.2 MeV from the same study would correspond to a minimum potential drop of only 10 MV and fluence of $10^{12}$ photons, consistent with the lower-end of GEANT4 simulations. The closer analysis of 'INTF' events from 2018, however, show an individual detection produced by a photon with energy of at least 6.4 MeV, giving a minimum potential drop of $\sim$50 MV, with the likely grazing-angle case suggesting values up to ∼150 MV. These then correspond to fluences of $10^{14}$–$10^{16}$ photons, respectively. Again, we settle for the ironclad lower-limits, but note that these minimums are only consistent with the very highest implied fluence values of Abbasi et al. (2018). Therefore, this discrepancy in fluence between satellite and ground detections remains unexplained alongside that of duration, though some hypotheses are discussed in section 8.5.

In summary, The TGFs at Telescope Array are reasonably consistent with satellite observations in terms of individual photon energies, but differ in both duration and implied fluence. Conversely, their durations and burst-like nature are similar to x-ray emissions, but consist of higher-energy photons produced at earlier stages in the leader stepping process. Rather than isolate these classes of events, the comparisons above may indicate a continuum of energetic radiation from lightning, in which the distinct types are merely different manifestations of the same phenomena. In this case, the varying features of radiation would simply be a result of the initial conditions. Greater statistics and more comprehensive observations will be needed in future to develop our understanding of these events.

## 8.4   Relation to Upward Lightning Development

The four events of this study all took place during downward negative lightning; TGFs A–C terminated in negative cloud-to-ground strokes, while TGF D was a low-level, negative intracloud flash. Although the TASD can only detect downward-directed TGFs, the INTF and sferic sensors still record data of upward lightning. The parent flashes of TGFs B and C were each immediately followed by upward positive IC flashes less than 2 minutes later (see Figure 8.3 and Figure 8.4 for comparison). The close proximity in time between each of these flashes suggests that the overall storm structure has not changed drastically, and therefore that the difference in each subsequent flash is due only to the upward development.

### 8.4.1   Intracloud Flash Observations

The first two milliseconds of breakdown is shown for each flash in Figure 8.5 and Figure 8.6, clearly highlighting their different characteristics. The more continuous nature of breakdown and accompanying IBPs is expected in –CG flashes as compared to upward +IC, and is largely due to a difference in altitude (after 1 ms of development, both downward flashes are at ∼3 km above ground and upward flashes are at ∼6.5 km). The larger charge separation and lower pressure in the upper regions of typical storm clouds result in fewer, longer, and more intermittent leader steps than those of low-altitude –CG flashes (Stock (2014); Edens et al. (2014)); INTF points associated with each grouping of IBPs can be distinguished as separate, linear segments in the upward development. This is more clearly seen in the 2 ms-timescale plots of Figure 8.5 and Figure 8.6 when compared to the relatively consistent downward progression during IBPs in the –CG flashes.

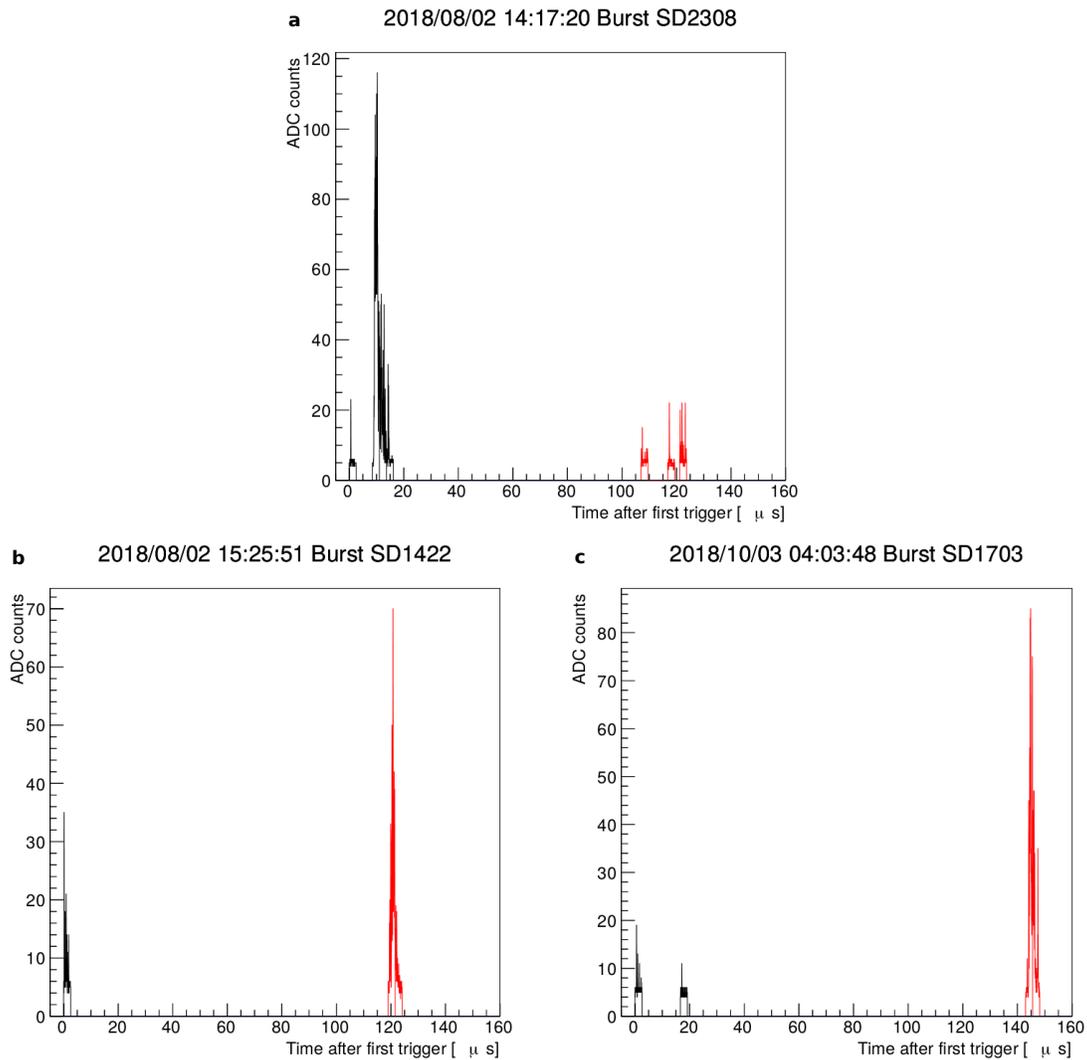

**Figure 8.2:** Combined waveforms for each of the TGFs that produced multiple triggers (TGFs A, C, and D). Signals given are from the detector with highest energy deposit. Individual durations are 5–10 $\mu$s with spacing between 10–120 $\mu$s.

Despite these cosmetic variations, the IBP-associated upward breakdown exhibits many of the same phenomena explored in section 8.1. Namely, the IBPs are each correlated with periods of higher-power VHF activity, strong sub-pulses, and fast negative breakdown. Above, the early breakdown in –CG flashes was described as consistent, but a closer look reveals that this actually consists of many distinct, closer-spaced IBPs, each with its own distinct interval of FNB. These are resolved in the 240 $\mu$s plots of Figure 8.7 and Figure 8.8, which again compare the activity surrounding the strongest IBP for both the upward IC and downward CG flashes. Just as the IBPs are more greatly separated, so are the sub-pulses themselves. In the upper panel (downward CG) of Figure 8.7, for example, sub-pulses in the main IBP are barely visible while those of the lower panel (upward IC) appear as separate, strong peaks. In fact, they are so separated that one might be tempted to characterize these as a group of three separate IBPs. However, the discerning feature of an IBP is the characteristic 'overshoot', or the opposite-polarity field change signaling a decrease in VHF amplitude and cessation of FNB. This occurs only after the final peak, defining this signals as a group of sub-pulses during a single IBP. In Figure 8.8, sub-pulses in the lower panel (upward IC) appear more obviously as part of the overall IBP and sub-pulses in the sharp IBP of the upper panel (downward CG) are virtually imperceptible due to the short duration (they can be identified in Figure 7.22 and panel d of Figure 7.15). Again, both instances are followed by a strong, opposite-polarity overshoot.

### 8.4.2   Intracloud Flash Interpretation

The comparison plots above (Figure 8.5–Figure 8.8) show that while upward and downward-developing IBPs may have different characteristics, they share the defining features associated with the production of TGFs in this study, namely that both types of IBP are produced by fast negative breakdown. We have also shown that the IBPs in IC flashes have longer durations and spatial extents due to their higher altitude and location within a storm's charge structure. In Figure 8.7, for example, the durations of both the IBP sferic and associated fast negative breakdown of the TGF-producing flash were about half that of the IC flash. A similar average ratio was reported in a study of 72 IC and CG flashes in Florida (Smith et al. (2018)). These differences in duration and spacing are also observed in leader steps (Edens et al. (2014)), strengthening the connection between IBPs and early leader steps. Also note from the azimuth-elevation plots that the strongest IBPs in upward flashes correspond to the transition between linear development and leader branching just as in the strong IBPs of downward flashes associated with TGF production. This reinforces the idea that the strong IBPs signal a maturing of leader development which can facilitate TGF production in both upward and downward-propagating lightning (section 8.5).

The relationship between IBPs in upward and downward lightning mimics that of upward and downward TGFs, in that they share many defining characteristics, but those associated with upward IC flashes tend to last longer. In fact, from section 8.3, the more common durations seen in upward TGFs of 20–200 $\mu$s are surprisingly well represented by the IBP cluster durations in the lower panel of Figure 8.6, each of which was accompanied by periods of continuous FNB. Recent ASIM observations of a TGF detected simultaneously by Fermi also appear very similar to this plot, consisting of 5 TGF pulses occurring over 2 ms and indicating their consistency with upward leader steps (Østgaard et al. (2019)).

Also discussed in section 8.3 is the discrepancy of inferred fluence values. The greater strength and impulsivity of sub-pulses, coupled with the longer durations of FNB, suggest that if the IC flashes were TGF producers, the resulting showers would be stronger and more developed than those of TGFS B and C, perhaps enough to produce the high photon fluences implied by satellite observations. These ideas pertaining to TGF production as a whole are explored in section 8.5. Unfortunately, the relatively low rate of storms (and flashes) over Telescope Array make it unlikely that an upward TGF from within the INTF's field of view would be detected by satellite, but we have begun searching for just such an occurrence in published ASIM events in hopes of providing detailed interferometry data of a confirmed upward-TGF-producing flash. In summary, although they have varying features based on altitude, these results support the idea that satellite-detected TGFs and those observed by TASD are produced by the same mechanism in leader steps, signaled by strong IBPs and fast negative breakdown.

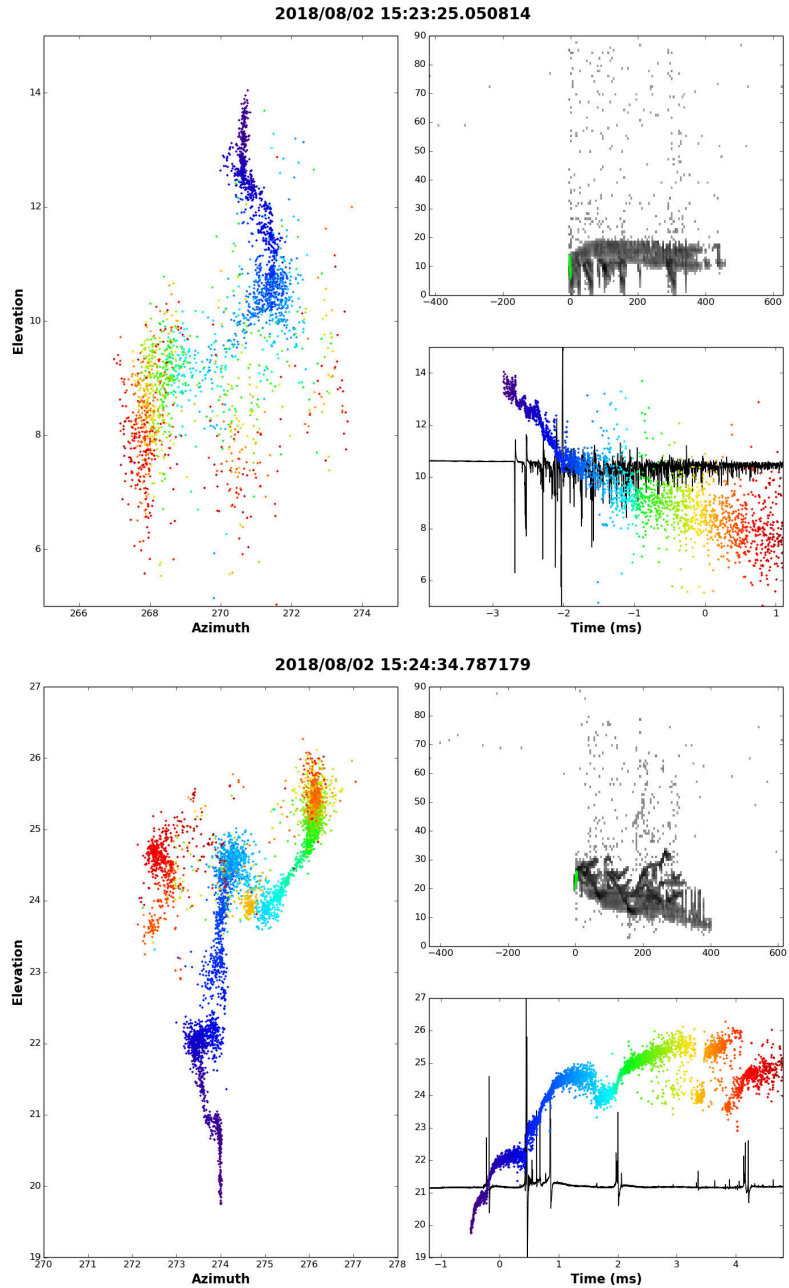

**Figure 8.3:** INTF data from the first few milliseconds of TGF B (upper panels) and a subsequent upward intracloud flash (lower panels), showing how breakdown was initiated. Point color represents time, as seen in the lower-right panel of each set. Fast antenna sferics are included in this panel, showing their spacing and correlation with steps in the leader development. The left panels of each set show the azimuth-elevation progression of the same initial few milliseconds of data, while the upper-right panels show all INTF activity from each entire flash.

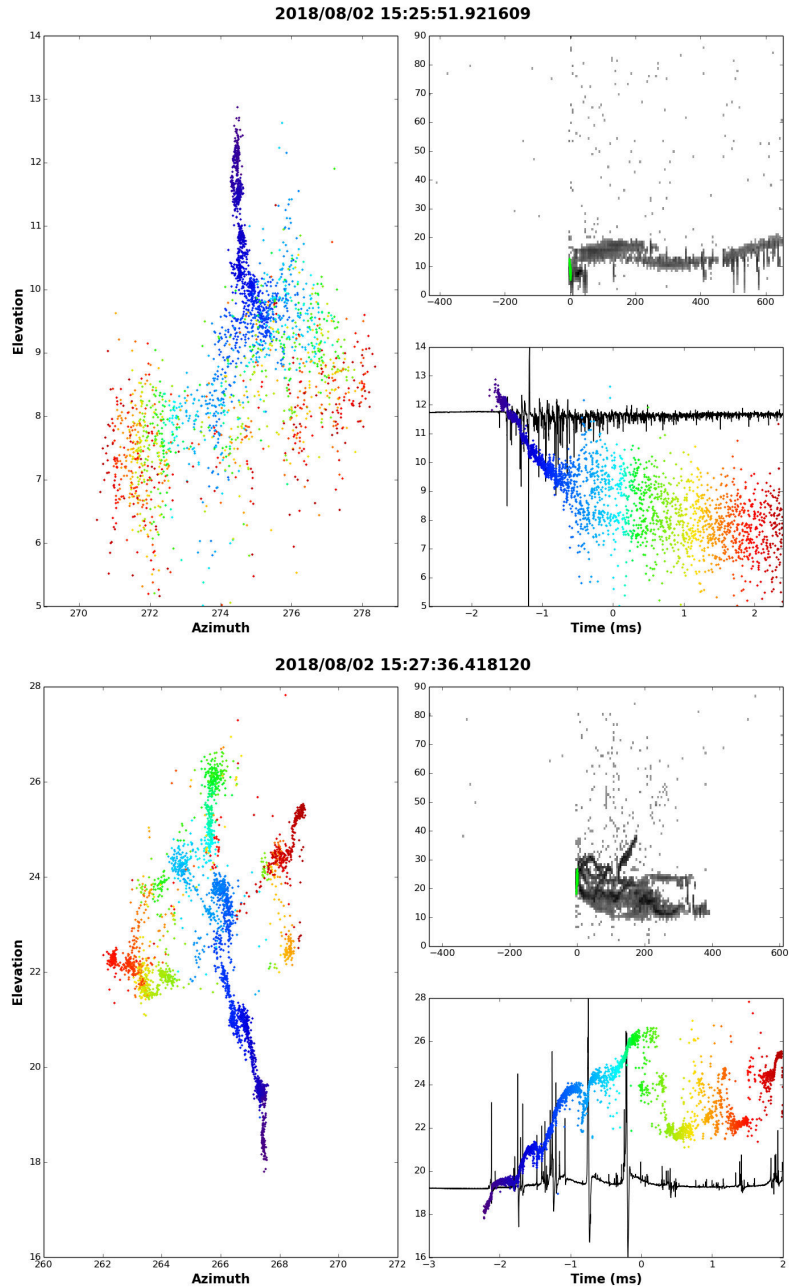

**Figure 8.4:** INTF data from the first few milliseconds of TGF C (upper panels) and a subsequent upward intracloud flash (lower panels), showing how breakdown was initiated. Point color represents time, as seen in the lower-right panel of each set. Fast antenna sferics are included in these panels showing their spacing and correlation with steps in the leader development. The left panels of each set show the azimuth-elevation progression of the same initial few milliseconds of data, while the upper-right panels show all INTF activity from each entire flash.

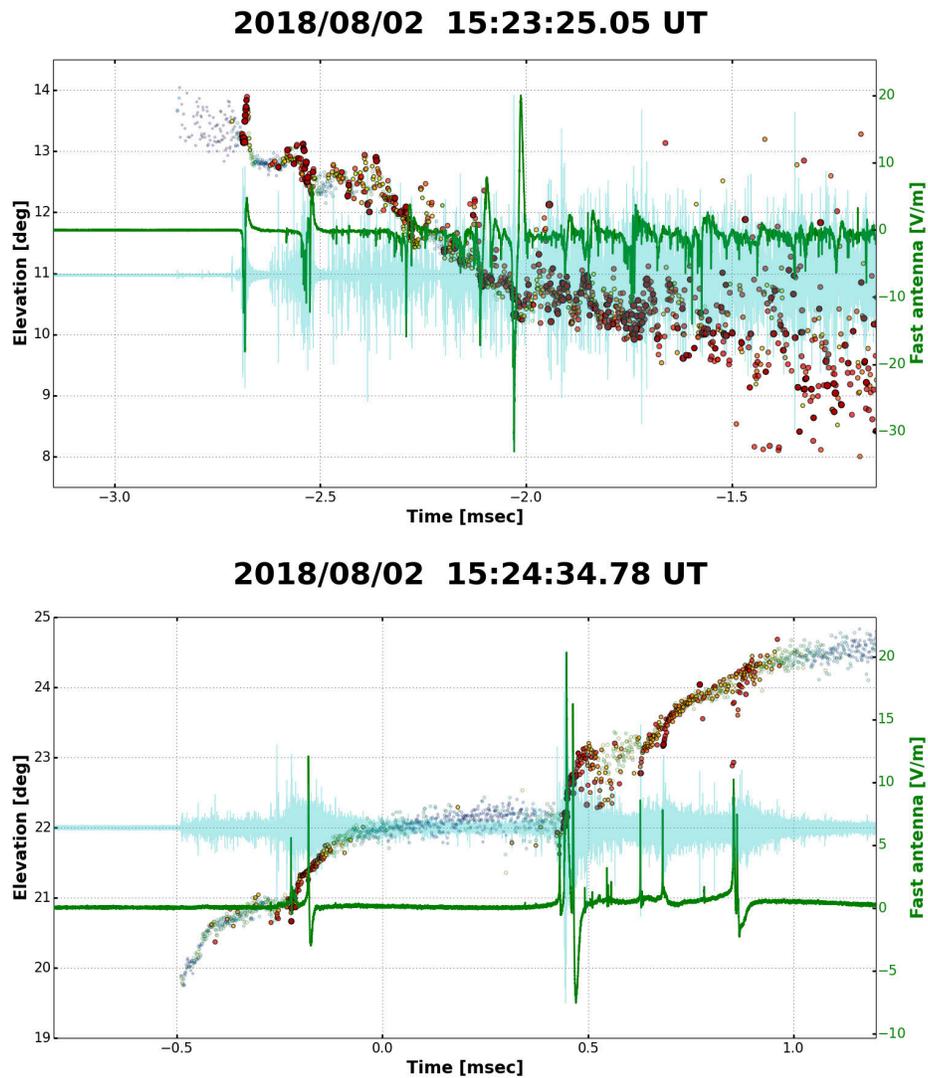

**Figure 8.5:** Zoomed-in views of the IBPs and associated VHF activity of the downward cloud-to-ground flash which produced TGF B and a subsequent upward intracloud flash from Figure 8.3. The cyan waveform is the raw VHF signal recorded by the INTF, with circles representing point solutions to that signal (color, size, and opacity referring to relative power). The green curve is the electric field change detected by the fast sferic sensor. The comparison shows the long durations and large spacing of IC IBPs versus the more continuous (closer spaced) IBPs of the CG flash. The steps during the CG IBPs averaged 180 m in length and lasted <10 $\mu$s each. The IC IBPs and related steps were more complex, with steps averaging 350 m and of varying durations. Steps from both flashes had peak speeds of 1–3×10$^7$ m/s.

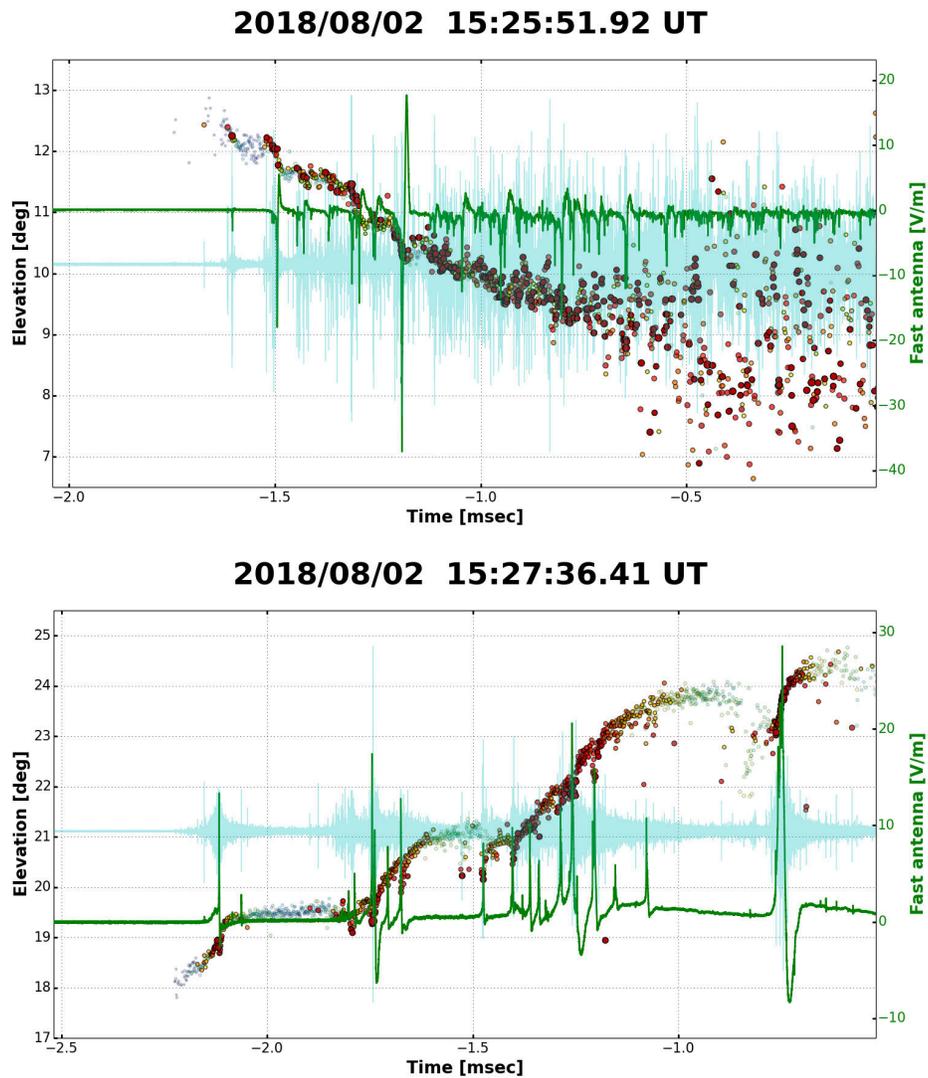

**Figure 8.6:** Zoomed-in views of the IBPs and associated VHF activity of the downward cloud-to-ground flash which produced TGF C and a subsequent upward intracloud flash from Figure 8.4. The cyan waveform is the raw VHF signal recorded by the INTF, with circles representing point solutions to that signal (color, size, and opacity referring to relative power). The green curve is the electric field change detected by the fast sferic sensor. The comparison shows the long durations and large spacing of IC IBPs versus the more continuous (closer spaced) IBPs of the CG flash. The steps during the CG IBPs averaged 140 m in length and lasted <10 μs each. The four IC steps varied, with lengths of ∼400, 570, 1000, and 450 m and lasting ∼20, 130, 300, and 25 μs, respectively. Steps from both flashes had peak speeds of 1–3×10⁷ m/s.

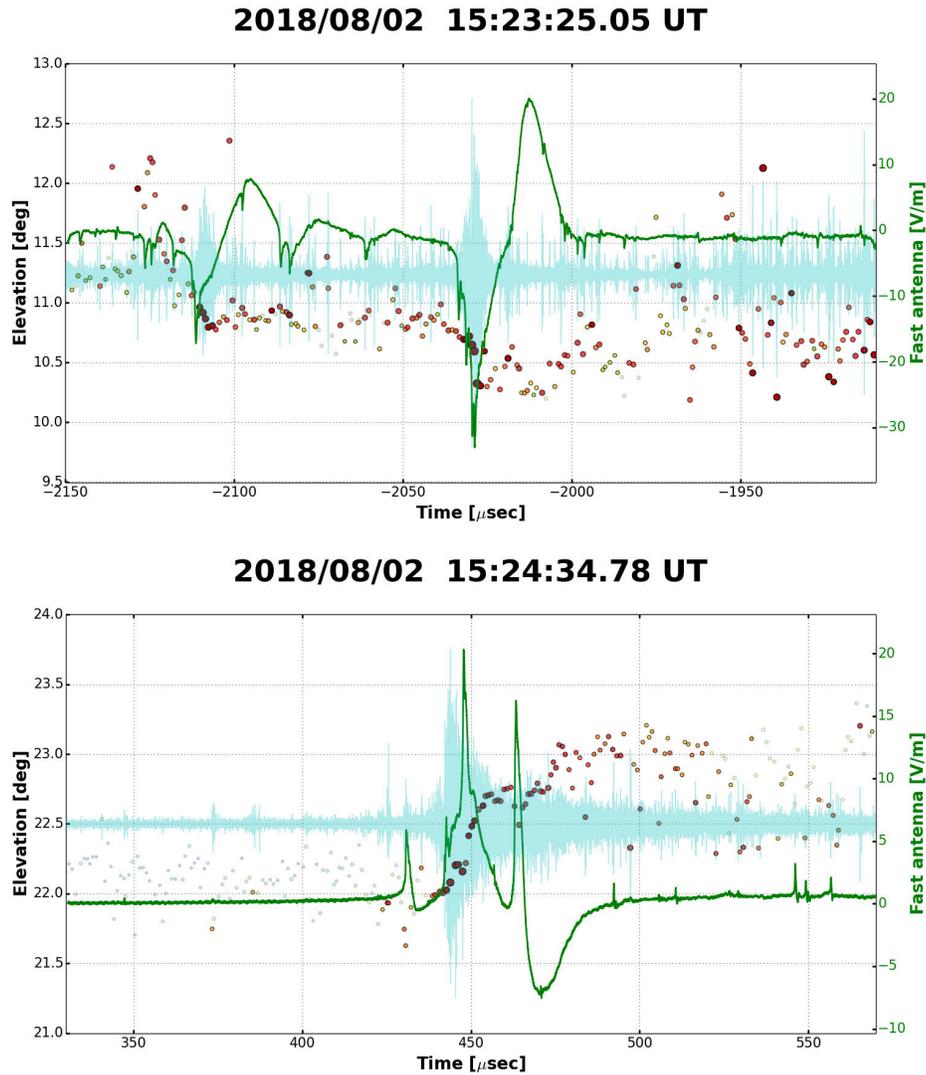

**Figure 8.7:** Further zoomed-in views of the strongest IBP and associated VHF activity of the downward cloud-to-ground flash which produced TGF B and a subsequent upward intracloud flash from Figure 8.3. The cyan waveform is the raw VHF signal recorded by the INTF, with circles representing point solutions to that signal (color, size, and opacity referring to relative power). The green curve is the electric field change detected by the fast sferic sensor. The comparison shows the relative strength and impulsiveness of the IC IBP and sub-pulses compared to those of the CG flash which initiated TGF B. In addition, the IC sub-pulses are further separated from the IBP peak, occurring both before and after, while those of the CG flash all occur along the IBPs leading edge. Both IBP peaks are correlated with strong VHF power, fast negative breakdown, and terminate in a strong, opposite-polarity field change. The IC sferic lasted $\simeq 70$ $\mu$s, twice as long as the $\simeq 35$ $\mu$s CG sferic.

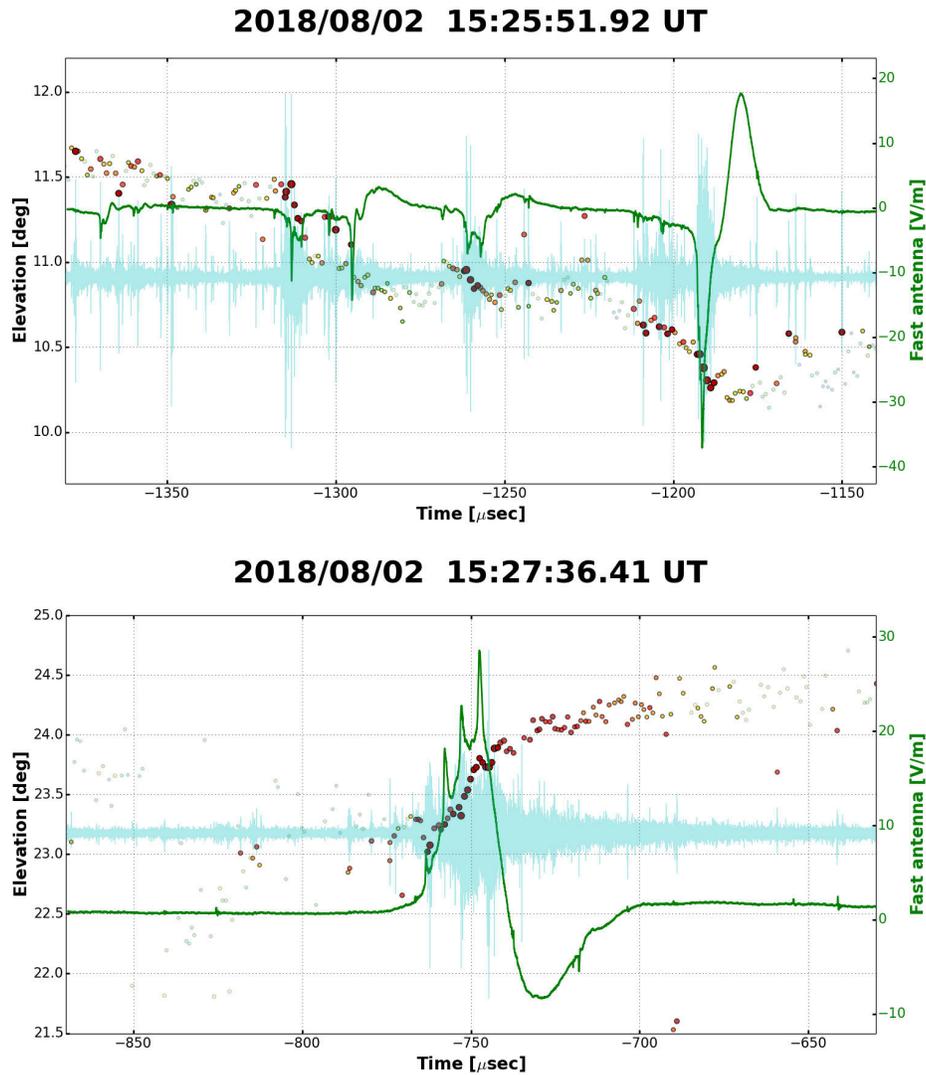

**Figure 8.8:** Further zoomed-in views of the strongest IBP and preceding VHF activity of the downward cloud-to-ground flash which produced TGF C and a subsequent upward intracloud flash from Figure 8.4. The cyan waveform is the raw VHF signal recorded by the INTF, with circles representing point solutions to that signal (color, size, and opacity referring to relative power). The green curve is the electric field change detected by the fast sferic sensor. This comparison emphasizes the difference in IBP duration and sub-pulse strength. The sub-pulses along the leading edge of the CG IBP are too subtle to be seen on this time scale, but are visible in Figure 7.22. The IC sferic is a near-perfect analog of the IBP correlated with TGF B, except being of longer duration and opposite polarity. Note that even the weaker sferic and FNB in the upper panel at –1,315 μs has very strong VHF peaks and was associated with the early gamma burst of TGF C (panels c and f of Figure 7.15). The IC sferic lasted nearly twice as long as that of the CG flash, being ∼70 and 40 μs, respectively.

## 8.5  TGF production

TGF studies over the past couple decades have proposed a few general theories for TGF production and development, as discussed in section 3.2. The most prevalent of these is RREA, in which energetic seed electrons cascade in the large-scale electric fields of thunderstorms (subsection 3.2.1). Early studies proposed that cosmic-ray air showers could provide these seed electrons, being relatively common at high altitudes (Daniel & Stephens (1974)). The Telescope Array is in a unique position to study this relationship, having detected both phenomena separately. While none of the TGFs of this study were detected in coincidence with ultra-high energy ($>10^{18}$ eV) cosmic rays, RREA seed electrons require only $\gtrsim$200 keV (depending on the ambient electric field) and could easily be produced by lower-energy showers that the experiment is not sensitive to.

An alternate source of seed electrons could be cold runaway, in which thermal electrons are accelerated by the exceptionally strong local electric fields ($\sim$2.4$\times10^7$ V/m) at the tips of negative leaders. Streamers form in the virgin air ahead of the existing leader, ionizing and heating the air to form the next leader step (section 2.3). Moss et al. (2006) shows that these streamers are capable of accelerating electrons to 2–8 keV or higher, sufficient for seeding RREA in these strong field regions.

Note that in the sferics presented of both downward CG and upward IC flashes, sub-pulses develop in IBPs until one causes the characteristic opposite-polarity field change, after which they and the associated FNB die out. Their strength and impulsivity suggest that these sub-pulses are a result of space leaders/space stems recombining in spark-like events, dubbed transient conducting events (TCEs) by Belz et al. (2020). When these disconnected space leaders eventually combine with the developed leader channel, the electric potential instantly shorts out and transfers into the new step, producing the opposite polarity overshoot characteristic of IBPs. Following this model, the recombination of leader and space leader could have been the source of TGFs A and B, which occur late in the FNB development during or after the main peak. By the same logic, the onsets of TGFs C and D each occur during strong sub-pulses earlier in FNB, suggesting they may have been produced by disconnected TCEs before recombination. Either way, results certainly indicate that the TGFs are seeded by runaway electrons during the streamer-to-leader transition and proceed to cascade into fully-fledged RREA.

The above interpretation of leader step formation is supported by optical observations of Biagi et al. (2010); Hill et al. (2011); Stolzenburg et al. (2013), in which bright spark-like events are seen ahead of the developed leader before traveling back into the leader tip during strong IBPs. While the latter's analysis did not consider TGFs, a further study of Stolzenburg et al. (2016) hypothesized the connection between TGF production and these visible, spark-like events. The results of section 8.1 show that (1): the TGFs of this study are produced during strong IBPs at the beginning of negative flashes, (2): that each IBP is accompanied (and likely produced) by instances of FNB during the streamer-to-leader transition, and (3): that IBPs grow in strength until reaching a maximum amplitude, at which point leaders tend to begin branching and IBPs become weaker. Considering that the strongest burst of gamma-rays from each TGF is also associated with the flash's strongest IBP, the combined results 1–3 above suggest suggest that early leaders mature until reaching a point that FNB is able to facilitate both TGF production and leader branching.

The FNB of these events propagated at speeds of $1.6$–$2.7 \times 10^7$ m/s, similar to that of fast positive breakdown (FPB), its reverse polarity analog. Rison et al. (2016) has shown that FPB is responsible for narrow bipolar events (NBEs) during early lightning breakdown as seen in TGF A's parent flash with an average speed of $1.7 \times 10^7$ m/s (Figure 7.17). Streamer modeling by Attanasio et al. (2019) shows that both FPB and FNB greatly intensify local electric fields, accelerating runaway electrons well into the regime of RREA (and therefore of TGF production). Although TGFs A–D take place at various points relative to their respective IBPs' main peaks, all take place during FNB, presumably in the regions of field intensification. Additionally note that once initiated, the TASD particle detections of each TGF in this study are sustained as long as the FNB continues. In subsection 7.2.3, we show that the TASD signals accurately reflect the source are not artificially extended by atmospheric scattering effects, suggesting that the TGF production is not sustainable without runaway seeding from active FNB. Since the resulting shower fronts propagate at the speed of light, the average 3–5 $\mu$s TGF durations imply that electron avalanches propagate ahead of the FNB over 1–1.5 km, many times the 100–240 m extent of FNB. This provides the avalanches with additional acceleration for as long as the ambient electric field stays above the avalanching threshold of $\sim 2.8 \times 10^5$ V/m (Dwyer et al. (2003)).

This proposed model of TGF production has some other interesting implications. From section 8.3, we note that the implied photon fluences of TGFs at TA are $10^1 2$–$10^1 4$, several orders of magnitude lower than those of satellite studies ($10^1 6$–$10^1 8$, section 3.1). The simulations of Celestin et al. (2015) rely on electron acceleration in the small local fields at leader tips, but the observations of sustained FNB over several microseconds at TA agree with the streamer-based field amplification of Attanasio et al. (2019). For TGFs A–D, the instances of FNB continued for 100–240 m (Table 7.3). Multiplied by the negative streamer stability field of $\sim 6 \times 10^5$ V/m required for streamer propagation at altitude (subsection 2.4.2), this results in a potential difference drop of 60–140 MV (corresponding to $10^14$–$10^16$ photons), many times higher than the 10–50 MV potentials ($10^12$–$10^14$ photons) implied by Abbasi et al. (2018).

Even under this consideration, implied fluences of TA events are a couple orders of magnitude below those of upward TGFs. Another popular model of TGF development involves relativistic feedback (subsection 3.2.2), originally developed to explain these extremely high fluences. In this model, traditional RREA is amplified even further in large-scale electric fields of thunderstorms. This happens via back-scattered positrons or photons which in turn seed new instances of RREA (Figure 3.2). Given electric fields large and stable enough, this process can amplify flux by a factor of up to $10^9$ (Dwyer et al. (2003)). The much lower inferred fluences of TASD events suggest that RFD does not play a dominant role in shower development. However, if satellite-detected TGFs do share the same production processes involving thermal runaway as suggested in section 8.4, their much greater fluences could be explained by the shower enhancement via RFD in the larger-scale fields of upward IC flashes. This would also extend shower durations as long as feedback showers continue to develop.

This integrated model of TGF production could additionally unite the x-ray observations of section 3.1. The spectra of these events are too soft to be a result of fully-developed RREA, but are within the range of thermally accelerated electrons enhanced by FNB. The similarities in lightning leader activity during production of x-ray observations, TASD downward TGFs, and satellite upward TGFs suggest that all could be sourced by cold runaway and be enhanced to varying degrees depending on the ambient thunderstorm conditions. Additional upgrades are underway at Telescope Array in order to further develop this overarching model of TGF production (section 9.3).

# CHAPTER 9

# CONCLUSION

The serendipitous discovery of TGFs at Telescope Array has demonstrated the potential of studying energetic radiation from lightning using cosmic ray detectors. Despite the difficulties of adapting analysis methods, these studies have provided a significant portion of all downward TGF observations and identified processes involved in lightning breakdown in general. Of course, these measurements have opened up a whole new set of questions, particularly in the differences between these and other observations.

This chapter summarizes the results of the 2018 TGF observations and their implications on our current understanding of atmospheric discharge as a whole. It shall additionally explore the discrepancies and unanswered questions that remain. The TASD study of lightning is expected to persist many years and, with more statistics, begin filling in the gaps in our current understanding of lightning and its energetic emissions. This will be continued in part by various detector upgrades and additions, briefly described in section 9.3.

## 9.1   Summary of Results

With the previous studies of Abbasi et al. (2017) and Abbasi et al. (2018), the strange burst events at Telescope Array were confirmed as downward TGFs. Current investigations have been able to take a closer look at the physical mechanisms taking place during TGF production and lightning discharge in general.

The four TGFs of 2018 had characteristics similar to those of previous TA studies. TGFs are produced in the first 1–2 milliseconds of breakdown activity, individually lasting $\leq$10 $\mu$s and separated by tens or hundreds of microseconds, though occasionally consisting of only a single gamma burst. The showers are typically sourced 2.8–3.2 km above ground and produce footprints on ground with diameters between 3–5 km, resulting in half-opening angles between 25–40°. The combined energy deposits are consistent with total fluences of $10^{12}$–$10^{14}$ photons, though there is still some uncertainty on the issue (see section 9.2). Individual photons have gamma-ray energies, with evidence of some photons having at least 6.4 MeV.

Detailed interferometer data, coupled with the analysis of section 7.2, show that the strongest gamma bursts in each downward TGF are produced during the strongest IBPs in the early stages of negative lightning flashes. The IBPs, in turn, are associated with periods of streamer-based fast negative breakdown during leader steps. These streamer systems produce strong, local electric fields which are able to greatly accelerate and eject electrons into the relativistic regime. In a sufficiently-strong ambient electric field, these electrons then seed RREA showers which make up the downward TGFs.

## 9.2 Unanswered Questions

Although we have built a nearly-complete model for downward TGF production, some aspects remain unclear. For example, these ground observations of TGFs differ in duration and fluence from those obtained via satellite. As discussed in section 3.1, upward TGFs have measured durations of 20–200 $\mu$s. Note that the more recent instruments have better timing resolution and also report shorter durations. A possible resolution to some cases lies in the study of Celestin & Pasko (2012), which concludes that a number of shorter TGFs are consistent with source durations of $\lesssim 10$ $\mu$s, consistent with TA observations.

As a whole, simulations show that the durations observed at TA are not significantly extended due to atmospheric scattering, and therefore reflect the source durations. Since the ground detections mimic the durations of FNB during strong IBPs, this suggests that the RREA seeding is continuously driven by the FNB streamer systems until they die out. This may help explain the discrepancy when considering the INTF observations of upward IC flashes discussed in section 8.4. In these examples which occurred in the same storm as TGFs B and C, FNB lasted tens or hundreds of microseconds. This IBP and breakdown activity is characteristic of upward, higher-altitude flashes, largely due to the decreased atmospheric pressure. If we assume that these were TGF-producing flashes and employ the same logic as for downward events, we would assume the resulting TGFs would also last tens or hundreds of microseconds, consistent with many satellite observations. Unfortunately we cannot yet confirm this theory, but detailed interferometer observations of an upward, TGF-producing flash may be able to resolve this inconsistency in the future.

In fact, the explanation above may also apply to differences in fluence. Simulations based on upward TGF observations imply showers over a wide range of $10^{15}$–$10^{19}$ photons (section 3.1), while those at TA imply $10^{12}$–$10^{14}$ photons (subsection 6.2.2). As above, the extended duration of upward FNB and IBPs may account for the much higher fluences, but it is no so simple as that. The nature of TGF observations means that these values are always model dependent and not directly observable. One issue with common models is the dependence on available potential. Simulations of Celestin et al. (2015), for example, allow gamma detectors to infer discharge potential and TGF fluence based on measured photon energies. According to the study, a photon of 20 MeV (not uncommon for upward detections), requires at least a 160 MV drop in potential. This is correlated with a fluence of at least $4 \times 10^{16}$ photons.

In the case of Telescope Array, inferred fluences of $10^{12}$–$10^{14}$ photons would be produced by potential drops between 10–50 MV. Resulting photons should subsequently have energies of $\lesssim 700$ keV. Even though the TASDs are not designed to measure exact energies of incident particles, we do show that there is evidence of photons of at least a few MeV, one in particular having a minimum of 6.4 MeV, implying a potential drop of $>60$ MV and fluence of $>6 \times 10^{14}$ photons. These values are more consistent with the observed periods of FNB (see section 8.5) and introduce some overlap between downward and upward TGFs. Although these complicated inconsistencies are important to discuss, it should also be noted that fluence enhancement may simply be a result of relativistic feedback which has more room to develop in upward TGFs. The mechanism was originally theorized to resolve this very issue (subsection 3.2.2).

## 9.3  Future Study

In the effort of pursuing these (and other) pieces of the puzzle, additional upgrades to both cosmic ray and lightning instrumentation are underway to continue advancing the capabilities of the Telescope Array. New lightning detectors include a second, more robust interferometer (subsection 9.3.2), electric field mill (subsection 9.3.3, and high-speed optical camera (subsection 9.3.4). In addition, the TA collaboration has been planning and implementing a significant expansion during the analysis of this project (subsection 9.3.1). Just as the original TASD's large area incidentally cast a wide net for catching TGFs, the expansion is expected to vastly increase the statistics of both UHECRs and lightning discharges for further study.

### 9.3.1  Telescope Array Expansion

At the end of 2018, the Telescope Array expansion began installing new surface detectors to fill the northern and southern lobes of the dubbed "TAx4" (Figure 9.1). As the name suggests, the expansion is meant to quadruple the total area of TA. In the Northern and Southern lobes, detector spacing has been doubled up to 2.4 km in order to get as much total coverage as possible. Four additional control towers and two fluorescence stations were also installed at locations shown in Figure 9.1 for the support and enhancement of TAx4. Data of the current study was recorded only by the original Telescope Array surface detectors — further information on the expansion can be found in Kido (2020); Abbasi et al. (2021c).

As far as its effect on TGF studies, the TAx4 expansion will be greatly beneficial. The existing INTF site just East of the main TA array (Figure 4.1) is positioned such that its range includes much of the expansion, though its detection efficiency will need to be tested for the most distant Southern and Northern extents. New LMA stations will also be deployed in the new lobes to give coverage over the entire expansion.

### 9.3.2   Second Broadband Interferometer

One substantial upgrade to the TGF study is that of a second, improved interferometer (Figure 9.2). The circle of seven antennas was temporarily installed outside of LRFD at the Southwestern end of the TASD in summer of 2020. Unfortunately, no TGFs were detected in the few months it was active. It was reinstalled June 2021 in a lesser capacity as a smaller, approximately equilateral triangle and is planned to be restored to its full configuration in 2022. The full configuration has greatly improved resolution and detection efficiency when operated individually, but can also be used in conjunction with the original INTF.

As discussed in section 7.3, the most significant contributor to uncertainty in this study's analysis is that of horizontal position, due largely to the LMA's lower resolution and data rate. With the addition of a second interferometer, any lightning activity detected by both stations could be tracked fully in three dimensions using the stereo method common in cosmic ray fluorescence analysis (Tameda et al. (2009)). For example, the uncertainty in horizontal distance between TGF A and the interferometer station was ∼150 m when relying on LMA data. Using the typical INTF angular error of 0.1° at a distance of 17 km, the result is ∼30 m, a factor of five improvement. Of course the true stereo location method is more complicated, but the benefits are clear.

### 9.3.3   Electric Field Mill

A new study at Telescope Array combines cosmic ray observations with lightning research in the form of cosmic ray rate variations due to overhead thunderstorms (Abbasi et al. (2021a)). Preliminary work shows fluctuations in the number of TASD triggers as storms pass above. While this is understood to be caused by fluctuations in the ambient electric field, TA can currently only monitor storm movement by cross-referencing LMA data, NLDN-recorded flashes, or public radar data. A specialized detector called an electric field mill (EFM) can measure these fields from ground level.

EFMs have been used for measuring thunderstorms for over a century, and remain useful to this day (Pellat (1890); Antunes de Sá et al. (2020)). The device consists of multiple separated plates which record voltage similar to a sferic sensor. In an EFM, however, the plate is alternated between being exposed and shielded, causing it to charge and discharge many times per second. This is typically done with a rotating, grounded plate. After processing, this results in a measurement of the ambient electric field, rather than the field change recorded by sferic sensors. An EFM of this design was recently added to the array in the summer of 2021 in an effort to better understand the effect of ambient electric fields on the propagation of cosmic ray extensive air showers observed by the Telescope Array.

### 9.3.4   Optical Camera

One of the earliest forms of lightning observation comes in the form of streak photography (Schonland & Collens (1934); Malan et al. (1935)) in which a revolving lens exposes stationary film, originally used to show the development of leaders prior to a flash's return stroke. Modern digital photography has more recently been useful for similar purposes (Biagi et al. (2010); Hill et al. (2011); Stolzenburg et al. (2013)). These studies give insight into the breakdown process of individual leader steps. In particular, they show that in early leader steps, luminosity first develops ahead of the existing channel in the form of space stems before propagating backward and reconnecting with the leader (section 2.3, Figure 2.3). Results of this study support these conclusions, and are discussed further in section 8.5.

Installation of a high-speed optical camera is currently underway at the INTF station at Telescope Array (Figure 5.7). The camera will be directed Westward from the on-site building overlooking much of TASD. At 40,000 frames per second, it will be able to resolve the development of individual leader steps which occur over several tens of microseconds. Alongside the existing interferometer, sferic sensors, and surface detectors, these observations will help to clarify the relationship between TGF production and the leader breakdown process.

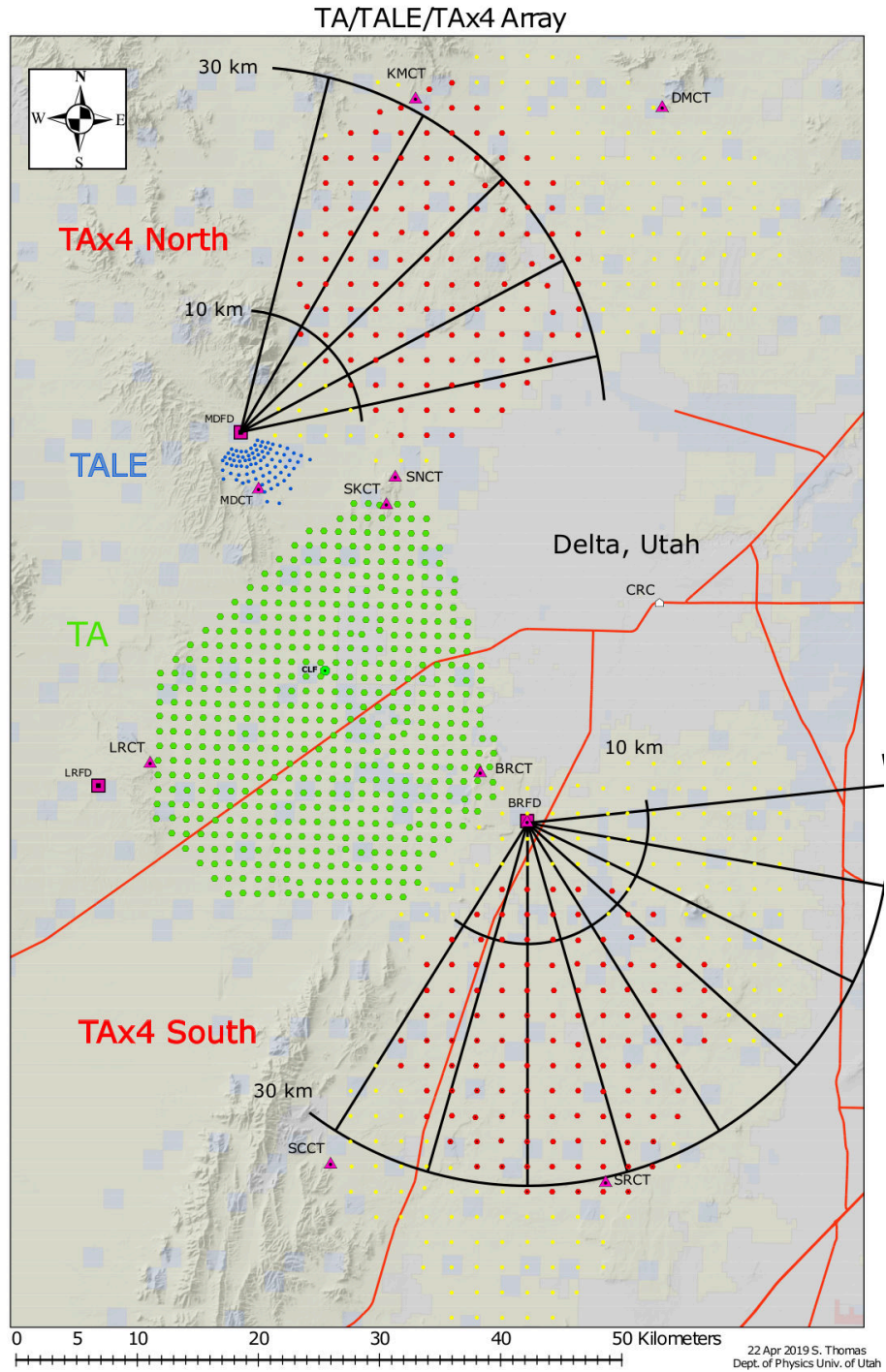

**Figure 9.1:** Map of the Telescope Array expansion. The original TASD (TA) is given in green with control towers (XXCT) as purple triangles and fluorescence detectors (XXFD) as purple squares. The smaller low energy extension (TALE) is shown in blue and the newly-deployed TAx4 SDs in red. Yellow points mark the planned sites of TAx4 detectors which have not yet been deployed. Additional telescopes were added to two of the existing fluorescence stations MDFD and BRFD with fields of view overlaid. The points labeled CLF and CRC represent the Central Laser Facility and Cosmic Ray Center, respectively.

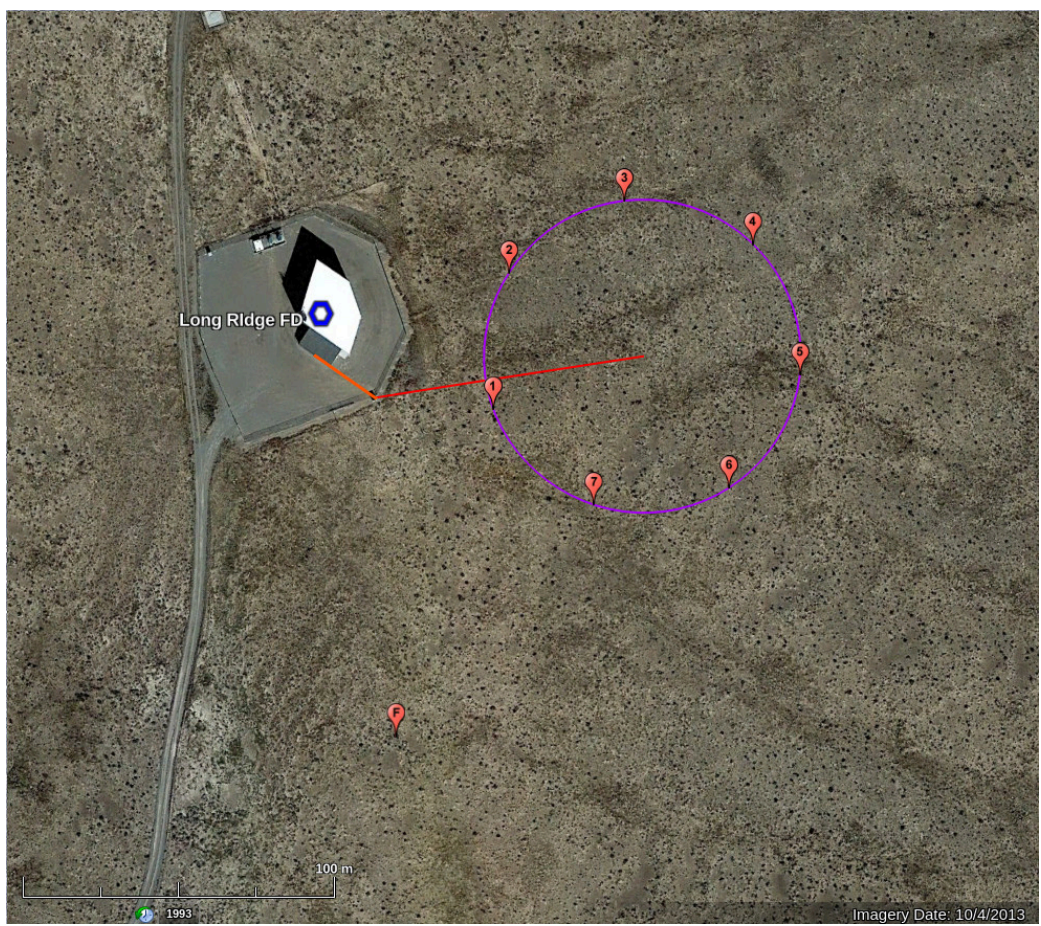

**Figure 9.2:** Overview of Long Ridge fluorescence station (LRFD) with overlaid locations of antennas (labeled 1–7) which make up the new interferometer (INTF2) installed in 2020. The nearby LMA station is also labeled 'F' south of LRFD.